\documentclass[aps,prx,floatfix,showpacs,twocolumn,nofootinbib]{revtex4-1}
\usepackage{dcolumn}
\usepackage{bm}
\usepackage{color}
\usepackage{amssymb}
\usepackage{amsmath}
\usepackage{graphicx}
\usepackage{amsfonts}
\usepackage{slashed}
\usepackage{pstricks}
\usepackage{float}
\usepackage{hyperref}
\usepackage{array}
\allowdisplaybreaks
\usepackage{dcolumn}
\usepackage{epsf}
\usepackage{float}
\usepackage{afterpage}
\usepackage[latin1]{inputenc} 

\begin{document}
\title{\textit{Ab initio} study of $\boldsymbol{(\nu_\ell,\ell^-)}$ and $\boldsymbol{(\overline{\nu}_\ell,\ell^+)}$ inclusive scattering in $^{12}$C: confronting the MiniBooNE and T2K CCQE data}
\author{
A.\ Lovato$^{\, {\rm a,b} }$,
J.\ Carlson$^{\, {\rm c} }$,
S.\ Gandolfi$^{\, {\rm c} }$,
N.\ Rocco $^{\, {\rm a, d}}$,
and R.\ Schiavilla$^{\, {\rm e,f} }$
}
\affiliation{
$^{\,{\rm a}}$\mbox{Physics Division, Argonne National Laboratory, Argonne, IL 60439}\\
$^{\,{\rm b}}$\mbox{INFN-TIFPA Trento Institute of Fundamental Physics and Applications, 38123 Trento, Italy}\\
$^{\,{\rm c}}$\mbox{Theoretical Division, Los Alamos National Laboratory, Los Alamos, NM 87545}\\
$^{\,{\rm d}}$\mbox{Theoretical Physics Department, Fermi National Accelerator Laboratory, P.O. Box 500, Batavia, Illinois 60510, USA}\\
$^{\,{\rm e}}$\mbox{Theory Center, Jefferson Lab, Newport News, VA 23606}\\
$^{\,{\rm f}}$\mbox{Department of Physics, Old Dominion University, Norfolk, VA 23529}
}
\date{\today}
\begin{abstract}
We carry out an {\it ab initio} calculation of the neutrino flux-folded inclusive cross
sections, measured on $^{12}$C by the MiniBooNE and T2K collaborations in the
charged-current quasielastic (CCQE) regime.  The calculation is based on
realistic two- and three-nucleon interactions, and on a realistic nuclear electroweak
current with  one- and two-nucleon terms that are constructed consistently
with these interactions and reproduce low-energy electroweak transitions.  Numerically exact quantum Monte Carlo methods are
utilized to compute the nuclear weak response functions, by fully retaining
many-body correlations in the initial and final states and interference effects
between one- and two-body current contributions.  
We employ a nucleon axial form factor of the dipole form with $\Lambda_A = 1.0
$ or $1.15$ GeV, the latter more in line with a 
	very recent lattice QCD determination.  The calculated
cross sections are found to be in good agreement with the neutrino data of
MiniBooNE and T2K, and antineutrino MiniBooNE data, 
yielding a consistent picture of nuclei and their electroweak properties across a wide regime of energy and momenta.

\end{abstract} 
\index{}\maketitle

\section{Introduction}

There is a large program of accelerator neutrino experiments in operation or in the
planning phase in the US and elsewhere to measure the parameters that characterize the
probabilities for flavor oscillations of these particles---mass differences, mixing angles, and the charge-conjugation
and parity violating phase.  These experiments do not directly measure oscillation probabilities, of course,
but rather event-rate distributions as  function of the observed energy $\overline{E}$ in the detector,
schematically
\begin{equation}
\label{eq:e1}
N^\alpha_\beta(\overline{E})\!\propto\!
 \!\int dE_\nu\, \phi_\alpha(E_\nu)\, P({\nu_\alpha \rightarrow \nu_\beta};E_\nu)\,\sigma_\beta(E_\nu,\overline{E}) \ ,
\end{equation}
where $\phi_\alpha(E_\nu)$ is the flux of neutrinos of flavor $\alpha$ ($\nu_\alpha$'s) at the source as function of
energy $E_\nu$, $P({\nu_\alpha \rightarrow \nu_\beta};E_\nu)$ is the probability for oscillation of a $\nu_\alpha$ into a
flavor $\nu_\beta$, and $\sigma_\beta(E_\nu,\overline{E})$ is the $\nu_\beta$-nucleus cross section. The
neutrino energy is reconstructed from the tracks in the detector of the outgoing lepton in an inclusive scattering
setting, and, additionally, the tracks of final hadrons in a semi-inclusive one.  As a consequence,
the determination of oscillation parameters depends strongly on neutrino interaction physics, since the interactions
observed in the detector result from the folding of the energy-dependent neutrino flux, energy-dependent cross
section, and energy-dependent nuclear (strong- and electroweak-interaction) effects.  

The appreciation of these difficulties has led, in the last decade or so, to a flurry of activity by nuclear theorists,
who have attempted to provide accurate estimates for neutrino-nucleus
($\nu$-$A$) inclusive (and semi-inclusive) cross
sections (for a summary of efforts in this area see Ref.~\cite{Alvarez-Ruso:2017oui}).  This is a very challenging task, primarily,
because neutrino fluxes in current (such as MiniBooNE, T2K, MicroBooNE, and Miner$\nu$a)
and future (DUNE) experiments extend over a rather wide energy range,
from threshold to, in several cases, multi-GeV energies.  Thus, observed $\nu$-$A$ cross sections,
resulting from the folding in Eq.~(\ref{eq:e1}), may include contributions from energy- and momentum-transfer
regions of the nuclear weak response where drastically different dynamical regimes are at play, from the
structure and collective behavior of low-lying nuclear excitations in the threshold region, to the quark
substructure of individual nucleons in the deep inelastic region.   Moreover, for some of the nuclear
targets employed in the detectors of these experiments, such as $^{40}$Ar (MicroBooNE), and $^{56}$Fe
and $^{208}$Pb (Miner$\nu$a), the full structure of the ground states is difficult to calculate exactly. 

Theoretical studies have attempted to provide a description of the nuclear weak response in this wide range
of energy and momentum transfers. They typically rely on a relativistic Fermi gas (RFG)~\cite{VanOrden:1978,Alberico:1988bv,VanOrden:1980tg} or relativistic mean
field (RMF)~\cite{Walecka:1974qa,Amaro:2006if,Amaro:2011qb,Gonzalez-Jimenez:2019qhq} picture of the nucleus.  
Some, notably those of Refs.~\cite{Martini:2009uj,Martini:2010ex,Nieves:2004wx,Nieves:2005rq,Nieves:2011}, include correlation effects in the
random-phase approximation (RPA) induced by effective particle-hole interactions in the $N$-$N$, $N$-$\Delta$,
$\Delta$-$N$, and $\Delta$-$\Delta$ sectors, use various inputs from pion-nucleus phenomenology, and lead
to predictions for electromagnetic and strong spin-isospin response functions of nuclei, as measured, respectively,
in inclusive electron scattering and in pion and charge-exchange reactions, in reasonable agreement with data.
Some utilize the phenomenological SuperScaling (SuSA)
approach---scaling with nuclear mass number~\cite{Barbaro:2006}---in which
a universal scaling function, derived from analyses of $(e,e^\prime)$ data on a number of nuclei,
is used to obtain estimates for the corresponding $\nu$-$A$ cross sections~\cite{Amaro:2013yna,Amaro:2011aa}.
Recently, SuSA, which has proven to be quite successful,
has been extended (SuSAv2) by incorporating elements from the RMF approach to account for
differences between the vector and axial components of the weak current, and between their
isoscalar and isovector content~\cite{Gonzalez-Jimenez:2014eqa,Megias:2016lke,Megias:2017cuh}.
Yet others rely on factorization of the hadronic final state and realistic spectral functions $S(p_m,E_m)$
to combine an accurate description of the nuclear ground state with relativistic currents and kinematics. 
Spectral functions of atomic nuclei are calculated either with microscopic methods---for example, the self-consistent Green's function
technique~\cite{Dickhoff2004ppnp,Barbieri2014QMBT,Rocco:2018vbf,Barbieri:2019ual}---or by combining inputs from $(e,e^\prime p)$ data 
to characterize the low missing-momentum and missing-energy region with accurate many-body calculations of the nuclear matter spectral 
function~\cite{Benhar:1989aw,Benhar:1991af,Benhar:1992cnb,Benhar:2006wy,Rocco:2015cil} folded with the single-nucleon density to 
describe the ``correlation region'', corresponding to high missing energies and momenta~\cite{Benhar:1994hw,Sick:1994vj}.

Many of the above models have achieved a remarkable phenomenological success and much improved agreement with experimental data,
compared to simple RFG calculations. However, it is fair to note that they rely on a somewhat approximate description of
nuclear dynamics that does not fully capture correlation effects in both the initial and final states
and does not generally use as inputs realistic nuclear interactions and consistent electroweak currents.
Hence, it is important to carefully assess their validity---especially in the
axial sector---by testing them against more microscopic calculations.


In the present study we report on an {\it ab initio} calculation of the nuclear weak response induced by
charged-current (CC) $(\nu_\ell,\ell^-)$ and $(\overline{\nu}_\ell,\ell^+)$ processes.  The strong
interactions among nucleons are represented by two- and three-body terms, while their
coupling to the electroweak field is accounted for by one- and two-body currents
(see the reviews~\cite{Carlson:1998,Carlson:2015} and references therein).  The two-body
interaction~\cite{Wiringa:1995} is constrained by fits to the nucleon-nucleon ($NN$) database up to lab energies of
350 MeV (albeit it provides a good description of the $NN$ cross section well beyond the pion production
threshold, up to 500 MeV or so).  The three body interaction~\cite{Pieper:2008} is calibrated by a fit to the energies
of a number of low-lying nuclear states in the mass range $A\,$=$\,$3--10.  

The one-body currents follow from a non-relativistic expansion of the covariant single-nucleon CC, including
nucleon electroweak form factors consistent with available experimental data.  In particular, results
reported in Sec.~\ref{sec:sec3} are obtained by using a dipole axial form factor with cutoff $\Lambda_A$ equal to either
1.0 GeV or 1.15 GeV.  The former has been extracted from
proton and deuteron experiments~\cite{Baker:1981su,Miller:1982qi,Kitagaki:1983px,Ahrens:1986xe}, while the
latter is obtained by recent lattice QCD calculations~\cite{Bhattacharya:2019,Park:2020axe}
that also reproduce the vector form factors measured in electron scattering. 

Two-body currents are derived
from meson-exchange phenomenology including pion and $\rho$-meson exchanges as well as $N$-to-$\Delta$
transition currents (with $\Delta$'s taken, however, in the static limit)~\cite{Shen:2012}.  The short-range behavior of these
currents is prescribed to be consistent with that of the two-nucleon interaction~\cite{Marcucci:2000xy, Marcucci:2005zc}.
In the vector sector, they
contain no free parameters, while in the axial sector the single unknown parameter present---the $N$-to-$\Delta$
axial coupling constant---is fixed by reproducing the experimental value of the tritium Gamow-Teller matrix element.

The theoretical framework outlined above (and discussed more expansively in Sec.~\ref{sec:sec2} below)
has been shown to provide, in numerically accurate quantum Monte Carlo (QMC) calculations, a quantitatively
successful description of a large body of experimental data on light nuclei ($A\leq 12$), including, among
others, energy spectra of low-lying states, static properties (magnetic and quadrupole moments),
low-energy radiative and weak transition rates, electromagnetic ground and transition form factors,
and electroweak dynamic response (for a review, see~\cite{Carlson:2015} and references therein).  Especially
relevant in the present context are the QMC studies of the $^{12}$C electromagnetic ground-state
structure~\cite{Lovato:2013}, and longitudinal and transverse response functions at intermediate momentum transfers
$q$ in the (300--700) MeV range, and for energy transfers $\omega$ in the quasielastic region~\cite{Lovato:2016gkq}.

However, it is also important to recognize the limitations inherent to the approach we have
adopted here: firstly, it addresses only inclusive scattering; secondly, it does not
account for explicit pion production mechanisms and therefore cannot describe the nuclear electroweak
response in the $\Delta$ resonance region and beyond; and thirdly, it relies on what is in essence a non-relativistic formulation
of the dynamics and electroweak currents.\footnote{Nevertheless, it could be argued that relativistic dynamical
effects are implicitly subsumed in the interactions, which are fitted to data; furthermore, the currents do include
corrections beyond the leading order~\cite{Shen:2012}.} 

These limitations notwithstanding, it should be emphasized that in the quasielastic regime
specified earlier, this approach includes all of the relevant physics for inclusive scattering and is expected to be quite accurate.
It is for this reason that we compare our predictions (in Sec.~\ref{sec:sec3}) for the $^{12}$C flux-averaged inclusive
cross sections---differential in the outgoing lepton energy and scattering angle---to the MiniBooNE and T2K
CC ``quasielastic'' (CCQE) data sets~\cite{AguilarArevalo:2010zc,Aguilar-Arevalo:2013dva,Abe:2016tmq}.
These data sets are characterized by the absence of pions in the
final state.  Clearly, their interpretation as purely ``quasielastic'' is complicated by pions that are created at the
interaction vertex and are subsequently reabsorbed in the nuclear medium~\cite{Leitner:2010kp}.  The
unambiguous identification of these contributions is problematic, and model-dependent at best,
requiring an accurate modeling of both the pion-production cross section and subsequent reabsorption
(and their interference).  Currently, they are estimated using Monte Carlo event generators.  As a consequence,
experimentally-extracted CCQE cross sections are accompanied by significant uncertainties.

\section{Calculation}
\label{sec:sec2}

The inclusive double-differential cross section for a charged-current scattering process initiated by a neutrino
off a nuclear target can be expressed as
\begin{align}
& \Big(\frac{d\sigma}{dT_\ell \,d\cos\theta_\ell}\Big)^{\rm CC}_{\nu/\bar{\nu}}=\frac{G_F^2\,
{\rm cos}^2\theta_c}{4\pi}\frac{|\mathbf{k}_\ell|}{ E_\nu}\Big[v_{00} R_{00}-v_{0z} R_{0z}\nonumber\\
&\qquad \qquad\qquad+ v_{zz} R_{zz}+v_{xx}R_{xx}\mp v_{xy}R_{xy}\Big]\ ,
\label{eq:cross_sec}
\end{align}
where the $-$ or $+$ sign corresponds to a neutrino ($\nu$) or antineutrino ($\overline{\nu}$) induced reaction.
We adopt the values $G_F\,$=$\, 1.1803 \times 10^{-5}\,\rm GeV^{-2}$, corrected for the bulk of the inner
radiative corrections~\cite{Nakamura:2002jg}, and $\cos\theta_c\,$=$\,0.97425$~\cite{PDG}. The initial $\nu$
(or $\overline{\nu}$) and final lepton four-momenta are, respectively, $k_\nu\,$=$\,(E_\nu,\mathbf{k}_\nu)$ and 
$k_\ell\,$=$\,(E_\ell,\mathbf{k}_\ell)$, $T_{\ell}$ is the kinetic energy of the lepton (rest mass $m_\ell$), and
$\theta_\ell$ is its scattering angle relative to the incoming neutrino direction.  The kinematical factors
$v_{\alpha\beta}$ associated with the contraction of the leptonic tensor, in the general case in which the
dependence on $m_\ell$ is kept, are reported in Appendix A of Ref.~\cite{Shen:2012}. 

The nuclear response functions encode all information on nuclear structure and dynamics, and are defined, in
a schematic notation, as (see Ref.~\cite{Shen:2012} for
explicit expressions)
\begin{align}
\label{eq:e2}
R_{\alpha\beta} (q, \omega) &= \sum_f \langle f | j_{\rm CC}^\alpha(\mathbf{q}, \omega) |i\rangle \langle f | j_{\rm CC}^\beta (\mathbf{q}, \omega) |i \rangle^*\nonumber\\
& \times \delta(\omega - E_f + E_i) \ ,
\end{align}
where $| i \rangle$ represents the $^Z\!A$ ground state of energy $E_i$,
$|f \rangle$ represents the bound or
scattering state of the final $^{Z+1}\!A$ or $^{Z-1}\!A$ nuclear system, depending
on whether the $(\nu_\ell,\ell^-)$ or $(\overline{\nu}_\ell,\ell^+)$ process is being
considered, of energy $E_f$, $j_{\rm CC}^\alpha(\mathbf{q}, \omega)$
are the relevant components of the weak charged current (CC), and
an average over the initial spin projections of $^Z\!A$ is understood (note, however,
that the $^{12}$C ground state has spin-parity assignments $J^\pi\,$=$\,0^+$). The dynamical
framework adopted in the calculations below has been described elsewhere in considerable
detail, most recently in the review~\cite{Carlson:2015}. Next, we provide a brief description for completeness.

\subsection{Interactions and currents}
\label{sec:sec2.a}
Strong interactions are described by two- and three-nucleon terms, respectively, the Argonne
$v_{18}$~\cite{Wiringa:1995} (AV18) and Illinois-7~\cite{Pieper:2008} (IL7) models.  The AV18
reproduced the nucleon-nucleon database available at the time (1995) with a $\chi^2$/datum close
to one~\cite{Wiringa:1995} for lab kinetic energy up to 350 MeV, slightly above the
pion production threshold. Even today that the database has increased in size considerably
(to over 5,200 data points over the energy range 0--300 MeV), the AV18 still gives (without a refit)
a very respectable $\chi^2$/datum of about 1.5~\cite{Piarulli:2015}. The IL7 three-nucleon
interaction model contains a small number (4)
of parameters, which characterize the overall strengths of two- and multi-pion exchange
terms involving $\Delta$-isobar excitations, and of a purely phenomenological (isospin-dependent)
central term.  These parameters are constrained by a fit to the energies of about 23 low-lying nuclear
states with mass number $A$ in the range 3--10~\cite{Pieper:2001ap}.  The resulting AV18+IL7 Hamiltonian
then leads, in accurate QMC calculations, to predictions for about 100 ground- and excited-state energies up
to $A\,$=$\, 12$, including the $^{12}$C ground- and Hoyle-state energies, in good agreement with the
corresponding empirical values~\cite{Carlson:2015}.

Electroweak probes couple to single nucleons (impulse approximation) as well as to clusters of
nucleons via one- and many-body currents.  The CC model adopted in the present study, identical
to that of Ref.~\cite{Shen:2012} and most recently employed to compute the muon-capture inclusive
rates on $^3$H and $^4$He~\cite{Lovato:2019fiw},  contains one- and two-body terms.
The former are derived from the covariant single-nucleon CC in a non-relativistic expansion that
retains corrections proportional up to the inverse square of the nucleon mass. Two-body (vector
and axial) terms arise from effective $\pi$- and $\rho$-meson exchanges, and $N$-to-$\Delta$
excitations, treated in the static limit.  A $\rho\pi$ transition mechanism is also included in the axial
component.  In GFMC calculations we utilize configuration-space representations of these
currents, regularized by a prescription which, by construction, makes their short-range
behavior consistent with the AV18 interaction~\cite{Carlson:1998}. The value for the
transition (axial) coupling constant $g_A^*$ in the $N$-to-$\Delta$ axial current is determined by
reproducing, within the present dynamical framework, the measured Gamow-Teller matrix
element contributing to tritium $\beta$-decay, and is listed in Table I (Set I) of Ref.~\cite{Shen:2012},
where explicit expressions for these currents can also be found.

The (isovector) nucleon form factors in the CC vector component
are taken as functions of the squared four-momentum transfer ($Q^2\,$=$\,q^2-\omega^2$)
from a modern fit to the available electron scattering data~\cite{Kelly:2004hm}
(in contrast to Ref.~\cite{Shen:2012}, in which we adopted a simple dipole parametrization
of these form factors).  
The axial form factor $G_A(Q^2)$ of the nucleon
is of a dipole form with a cutoff mass of either 1 GeV or 1.15 GeV, while its induced pseudoscalar
form factor, derived from the PCAC constraint and pion-pole dominance, is in accord
with values extracted from precise measurements of the muon-capture rate on hydrogen
and $^3$He~\cite{And07-all} as well as with predictions based on chiral perturbation theory~\cite{Ber94,Bernard:2001rs}.
Lastly, the $N$-to-$\Delta$ transition form factor in the vector sector is as
obtained in an analysis of $\gamma N$ data in the $\Delta$-resonance region~\cite{Carlson:1985mm},
while that in the axial sector, because of the lack of available experimental data,
is simply taken to have the same functional form of $G_A(Q^2)$, namely
$G^*_A(Q^2)/g^*_A\,$=$\,G_A(Q^2)/g_A$, where $g^*_A$ is the (fitted)
$N$-to-$\Delta$ axial coupling constant mentioned earlier.  Values for the parameters
entering these axial form factors are specified in Ref.~\cite{Shen:2012}.

\subsection{Electroweak response functions} 
\label{sec:sec2.b}
The calculation of the response functions in Eq.~(\ref{eq:e2}) proceeds in two steps.
The first consists in Laplace-transforming $R_{\alpha\beta}(q,\omega)$ with respect
to $\omega$, which reduces to the following current-current correlator (Euclidean response
function)
\begin{equation}
\label{eq:e3}
E_{\alpha\beta}(q,\tau)=  \langle i | j_{\rm CC}^{\beta \dagger} (\mathbf{q},\omega_{\rm qe}) \,
{\rm e}^{-\tau(H-E_i)} j_{\rm CC}^\alpha (\mathbf{q},\omega_{\rm qe}) |i \rangle\ ,
\end{equation}
where $H$ is the Hamiltonian (here the AV18-IL7 model). 
The energy dependence of $j^\alpha_{CC}({\bf q},\omega)$
comes in via the nucleon and $N$-to-$\Delta$ transition form factors, which are taken as
functions of $Q^2$, as noted above.  We freeze the $\omega$-dependence by fixing $Q^2$ at the value
$Q^2_{\rm qe}\,$=$\, q^2-\omega_{\rm qe}^2$ with the quasielastic energy transfer $\omega_{\rm qe}$
given by $\omega_{\rm qe}\,$=$\,\sqrt{q^2+m^2}-m$ ($m$ is the nucleon mass).  This is needed in
order to exploit the completeness over the nuclear final states in evaluating the Laplace transforms
of $R_{\alpha\beta}(q,\omega)$.
The correlator is then computed with Green's function Monte Carlo (GFMC)
methods~\cite{Carlson:1992,Carlson:1994zz,Lovato:2016gkq,Lovato:2013,Lovato:2015,Lovato:2017cux}.  It should be stressed that no 
additional approximations are made beyond those inherent to the modeling of the interactions and currents.
The response is thus calculated {\it ab initio} by treating completely correlations in the initial state, by accounting consistently through the imaginary-time propagation for interaction effects in the final states, and, in particular, by
retaining in full the important interference between one- and two-nucleon
currents.

Because of the computational cost of the
present study (of the order of 130 million core hours on the massively
parallel computer MIRA at ANL), however, we only propagate the $^{Z-1}\!A$ system, i.e.,
$j_{\rm CC}^\alpha$ in Eq.~(\ref{eq:e3}) is the charge lowering current corresponding
to the process $(\overline{\nu}_\ell,\ell^+)$. If electromagnetic interactions and
isospin-symmetry-breaking terms in the strong interactions were to be ignored,
the final states $|f; \,^{Z+1}\!A\rangle$ and $|f; \,^{Z-1}\!A\rangle$ of the
$^{Z+1}\!A$ and $^{Z-1}\!A$ nuclear systems would simply be related to each other
via $|f;\,^{Z+1} \! A\rangle\,$=$\, \left(\prod_i \tau_{i,x}\right) |f;\,^{Z-1}\!A\rangle$, where
$\tau_{i,x}$ is the isospin flip operator converting proton $i$ into a neutron or viceversa.
Matrix elements of the charge-raising and charge-lowering current between
the $^Z\!A$ state and, respectively, the $^{Z+1}\!A$ and $^{Z-1}\!A$ states would
then be identical.  We will assume here this is the case for $^{12}$C, and obtain the response functions corresponding
to the $(\nu_\ell,\ell^-)$ process from those corresponding to the $(\overline{\nu}_\ell,\ell^+)$
process by correcting the final state energies of the $^{12}$B system by
the difference in ground-state energies between $^{12}$N and $^{12}$B---in practice,
by shifting the response functions by about 5.5 MeV.
We expect this approximation to be inaccurate in the threshold region; however, in quasielastic
kinematics and beyond, it should be of little import.

\afterpage{\clearpage}
\begin{figure}[!h]
\centering
\includegraphics[width=\columnwidth]{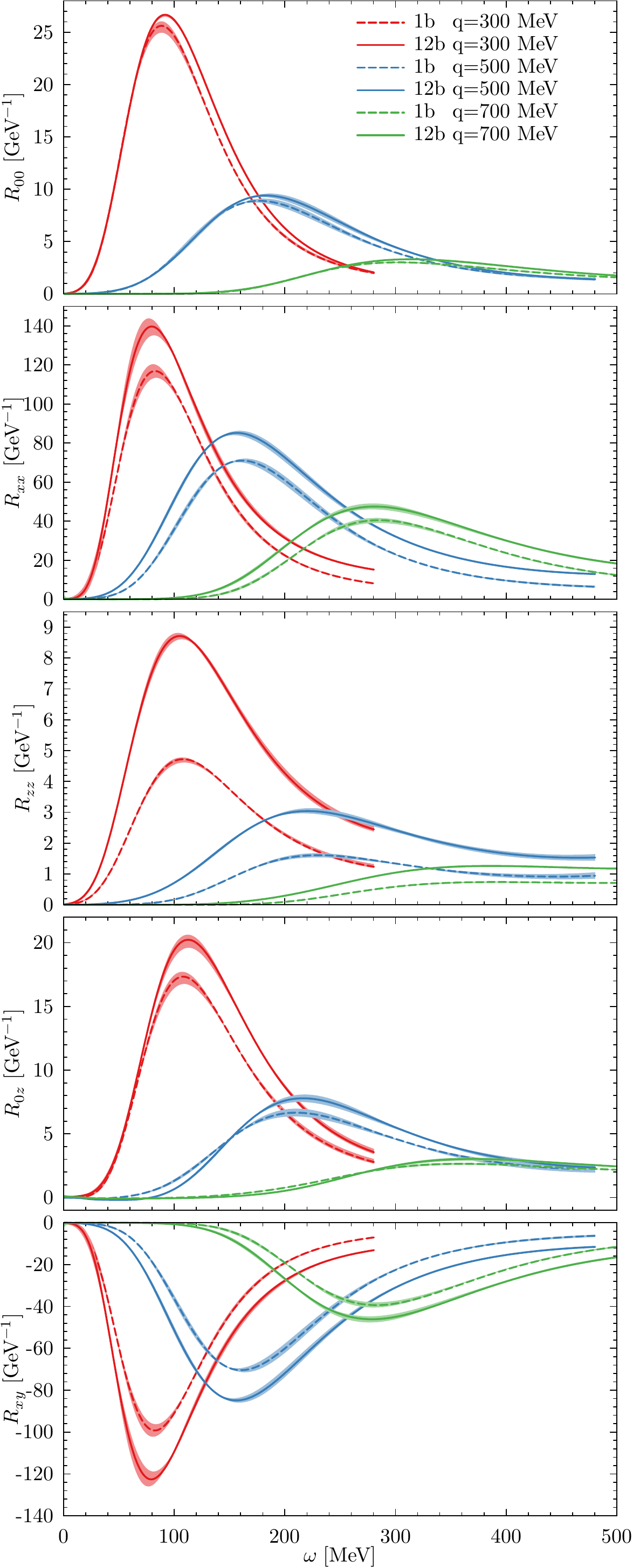}
\caption{GFMC response functions at $q\,$=$\,300$ (red), 500 (blue), and 700 (green) MeV.
Predictions obtained with one-body (one- and two-body) currents are shown by dash (solid)
lines. Shaded areas result from a combination of GFMC statistical errors and uncertainties
associated with the maximum-entropy inversion.}
\label{fig:res}
\end{figure}

The second step employs maximum-entropy techniques, developed specifically for this type of
problem in Ref.~\cite{Lovato:2015} (a fairly complete account of them is
given in that work), to perform the analytic continuation of the Euclidean response functions,
corresponding to the ``inversion'' of the Laplace transforms.  The resulting $R_{\alpha\beta}(q,\omega_{\rm qe})$
are rescaled as follows to account for the correct $\omega$-dependence of the various form factors.
The $00$, $0z$, $zz$, and $xx$ response functions are given by the incoherent
sum of the (squared) matrix elements associated with the CC vector ($V$) and axial ($A$) components,
while the $xy$ response function involves interference between these components.  The $00$, $0z$, and
$zz$ $V$ contributions are multiplied by the factor $\left[G_E^V(Q^2)/G_E^V(Q^2_{\rm qe})\right]^2$
and the $xx$ $V$ contribution by $\left[G_M^V(Q^2)/G_M^V(Q^2_{\rm qe})\right]^2$, where $G_E^V$ and
$G_M^V$ are the isoscalar and isovector combinations of the proton ($p$) and
neutron ($n$) electric ($E$) and magnetic ($M$) form factors (in the parametrization
of Ref.~\cite{Kelly:2004hm}).  These multiplicative
factors naturally emerge by considering the dominant one-body terms in the CC $V$ current.

The $0$, $0z$, $zz$, and $xx$ $A$ contributions are multiplied by the factor
$\left[G_A(Q^2)/G_A(Q^2_{\rm qe})\right]^2$.  For these contributions such a rescaling turns
out to fully restore the correct $\omega$-dependence, since the one- and two-body axial
currents, including those associated with $\Delta$-isobar intermediate states, are
proportional to $G_A(Q^2)$ in the present modeling~\cite{Shen:2012}.  Lastly, the
interference response is rescaled by the factor $\left[G_A(Q^2)/G_A(Q^2_{\rm qe})\right]\times 
\left[ G_M^V(Q^2)/G_M^V(Q^2_{\rm qe})\right]$.  Below, we show that the procedure above
essentially accounts for the correct $\omega$-dependence implicit in the complete CC response.

The five response functions entering the CC cross section have been calculated with
GFMC methods for momentum transfers in the range (100--700) MeV in steps of 100 MeV.
To reduce clutter, we present in Fig.~\ref{fig:res} only those obtained at $q\,$=$\,300$, 500, and 700 MeV
(note that the scales for $R_{\alpha\beta}$ are different in each panel).\footnote{Tabulations of
GFMC-calculated $R_{\alpha\beta}(q,\omega)$ for $q$ in the range (100--700) MeV and
$\omega$ from threshold to $\omega \lesssim q$ are available upon request.}
The transverse ($xx$) and interference ($xy$) response functions are largest
but of opposite sign (the $xy$ response as defined here is negative).  Consequently,
the contributions $v_{xx}\, R_{xx}$ and $v_{xy}\, R_{xy}$ in the CC cross section
add up for neutrino scattering and tend to cancel each other out for antineutrino scattering
(the kinematical factors $v_{xx}$ and $v_{xy}$ are positive~\cite{Shen:2012}).

Two-body terms in the CC significantly increase the magnitude of the response
functions obtained in impulse approximation (i.e., with one-body currents),
over the whole quasielastic region, except for $R_{00}$ at low
$\omega$.  This increase in strength mostly comes about
because of constructive interference between the one- and two-body current
matrix elements, and is consistent with that expected on the basis of sum rule
analyses~\cite{Lovato:2014}.  Two-body contributions are found to be especially
large---accounting for more than 50\% of the total calculated strength---in $R_{zz}$,
which involves the longitudinal components (along the direction of the three-momentum
transfer) of the CC.

\subsection{Scaling analysis}
\label{sec:sec2.c}

\begin{figure}[t!]
\centering
\includegraphics[width=\columnwidth]{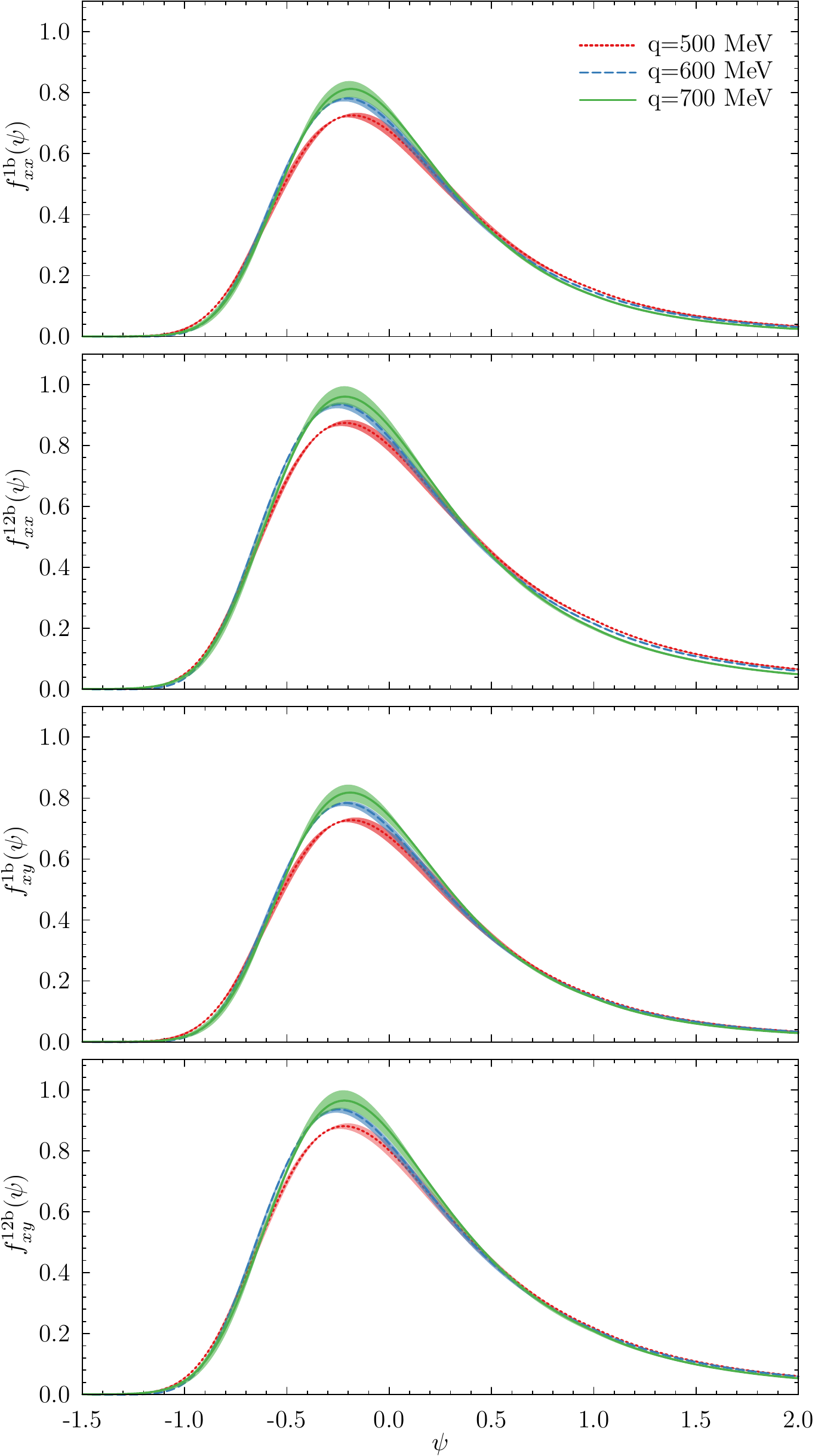}
\caption{Transverse scaling functions obtained from $R_{xx}$ and $R_{xy}$ including only one-body,
and one- and two-body, terms in the CC, denoted respectively as $f^{\rm 1b}_{xx}$ and $f^{\rm 1b}_{xy}$,
and $f^{\rm 12b}_{xx}$ and $f^{\rm 12b}_{xy}$. 
The different curves have been obtained for three different values of the moment transfer. }
\label{xx:xy:scaling:func}
\end{figure}

The analysis of scaling properties of nuclear response functions has proven to be a useful tool to elucidate important aspects of the many-body dynamics in the quasielastic region.  Scaling occurs when the electroweak response functions, divided by appropriate pre-factors describing single-nucleon physics, no longer depend upon the momentum $q$ and energy transfer $\omega$, but only on a specific function of them
$\psi(q,\omega)$, yielding
\begin{equation}
\frac{R_{\alpha\beta}}{G_{\alpha\beta}} \simeq \frac{1}{k_F} f_{\alpha\beta} (\psi)\,,
\label{scal:func:eq}
\end{equation}
where $k_F$ is the Fermi momentum of the system. In the non-relativistic limit, the scaling variable is given by~\cite{Rocco:2017hmh} 
\begin{equation}
\psi =  \frac{m}{q\, k_F}\Big(\omega - \frac{q^2}{2m} - \epsilon\Big)\, ,
\end{equation}
where $\epsilon$ is introduced to account for nuclear binding effects. 

The pre-factors associated with the electromagnetic longitudinal and transverse responses can be found in
Ref.~\cite{Rocco:2017hmh}. Here we extend the scaling analysis to the five response functions relevant for
neutrino-nucleus scattering induced by CC transitions. The (longitudinal and transverse) pre-factors associated
with vector currents are related to those of (isovector)  electromagnetic currents by the CVC
constraint; the pre-factors associated with axial
currents bring about additional terms, whose relativistic expressions can be found in Ref.~\cite{Amaro:2004bs}. 

Within the Fermi gas model~\cite{Alberico:1988bv}, the following scaling function can be analytically derived,
 \begin{equation}
f^{\rm FG}_{\alpha\beta} (\psi) = \frac{3}{4} (1-\psi^2)\theta(1-\psi^2)\ ,
\end{equation}
by assuming one-body currents only.  However, unlike the latter expression, which is symmetric and centered
around $\psi\,$=$\,0$, the scaling functions extracted from experimental data and those inferred from more realistic
models of nuclear dynamics exhibit a clearly asymmetric shape, with a tail extending in the $\psi>0$ region~\cite{Caballero:2007}. Moreover, while the Fermi gas scaling function is universal and does not depend upon the specific transition operator,
such is not the case when the spin and charge dependence of nuclear interactions in the final states are taken into
account~\cite{Carlson:1994zz}.  

The  $xx$ ($xy$) scaling functions displayed in the upper two (lower two) panels of
Fig.~\ref{xx:xy:scaling:func} have been obtained as in Eq.~\eqref{scal:func:eq}, i.e., by
dividing the GFMC electroweak response functions in the transverse (interference)
channel by the appropriate pre-factors for the CC vector and axial components. The upper and lower
panels for each set ($xx$ and $xy$) correspond to including one-body only, and one-and two-body,
current operators. 
The dotted (red), dashed (blue), and solid (green) lines show the $xx$ and $xy$ scaling functions for $q\,$=$\,$500, 600,
and 700 MeV, respectively. The shaded area indicates the uncertainty in the maximum-entropy inversion
procedure and also reflects the statistical errors of the GFMC calculations. The $xy$ scaling functions,
shown in the lower two panels of Fig.~\ref{xx:xy:scaling:func}, 
are almost identical to the $xx$ ones in both cases (one-body only,
and one- and two-body currents).  In contrast to the Fermi gas model, nuclear correlations in the initial
and final states, which are exactly treated in the GFMC method, yield asymmetric scaling functions, with
tails that extend well beyond $\psi > 1$. Note that the scaling functions can be significantly different
in the other channels, for example the longitudinal and transverse response in electron scattering~\cite{Benhar:2006wy}.

The different curves clearly exhibit a scaling behavior, as they are almost independent of momentum transfer.
This is expected to be even more accurate at larger  $q$ values.  More interesting is the observation that scaling
persists even when two-body current contributions are included in the response functions, as shown in the second and fourth 
 panels of Fig.~\ref{xx:xy:scaling:func}.  While these contributions generate
significant excess strength in $f_{xx}$ and $f_{xy}$, they do not spoil their scaling properties.  An
explanation of these features can be found in Ref.~\cite{Pastore:2019urn} for the case of the electromagnetic
response, and similar considerations remain valid here.  In essence, in the $xx$ and $xy$
responses the excess strength seen in the quasielastic region comes about because
two-body currents lead to final states which are very similar to those produced by an electroweak interaction
vertex on a single nucleon followed by the subsequent high-momentum strong interaction of this nucleon
with another nucleon.  The resulting (constructive) interference between the corresponding matrix
elements generates excess strength which is spread out over the quasielastic peak region in
a way very similar to the response arising from the high-momentum part of the single-nucleon
currents associated with pion exchange interactions.
We defer to Ref.~\cite{Pastore:2019urn} for a more comprehensive discussion
of scaling in the present context of microscopic Hamiltonians and currents.  This reference
also discusses superscaling \cite{Barbaro:2006}---scaling with respect  to the mass
number---and the absence of scaling observed in the $\Delta$-resonance region.    

An analogous scaling behavior is also seen in the $00$, $0z$, and $zz$ channels.  We capitalize
on this feature in order to extrapolate the response functions at large momentum transfers
$q > \overline{q}\,$=$\,700$ MeV, that is, beyond the range of those calculated with GFMC methods.
It turns out they are needed when computing flux-folded cross sections (see Sec.~\ref{sec:sec3} below).
We parametrize them as
\begin{equation}
R_{\alpha\beta}(q> \overline{q},\omega) = G_{\alpha\beta} (q,\omega) f_{\alpha\beta}(\psi) \ ,
\end{equation}
where $f_{\alpha\beta}(\psi)$ are the scaling functions determined from the GFMC-calculated responses
at $\overline{q}$.  The underlying assumption is that the $f_{\alpha\beta}(\psi)$ for $q>\overline{q}$ coincide
with those at $\overline{q}$.  To account for the small scaling violations, we conservatively associate an
uncertainty to this extrapolation procedure corresponding to twice the difference between the scaling
functions at $q\,$=$\,600$ and $700$ MeV.

\section{Results}
\label{sec:sec3}
\begin{figure}[bth]
\centering
\includegraphics[width=\columnwidth]{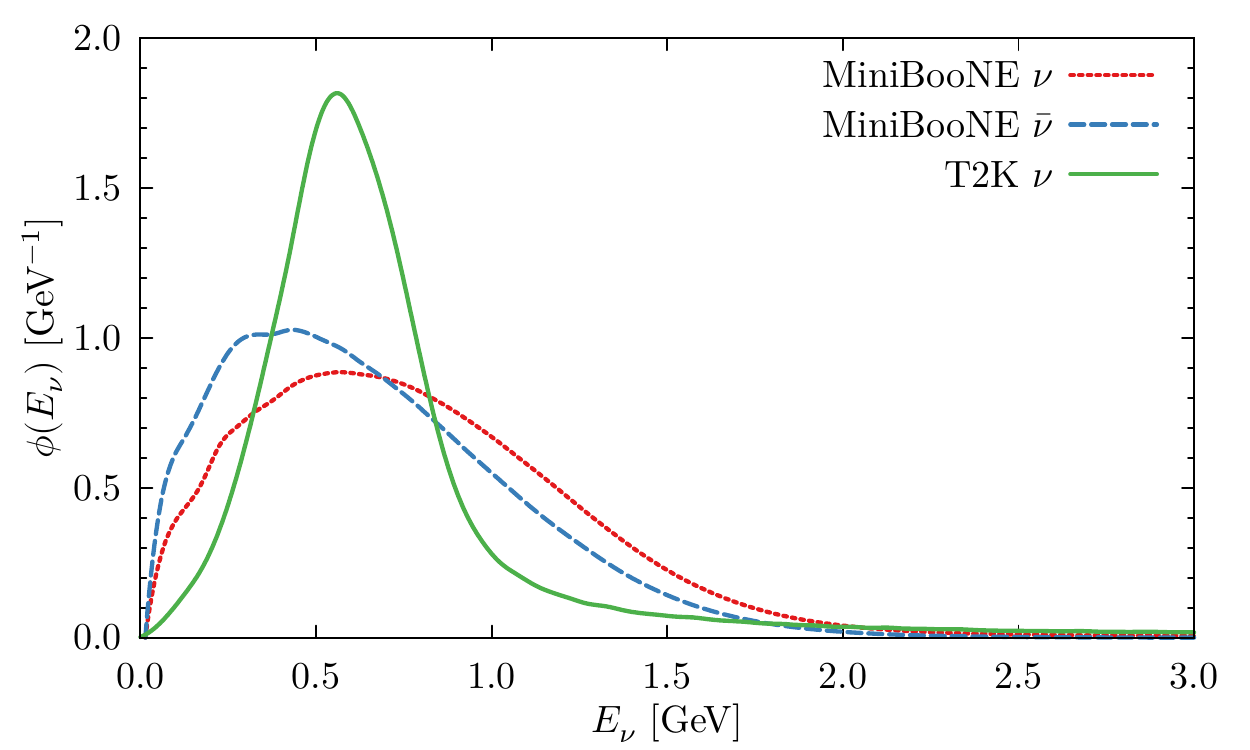}
\caption{Normalized $\nu_\mu$ fluxes of MiniBooNE and T2K, and normalized $\overline{\nu}_\mu$ flux of
MiniBooNE.}
\label{fig:flux}
\end{figure}

Muon neutrino and antineutrino flux-averaged cross sections are obtained from
\begin{equation}
\left \langle \frac{d\sigma}{dT_\mu \, d\cos\theta_\mu} \right\rangle=
\int dE_\nu \, \phi (E_\nu) \,\frac{d\sigma(E_\nu)}{dT_\mu \, d\cos\theta_\mu} \ ,
\end{equation}
where $\phi(E_\nu)$ is the normalized $\nu_\mu$ or $\overline{\nu}_\mu$ flux---those
for MiniBooNE and T2K are shown in Fig.~\ref{fig:flux}---and $d\sigma(E_\nu)/( dT_\mu\, d\cos\theta_\mu)$
are the corresponding inclusive cross sections of Eq.~(\ref{eq:cross_sec}).  The experimental data
are binned in $\cos\theta_\mu$ bins of constant width (0.1) for MiniBooNE, and
varying widths for T2K; when comparing to these data, the calculated cross sections
are averaged over the relevant $\cos\theta_\mu$ bin.

\begin{figure*}[]
\centering
\includegraphics[width=0.6\columnwidth]{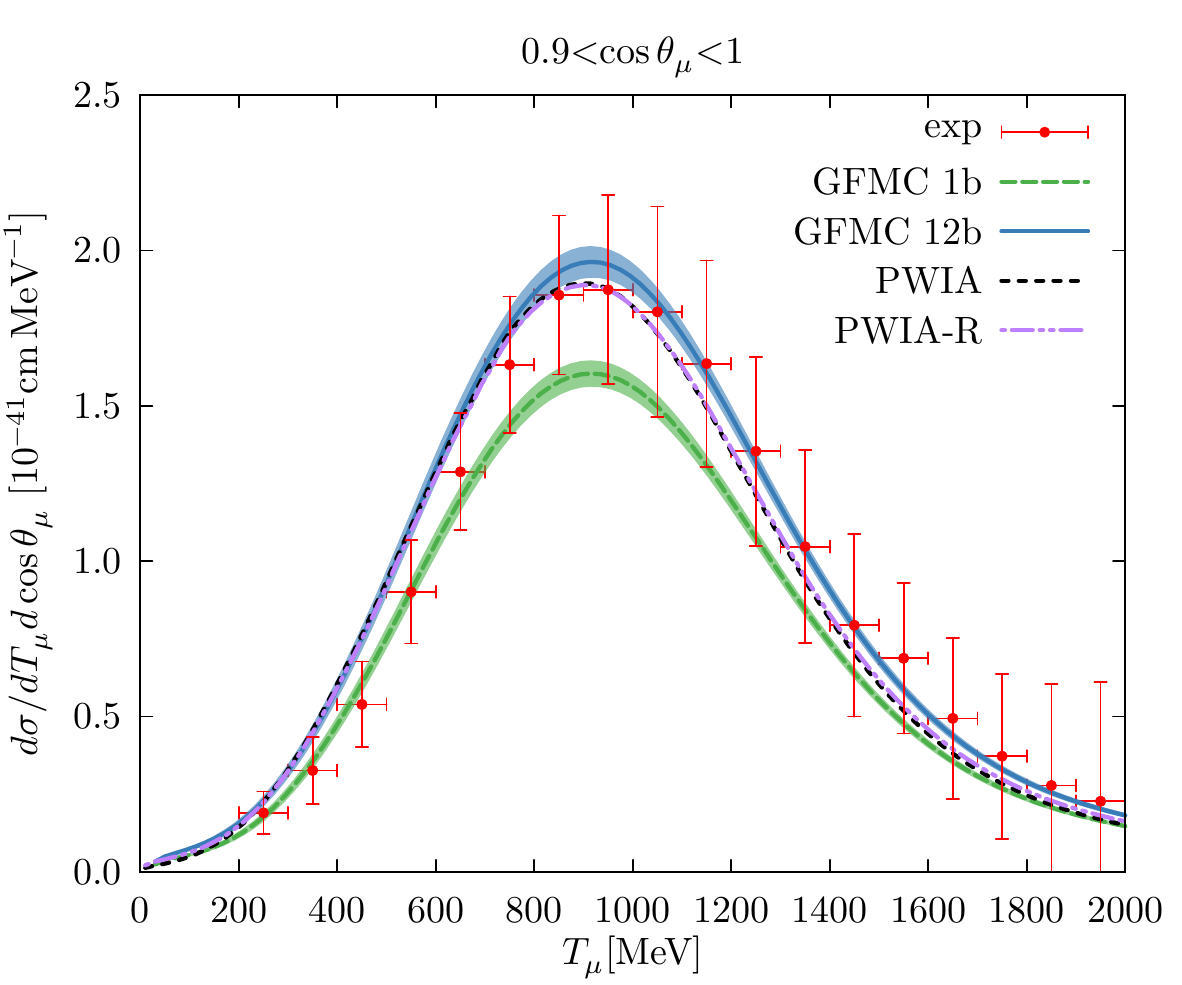}
\includegraphics[width=0.6\columnwidth]{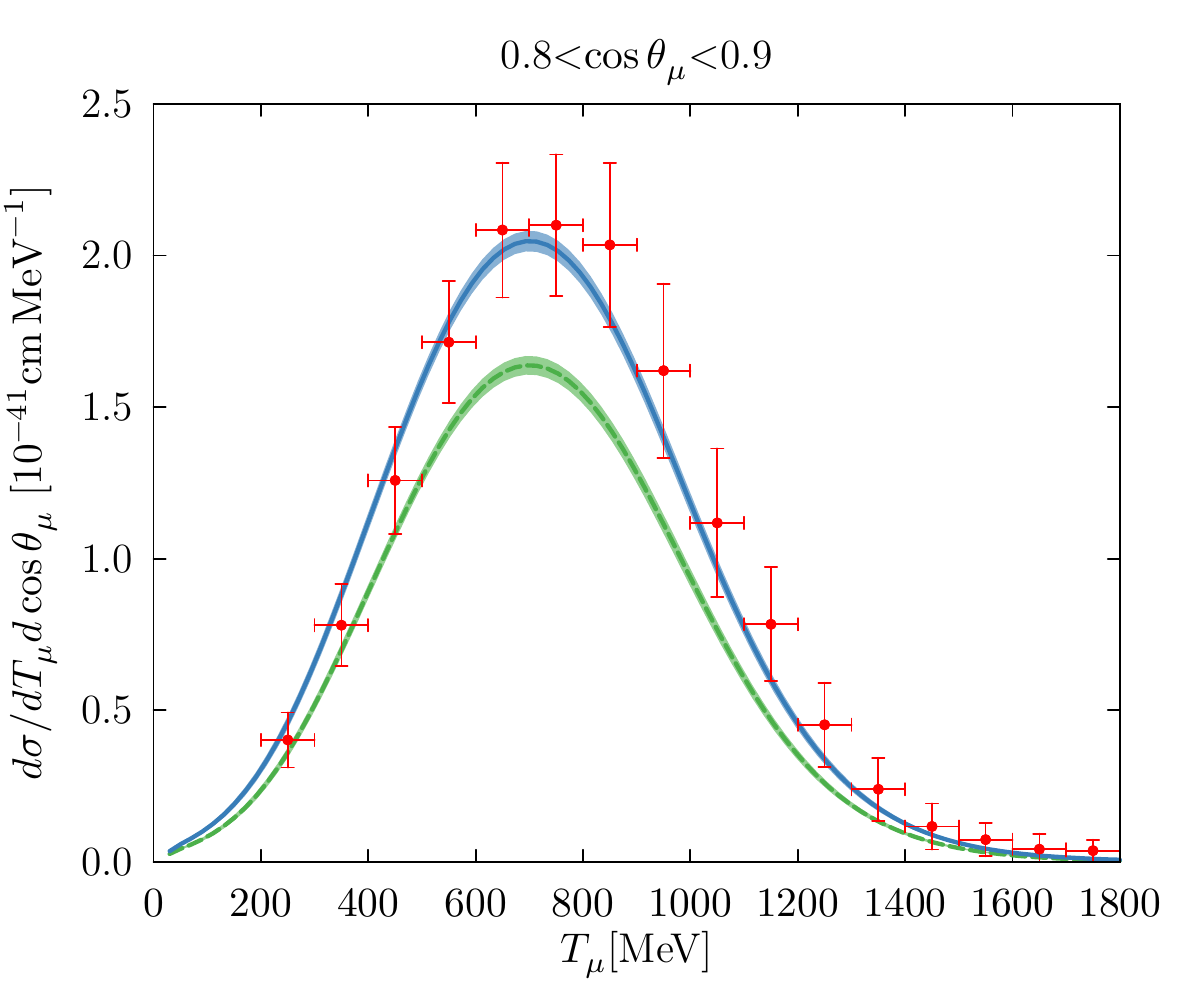}
\includegraphics[width=0.6\columnwidth]{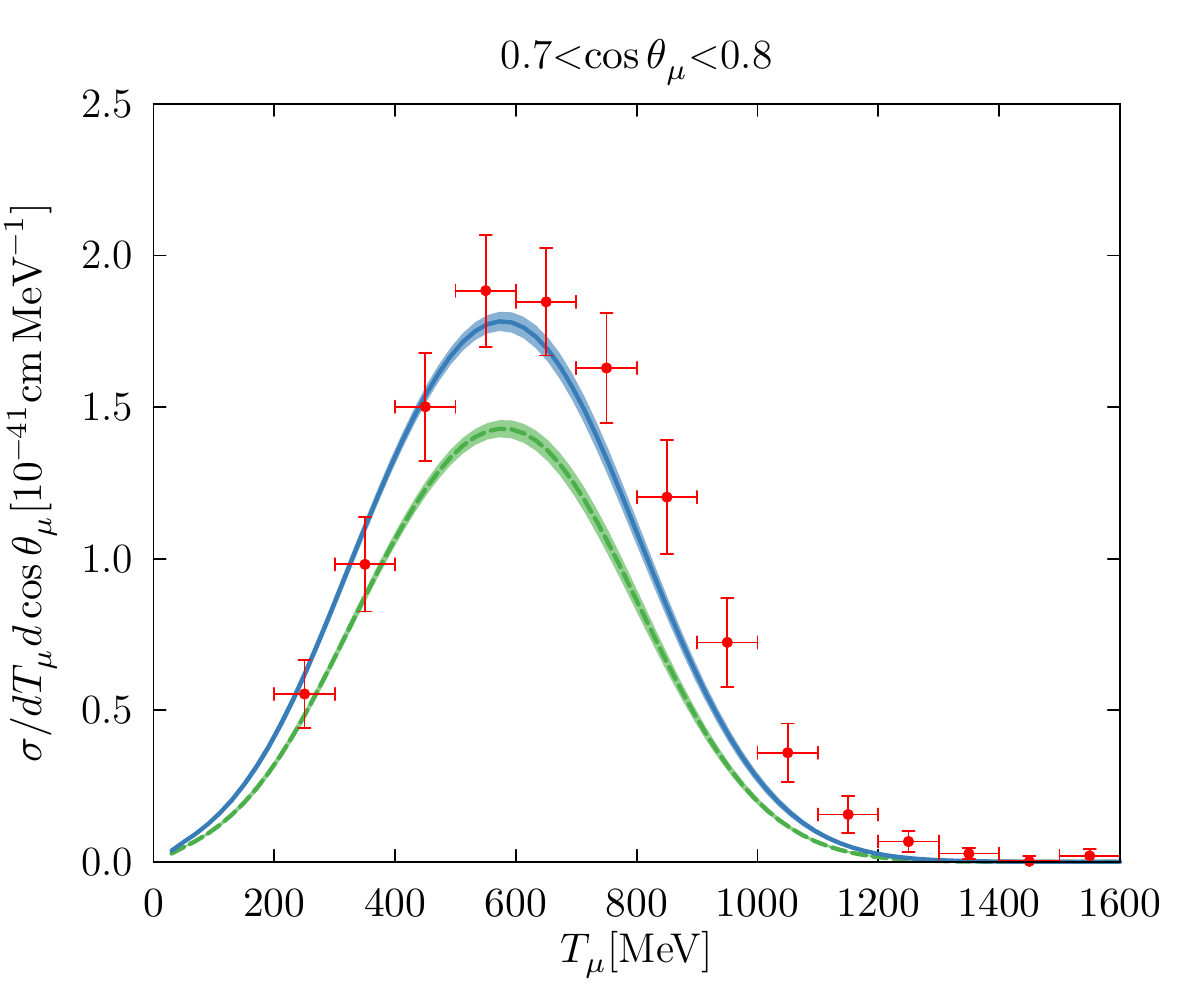}
\includegraphics[width=0.6\columnwidth]{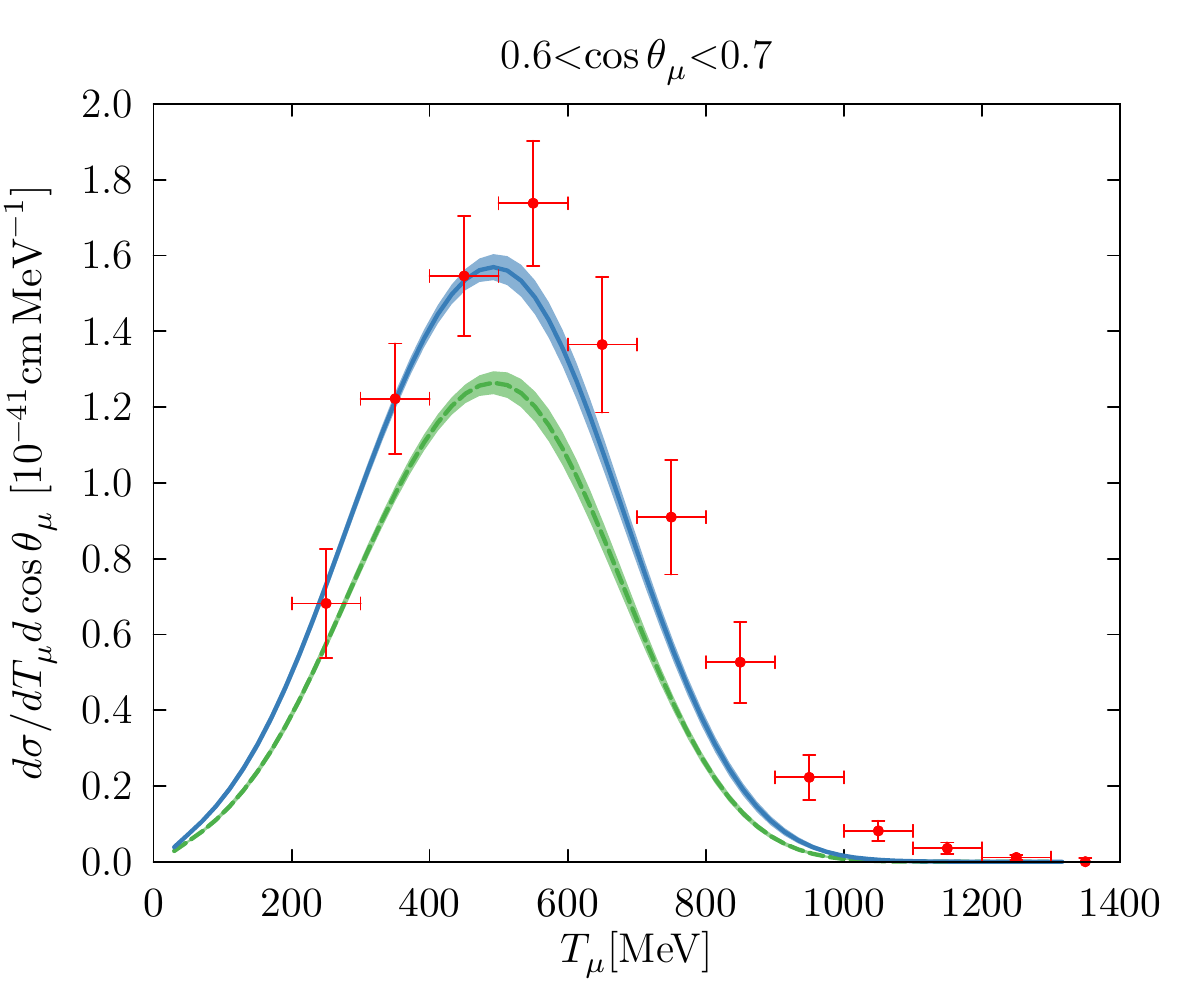}
\includegraphics[width=0.6\columnwidth]{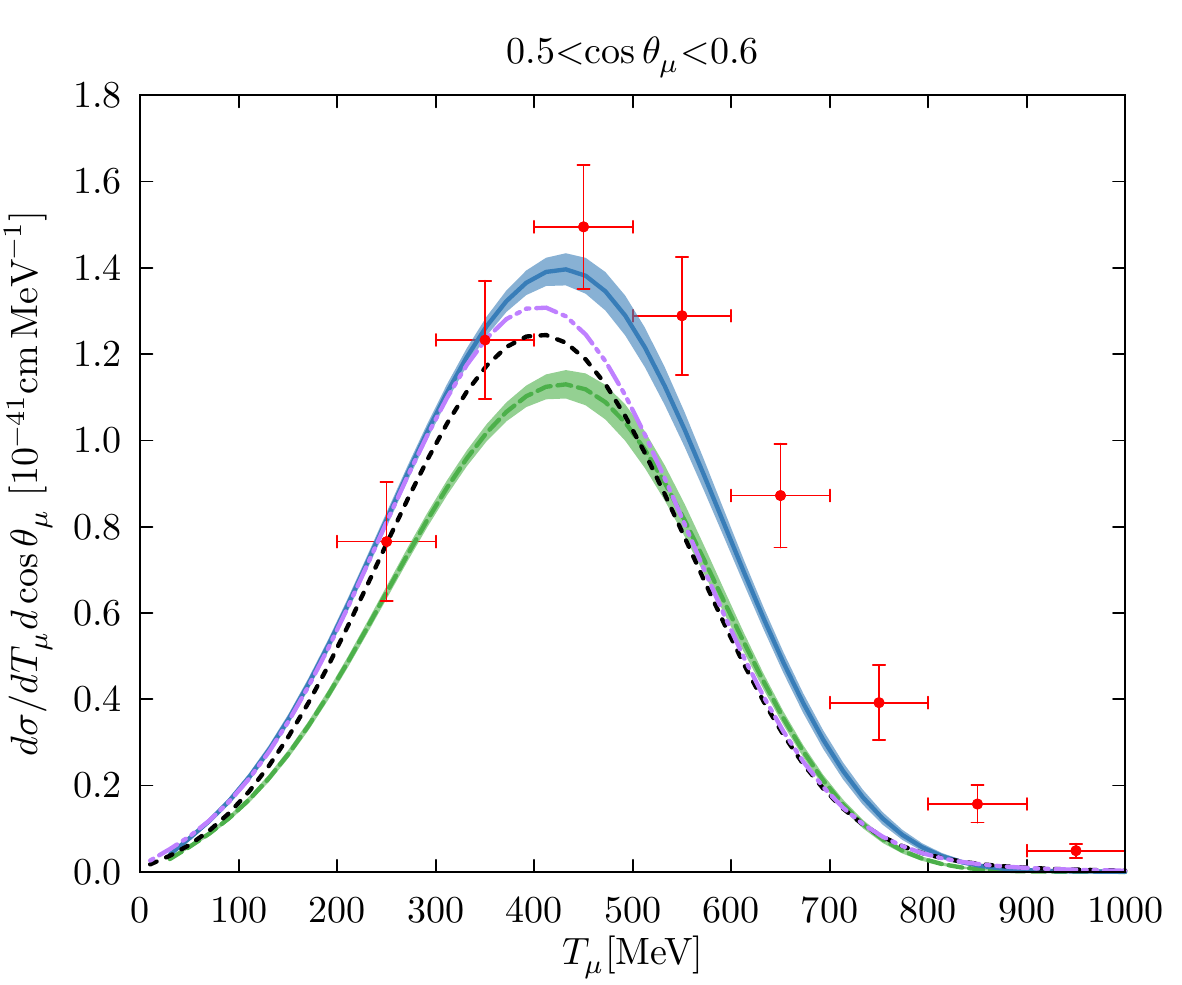}
\includegraphics[width=0.6\columnwidth]{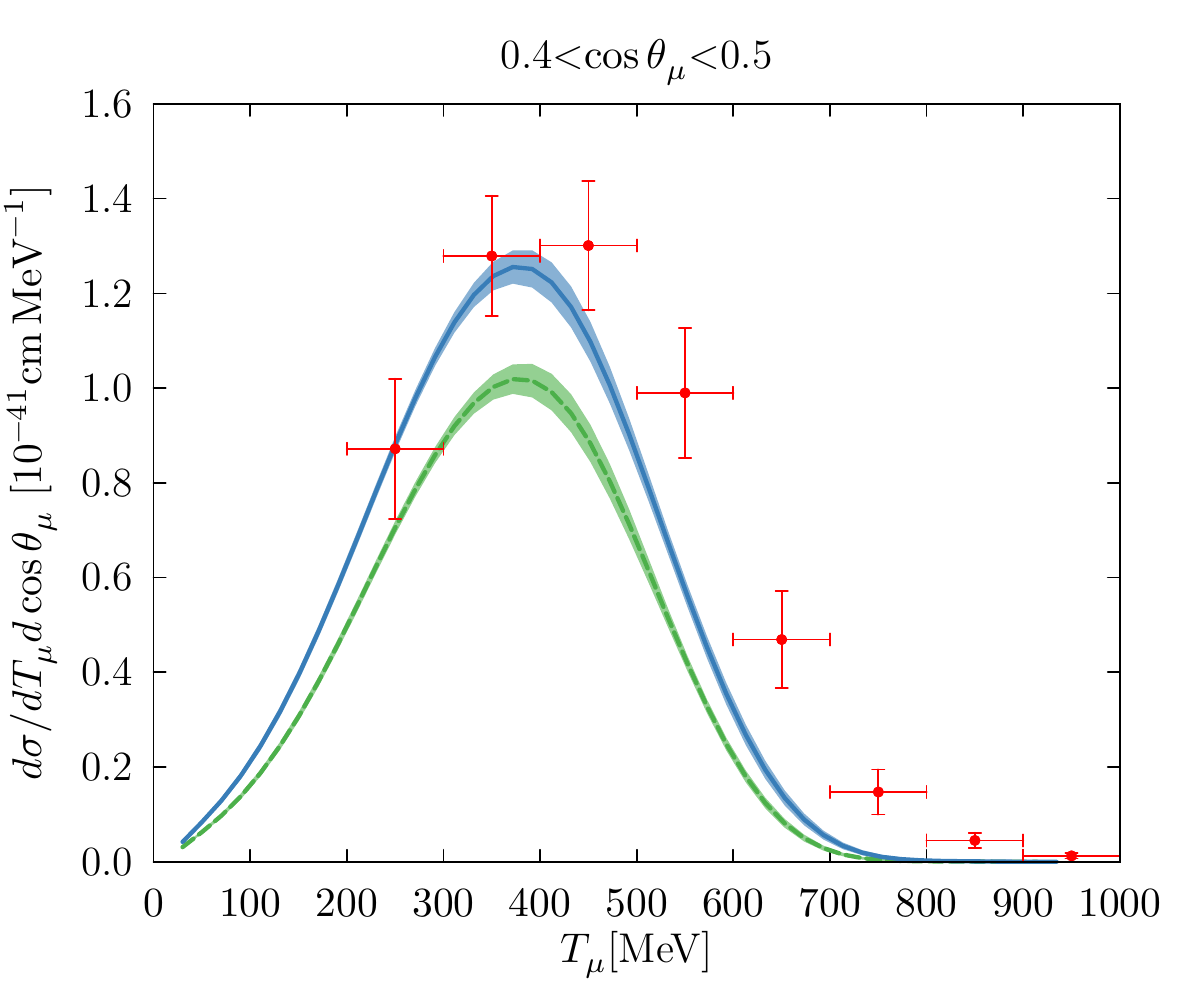}
\includegraphics[width=0.6\columnwidth]{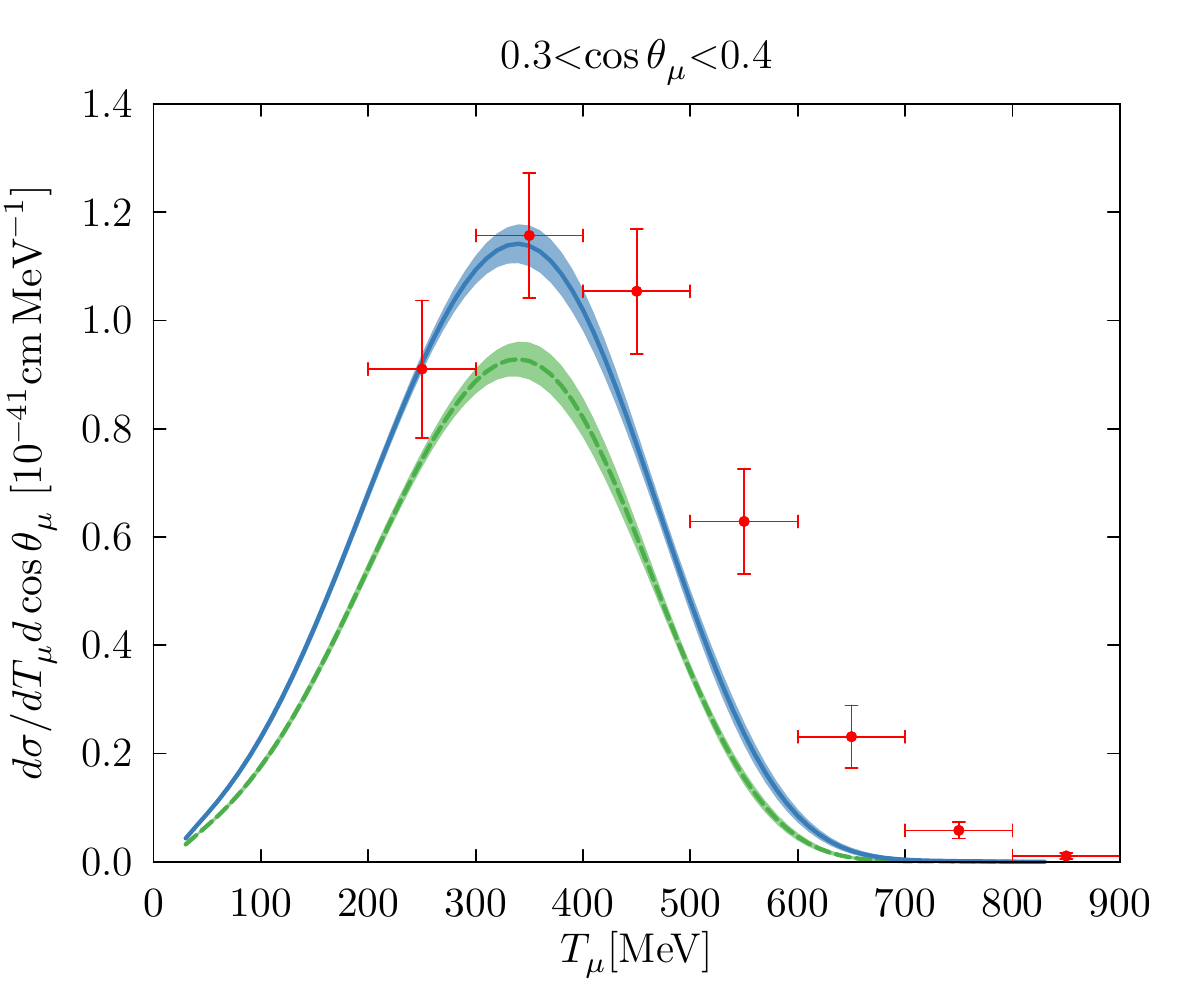}
\includegraphics[width=0.6\columnwidth]{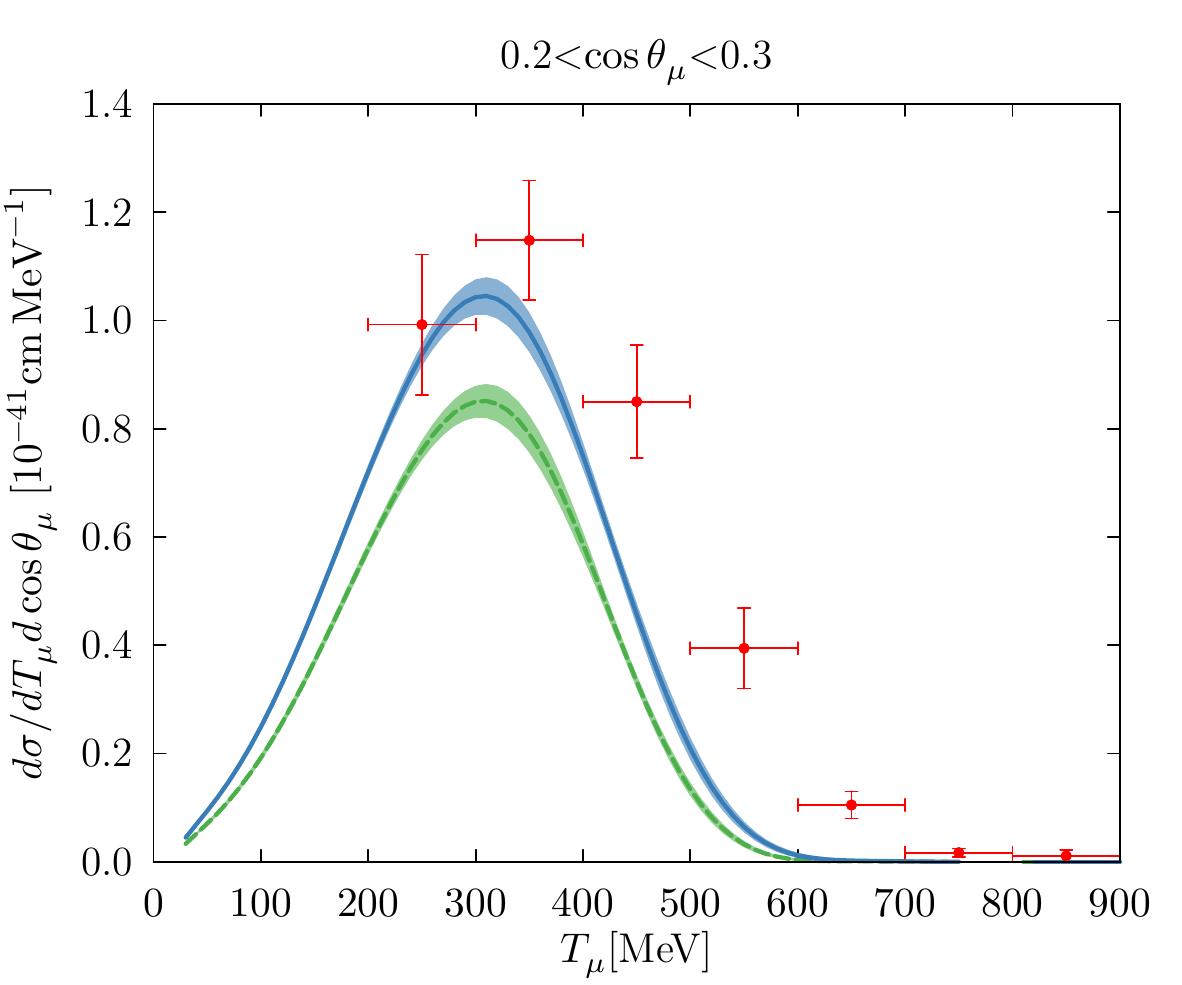}
\includegraphics[width=0.6\columnwidth]{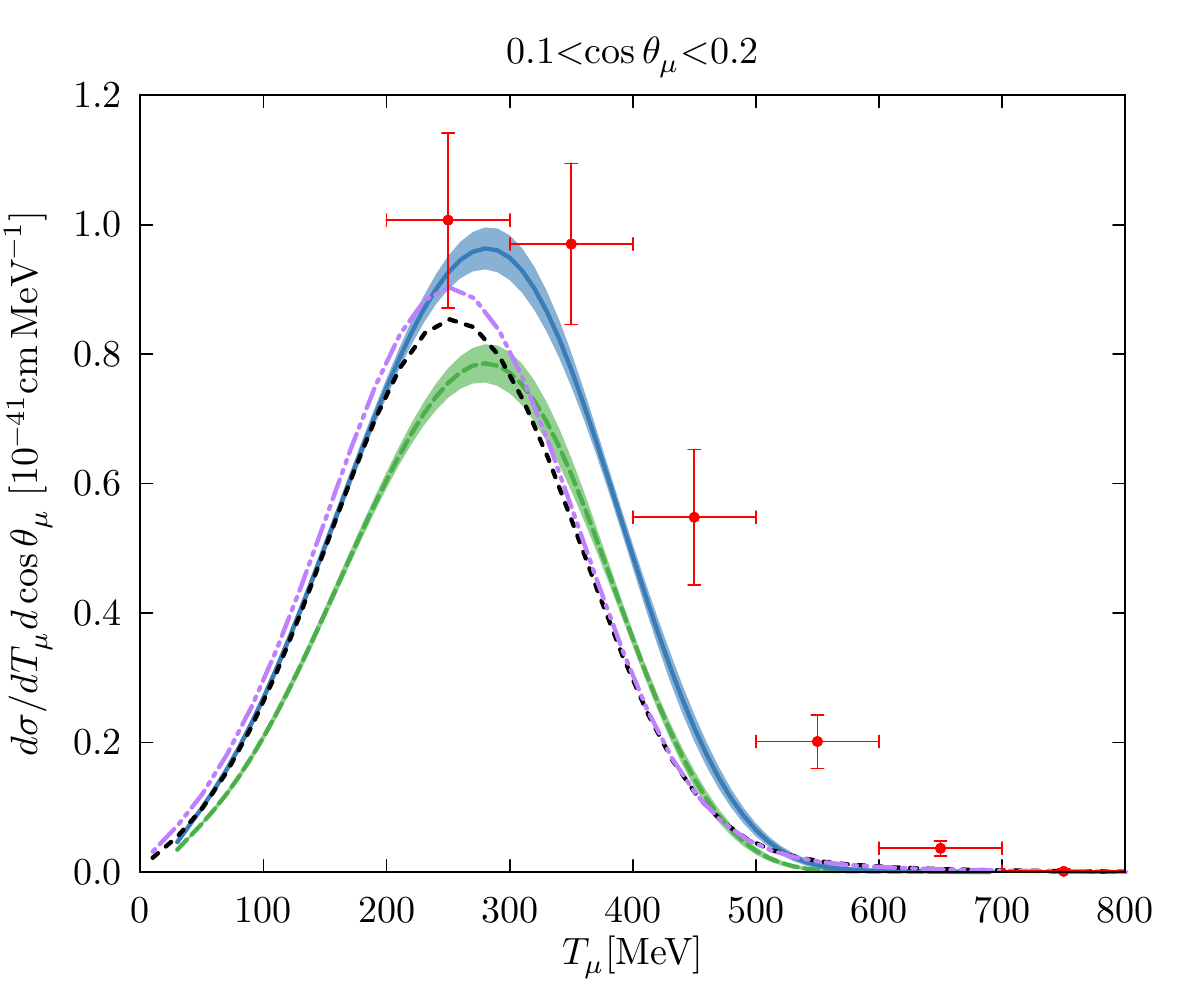}
\caption{MiniBooNE flux-folded double differential cross sections per target neutron for $\nu_\mu$-CCQE scattering on $^{12}$C,
displayed as a function of the muon kinetic energy ($T_\mu$) for different ranges of $\cos\theta_\mu$. The experimental data and their shape uncertainties are from Ref.~\cite{AguilarArevalo:2010zc}.  The additional $10.7\%$ normalization uncertainty is not shown here.
Calculated cross sections are obtained with $\Lambda_A\,$=$\, 1.0$ GeV.}
\label{fig:MiniBooNE_nu}
\end{figure*}

\begin{figure*}[]
\centering
\includegraphics[width=0.6\columnwidth]{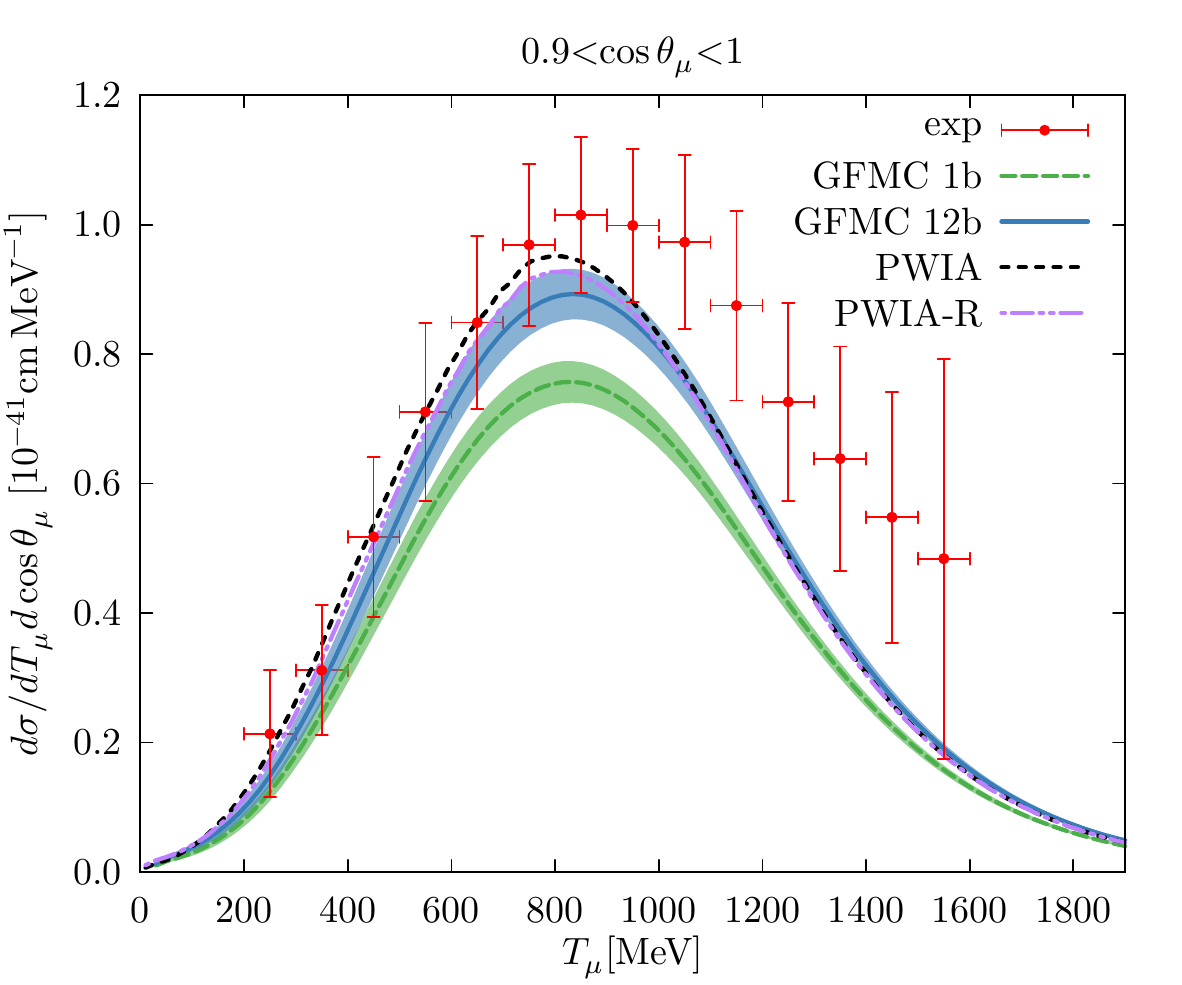}
\includegraphics[width=0.6\columnwidth]{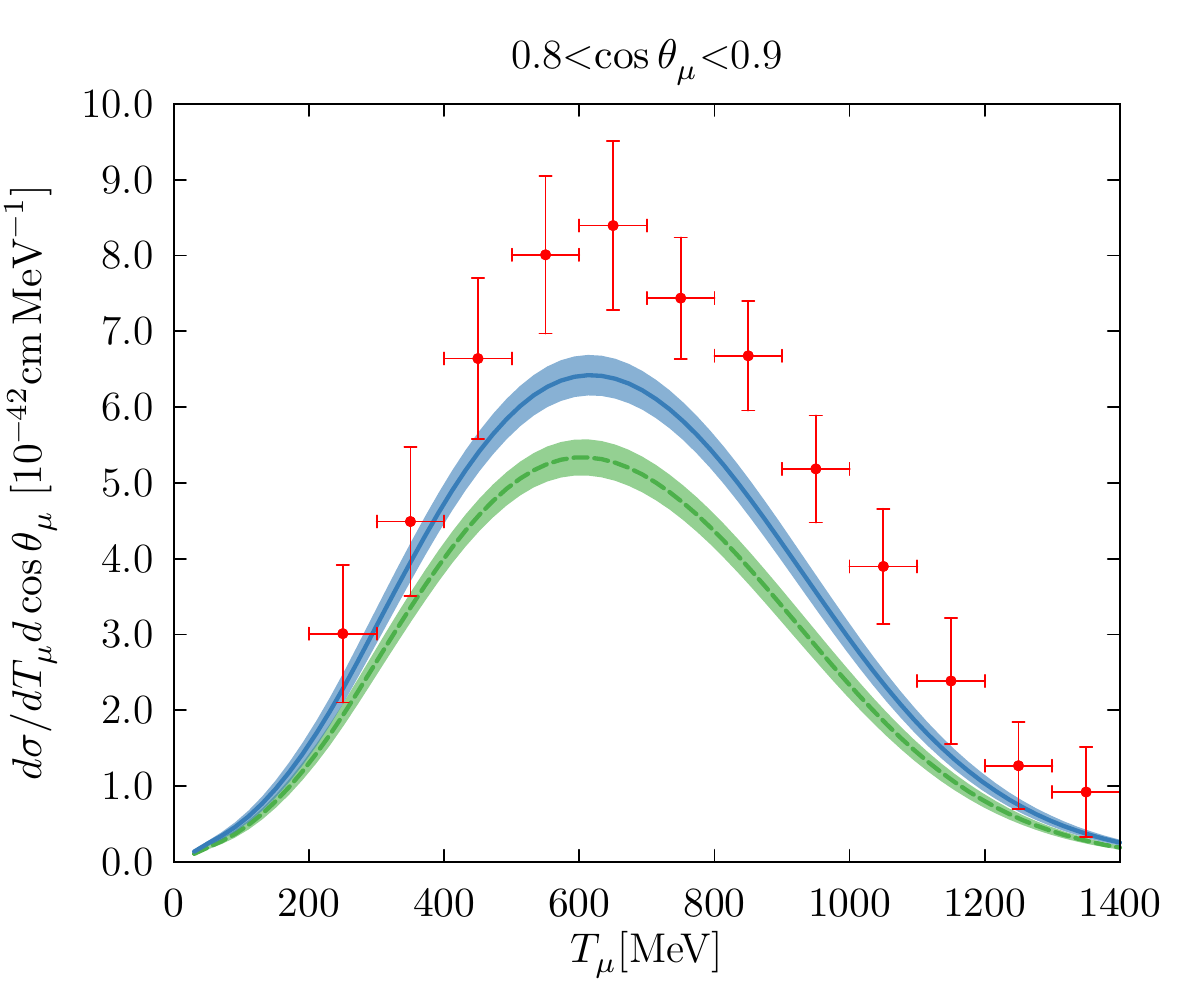}
\includegraphics[width=0.6\columnwidth]{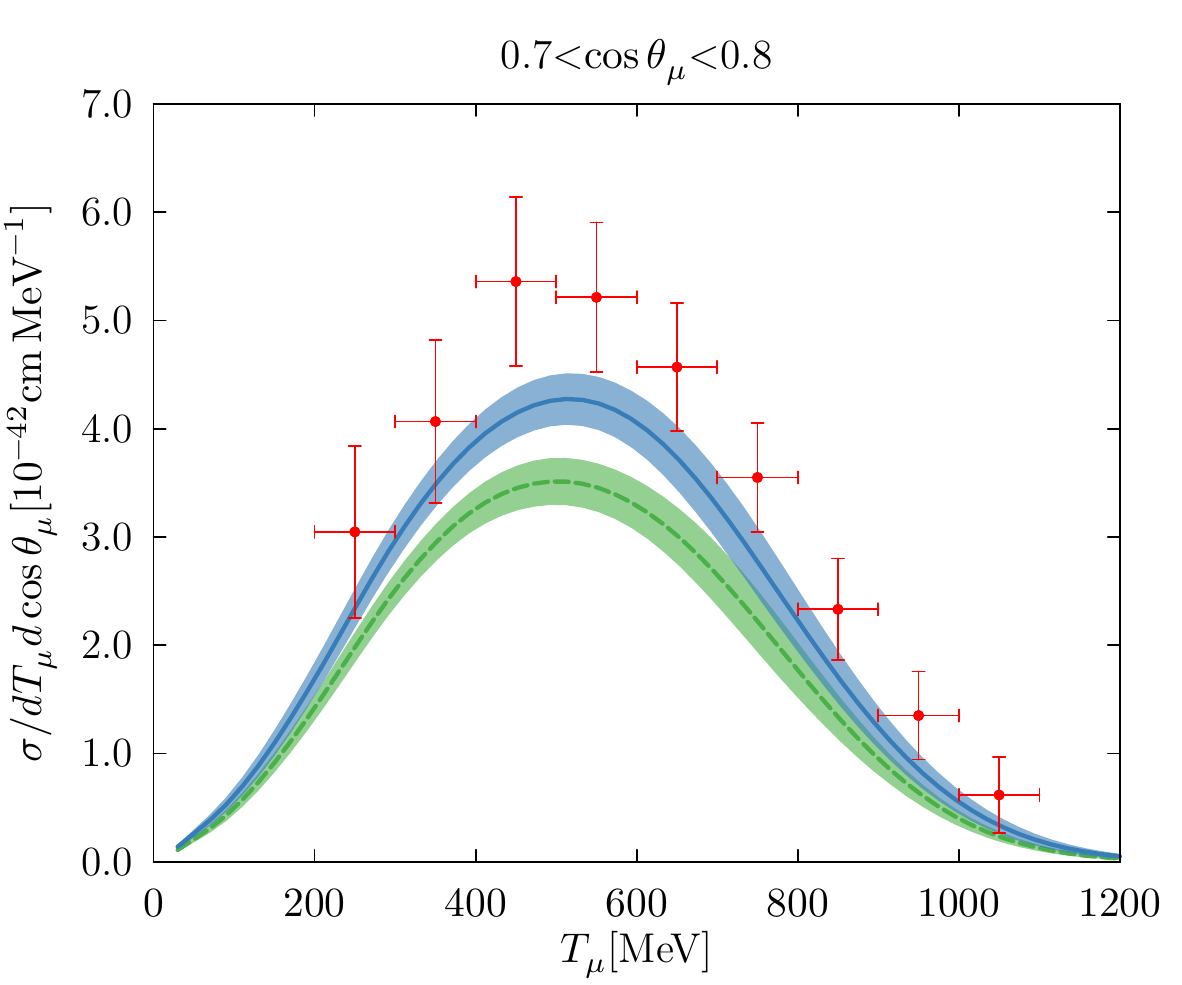}
\includegraphics[width=0.6\columnwidth]{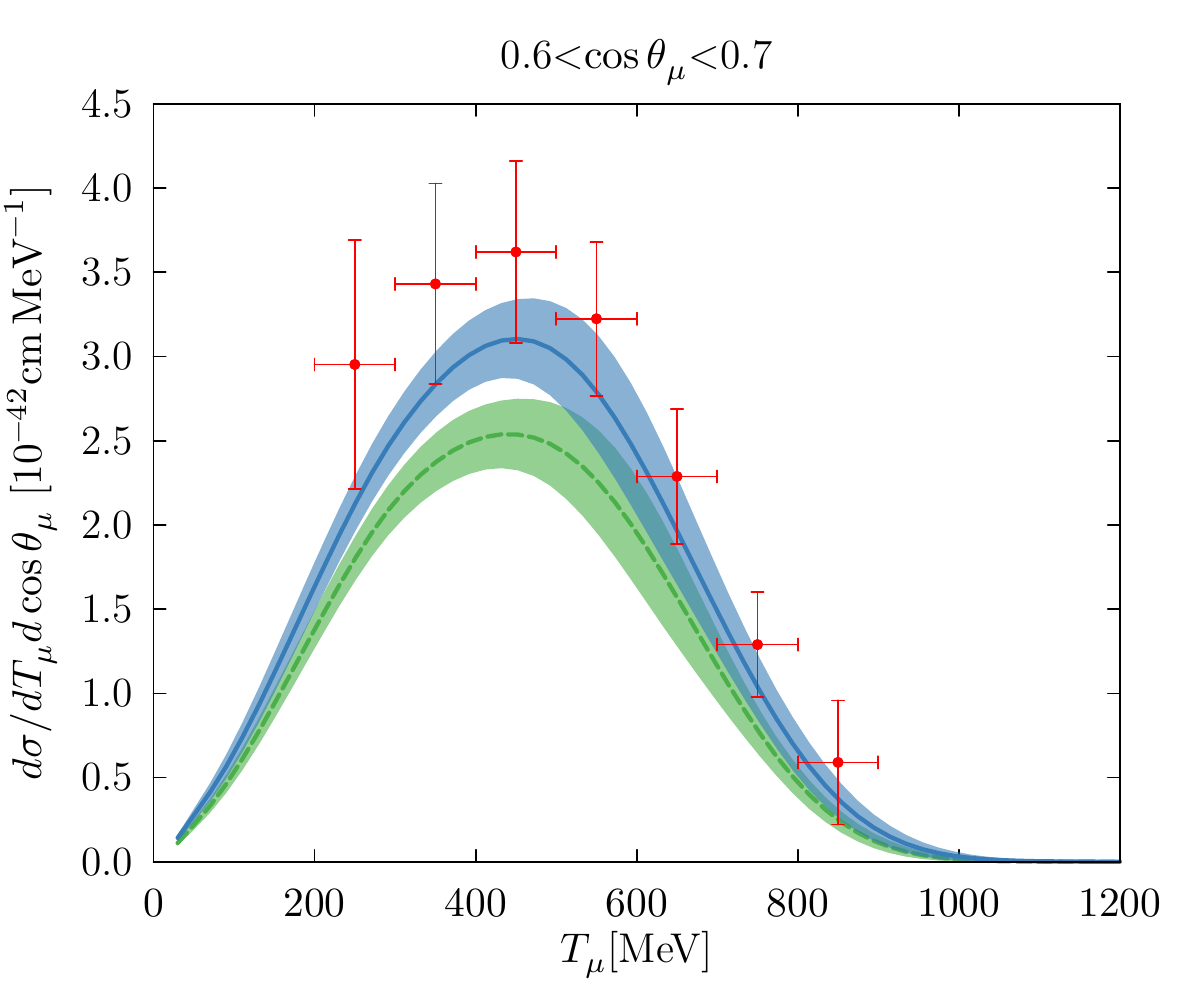}
\includegraphics[width=0.6\columnwidth]{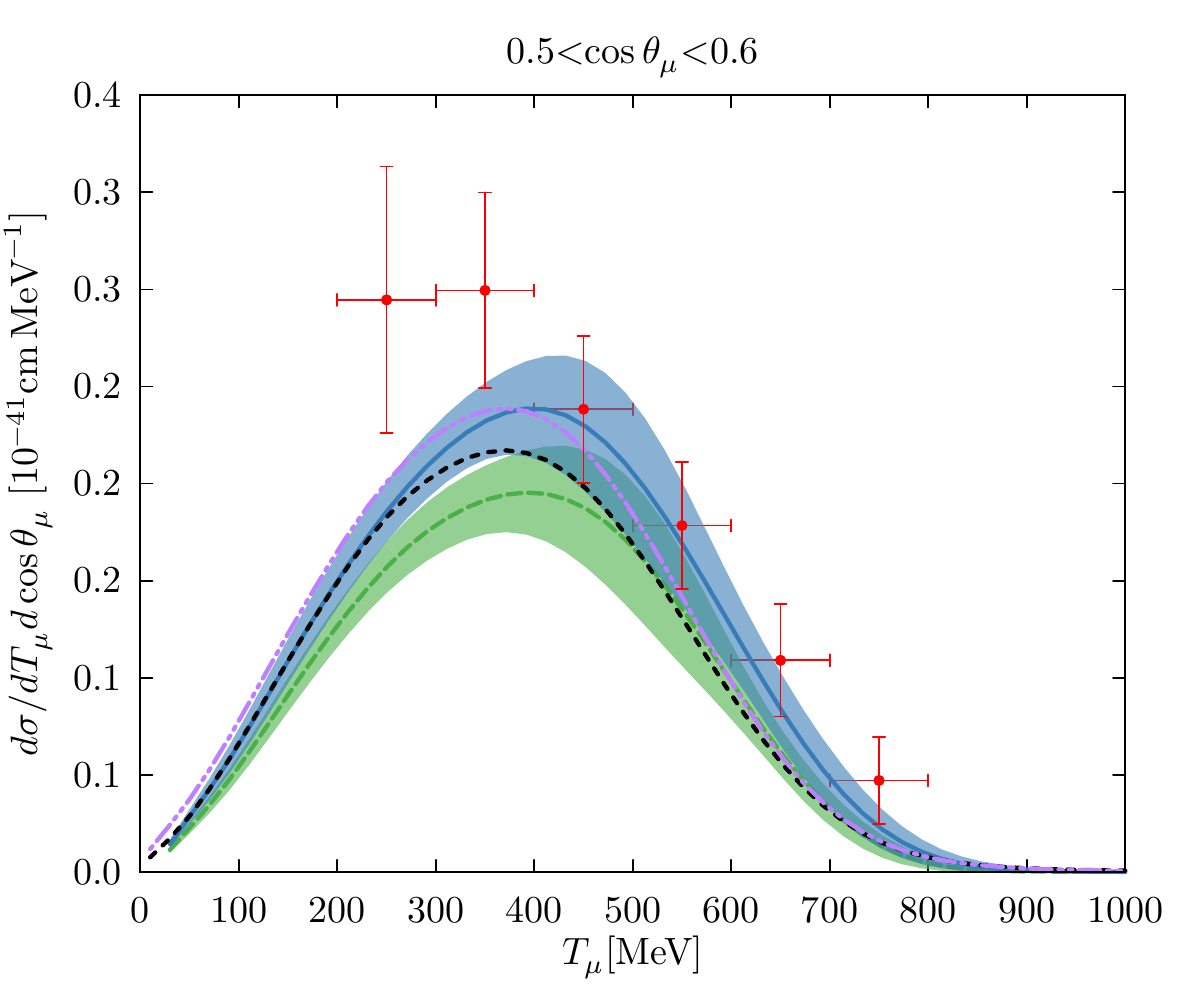}
\includegraphics[width=0.6\columnwidth]{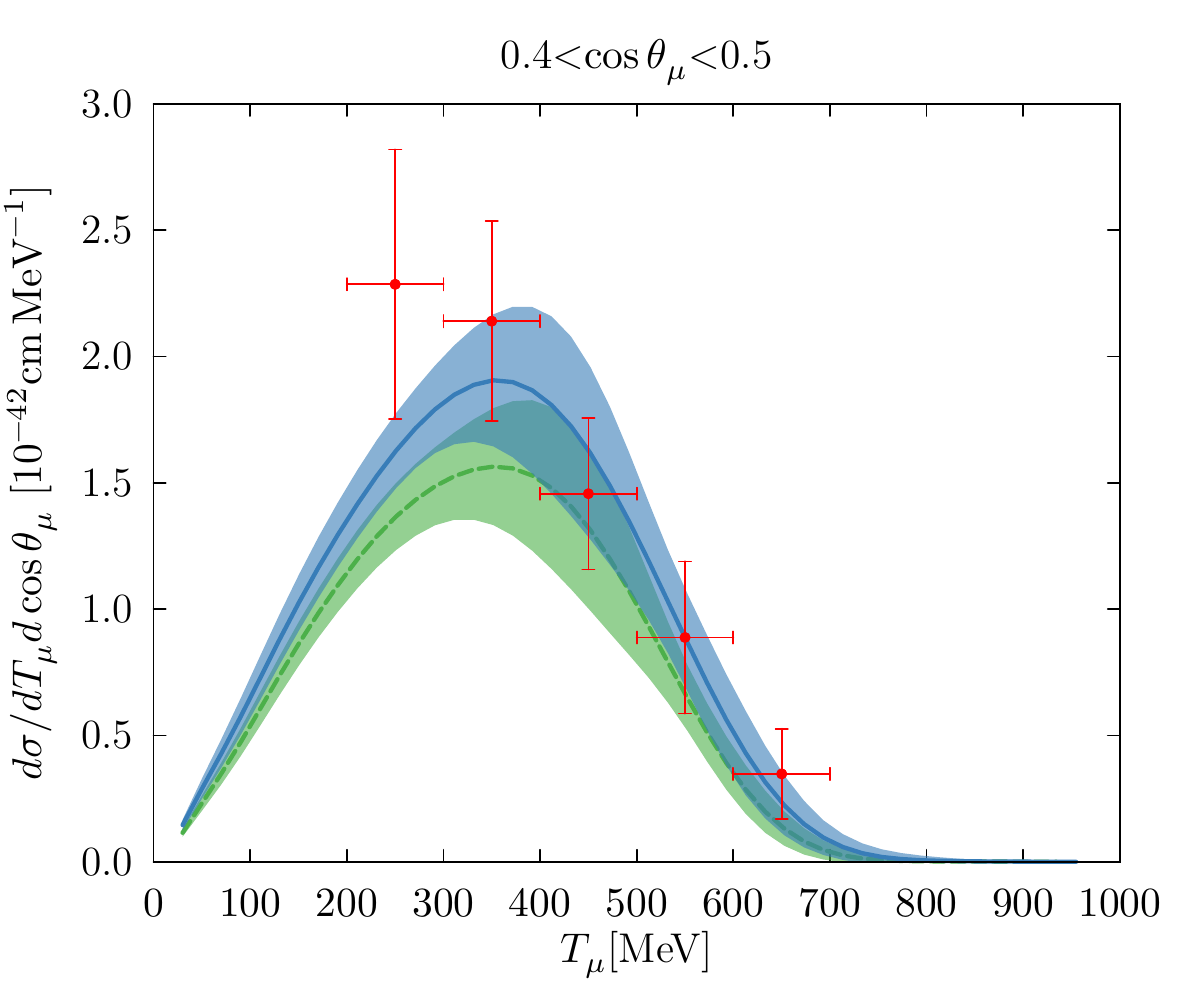}
\includegraphics[width=0.6\columnwidth]{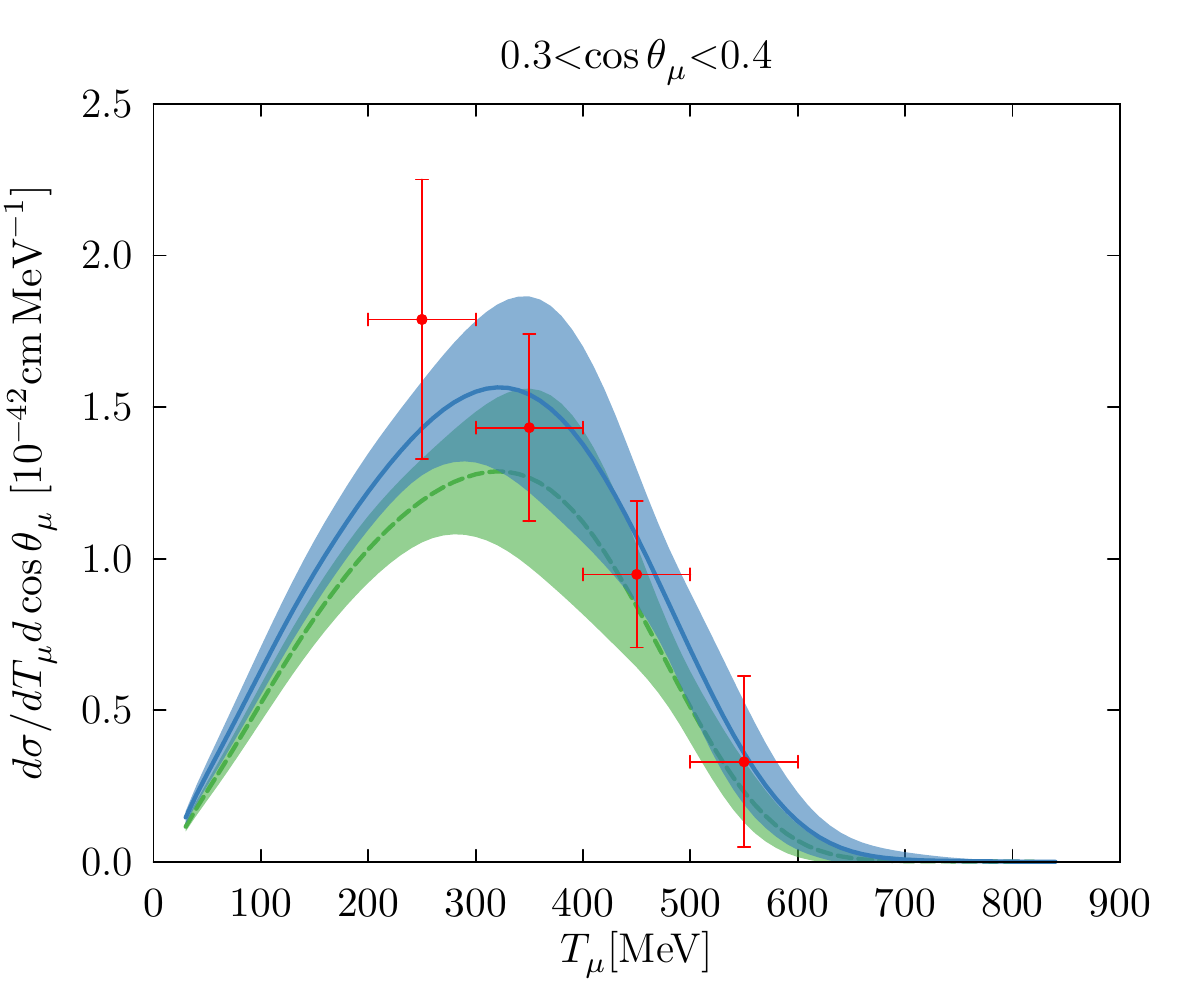}
\includegraphics[width=0.6\columnwidth]{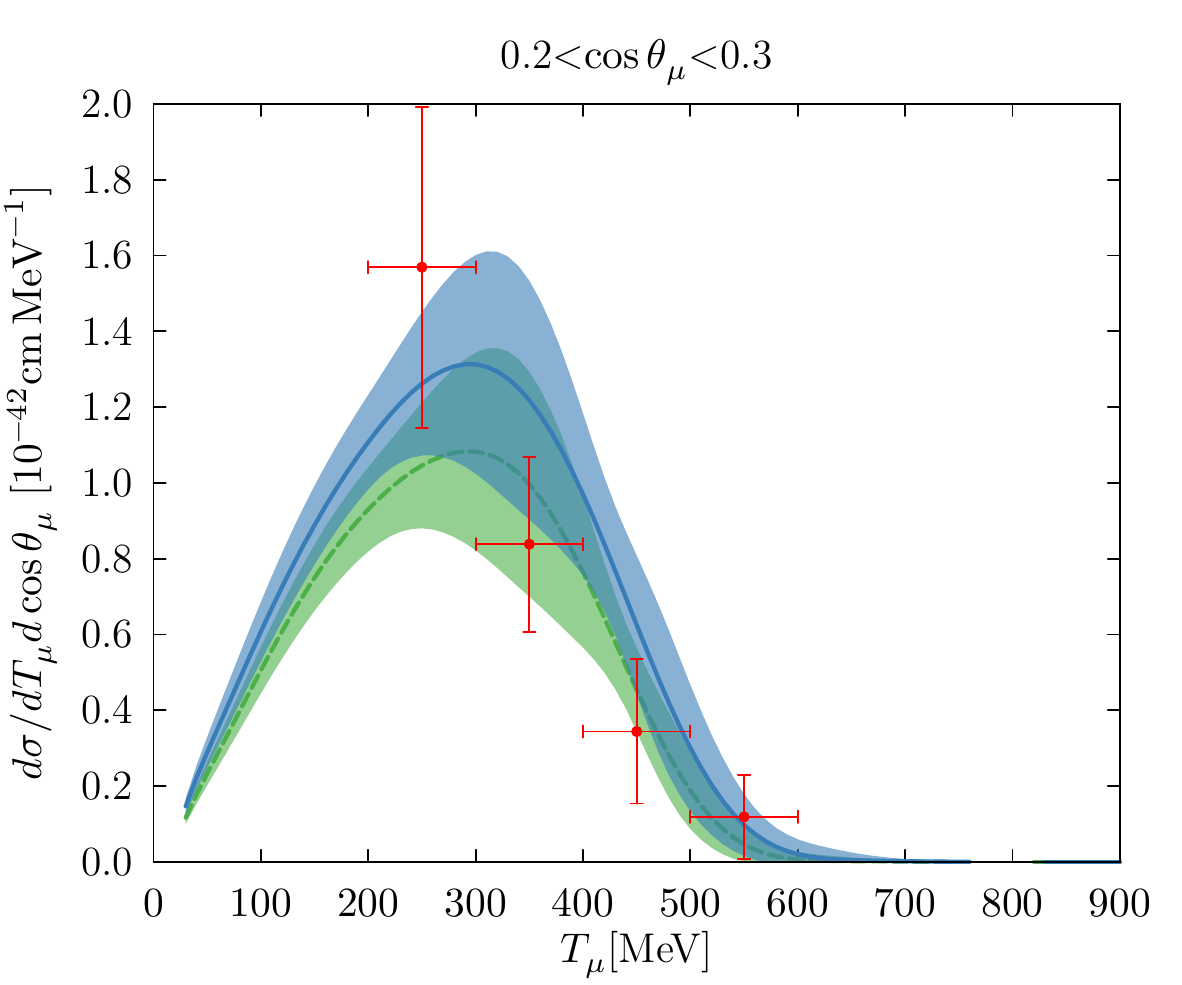}
\includegraphics[width=0.6\columnwidth]{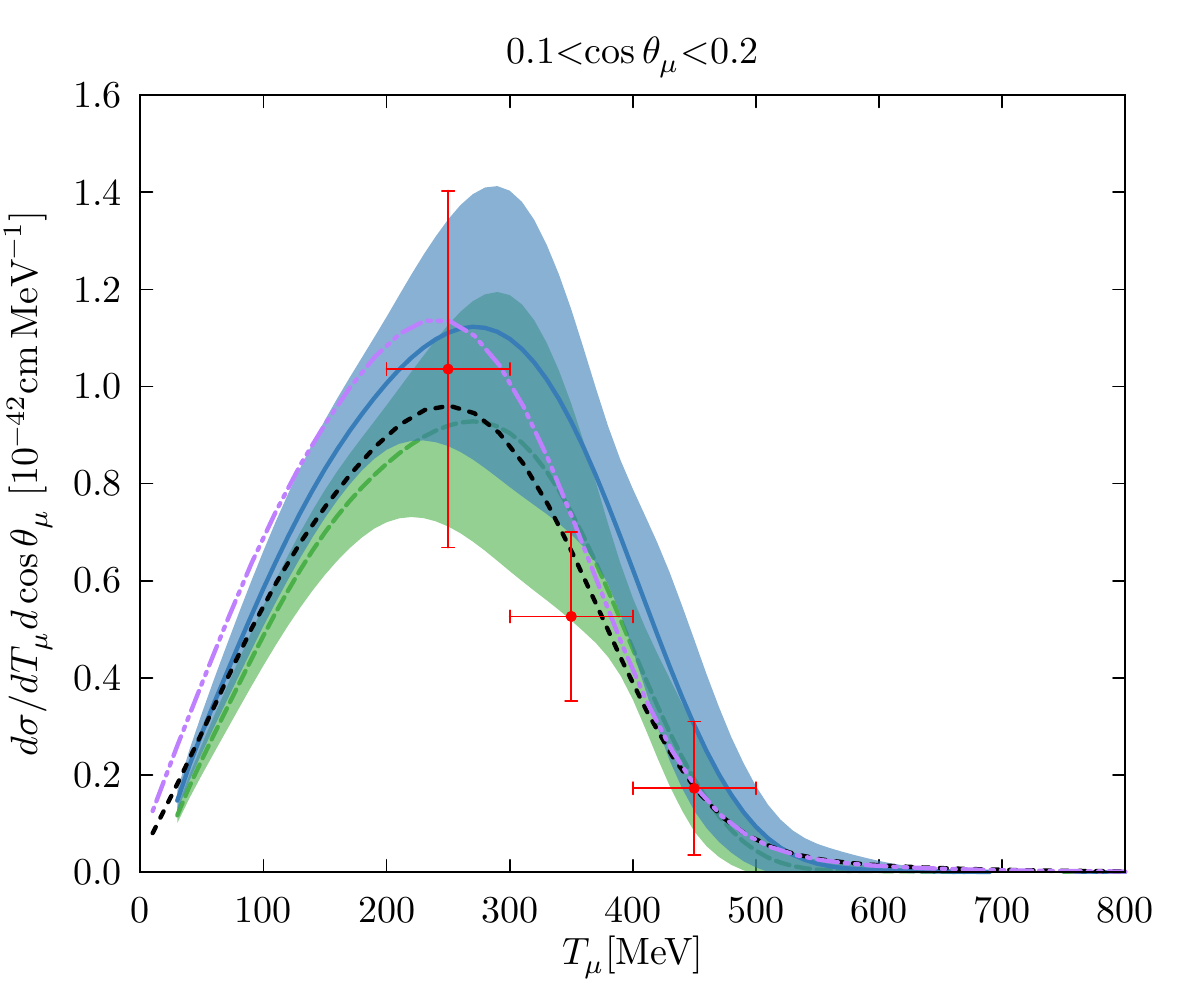}
\caption{Same as Fig.~\ref{fig:MiniBooNE_nu} but for $\overline{\nu}_\mu$-CCQE scattering. The experimental data and their shape uncertainties are from Ref.~\cite{Aguilar-Arevalo:2013dva}. The additional $17.4\%$ normalization uncertainty is not shown here.}
\label{fig:MiniBooNE_nubar}
\end{figure*}

\begin{figure*}[]
\centering
\includegraphics[width=0.6\columnwidth]{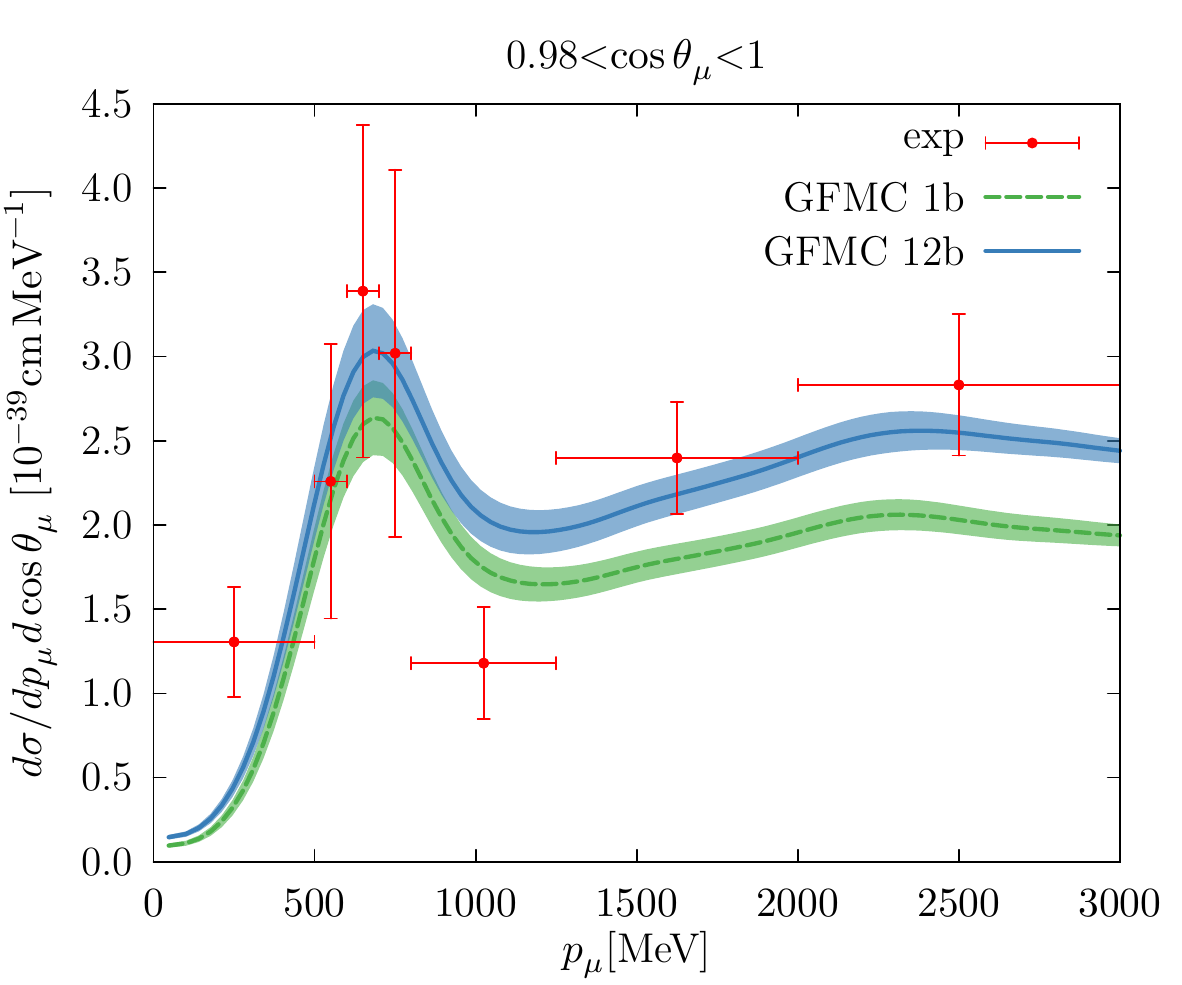}
\includegraphics[width=0.6\columnwidth]{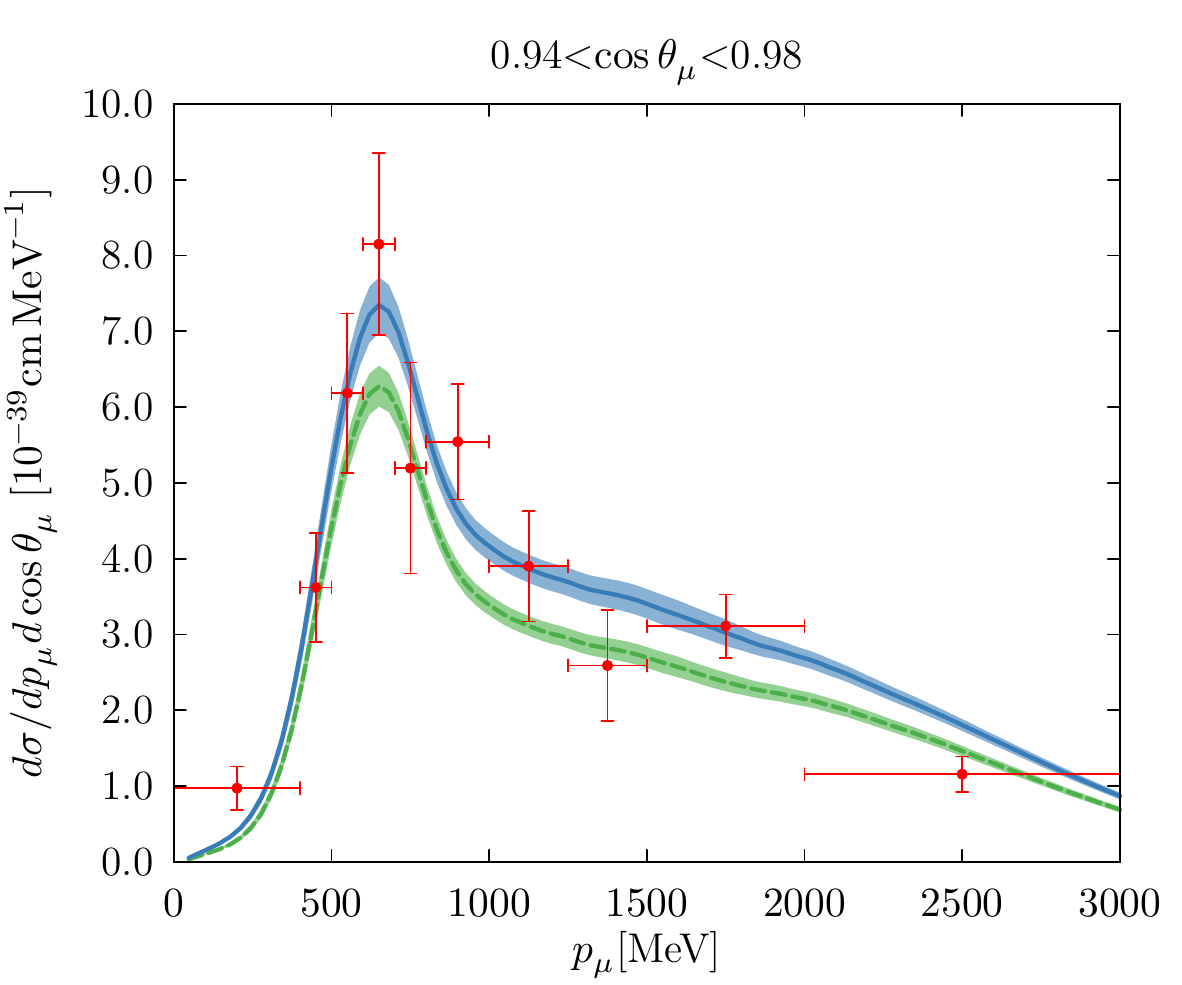}
\includegraphics[width=0.6\columnwidth]{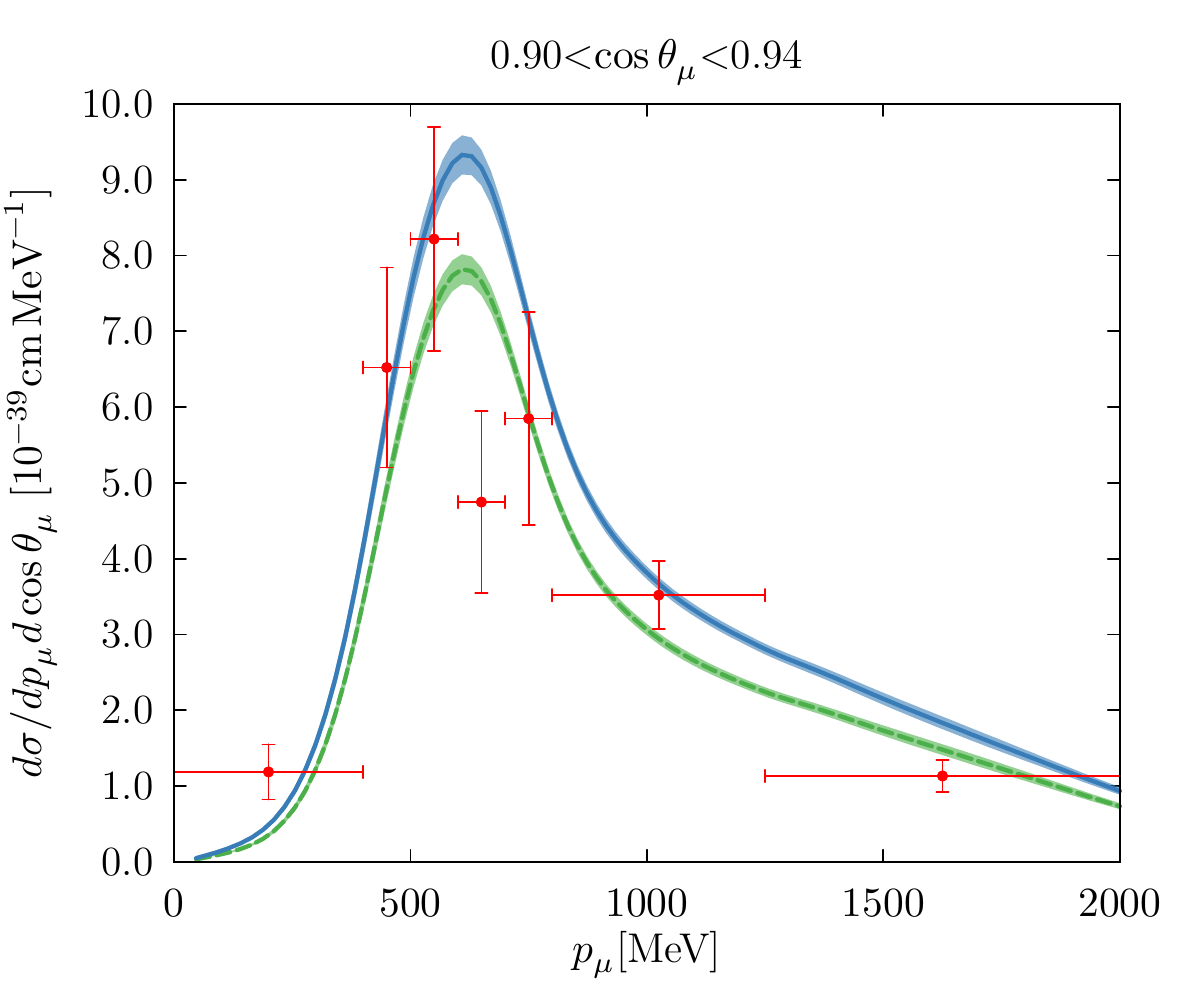}
\includegraphics[width=0.6\columnwidth]{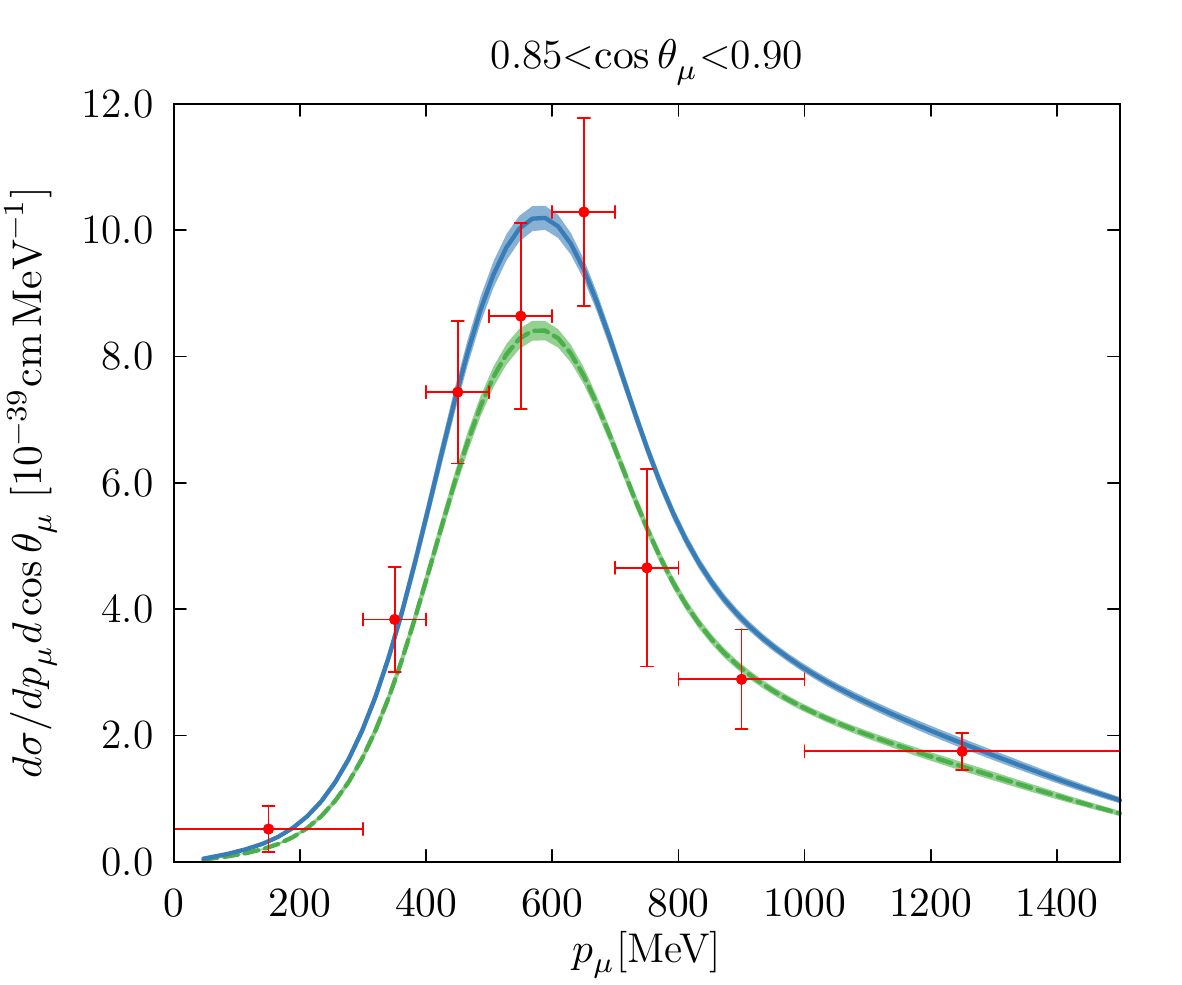}
\includegraphics[width=0.6\columnwidth]{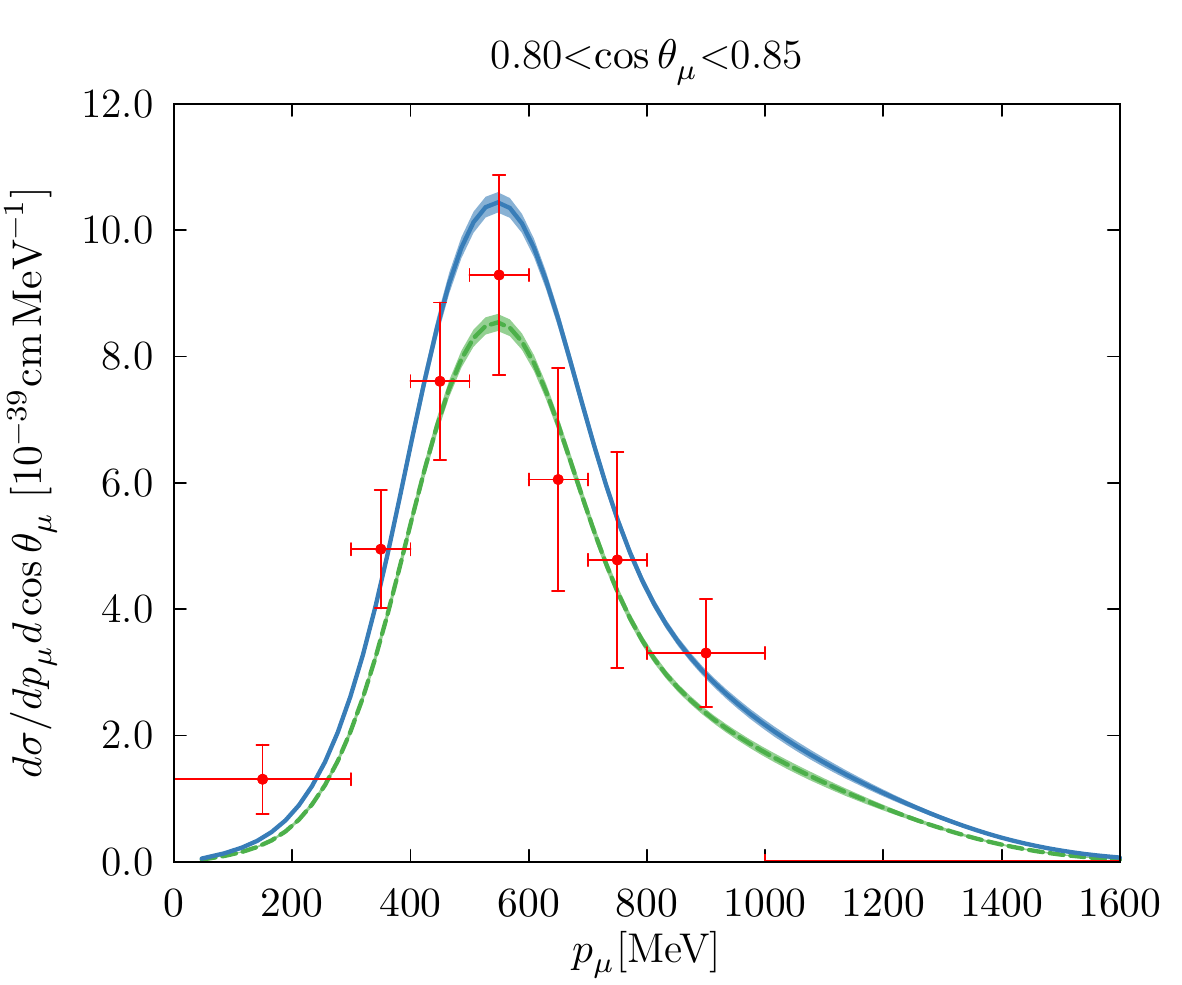}
\includegraphics[width=0.6\columnwidth]{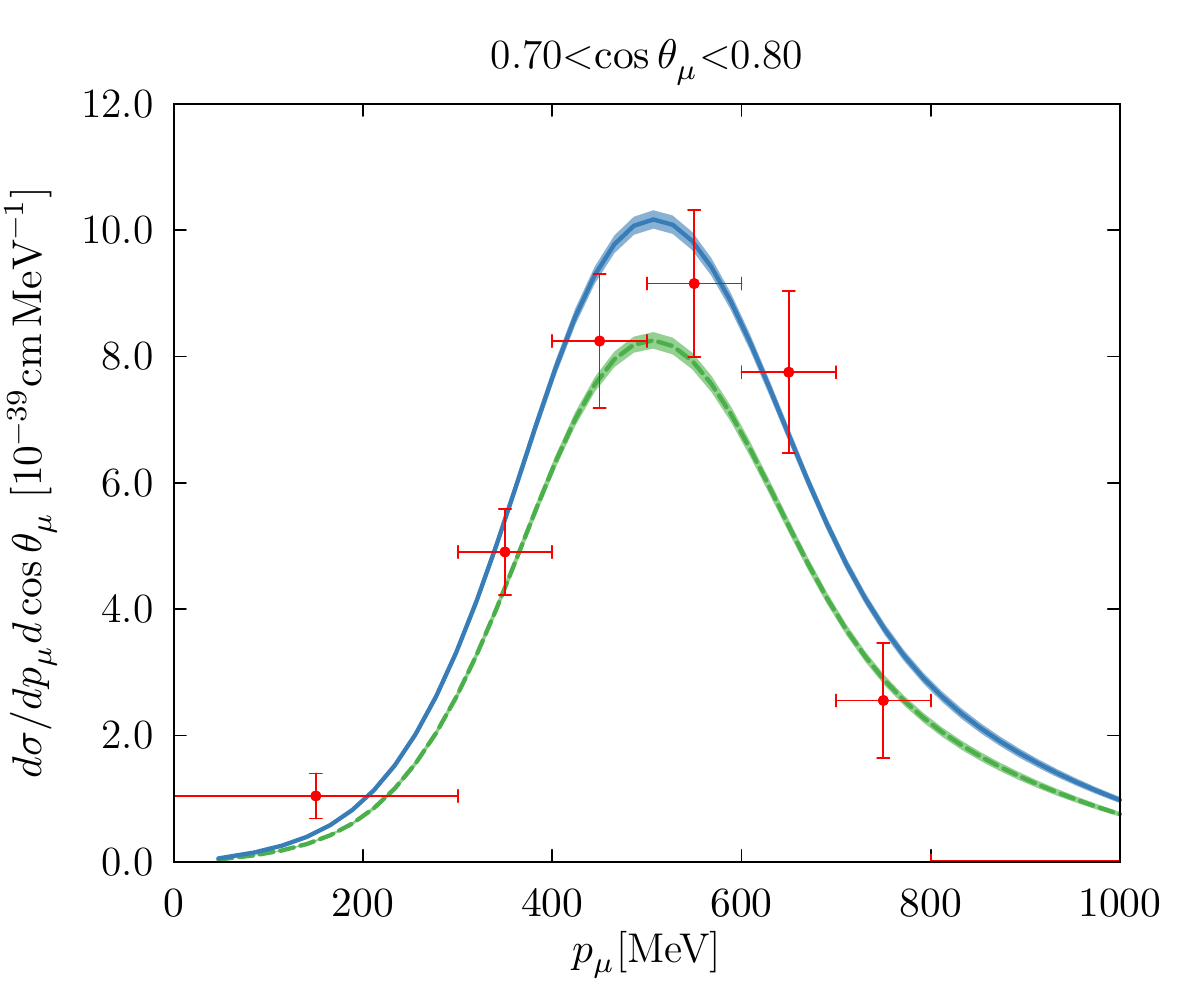}
\includegraphics[width=0.6\columnwidth]{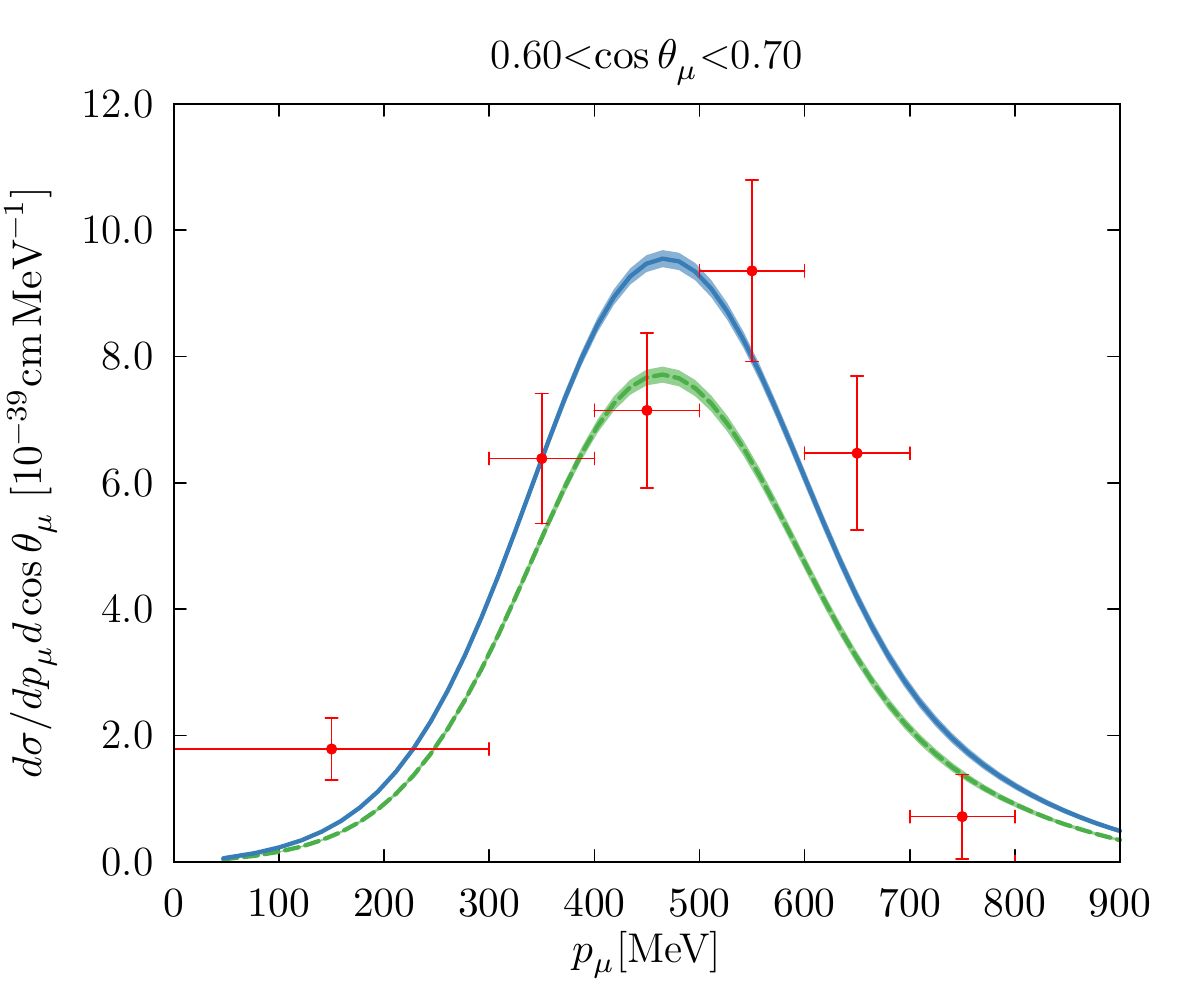}
\includegraphics[width=0.6\columnwidth]{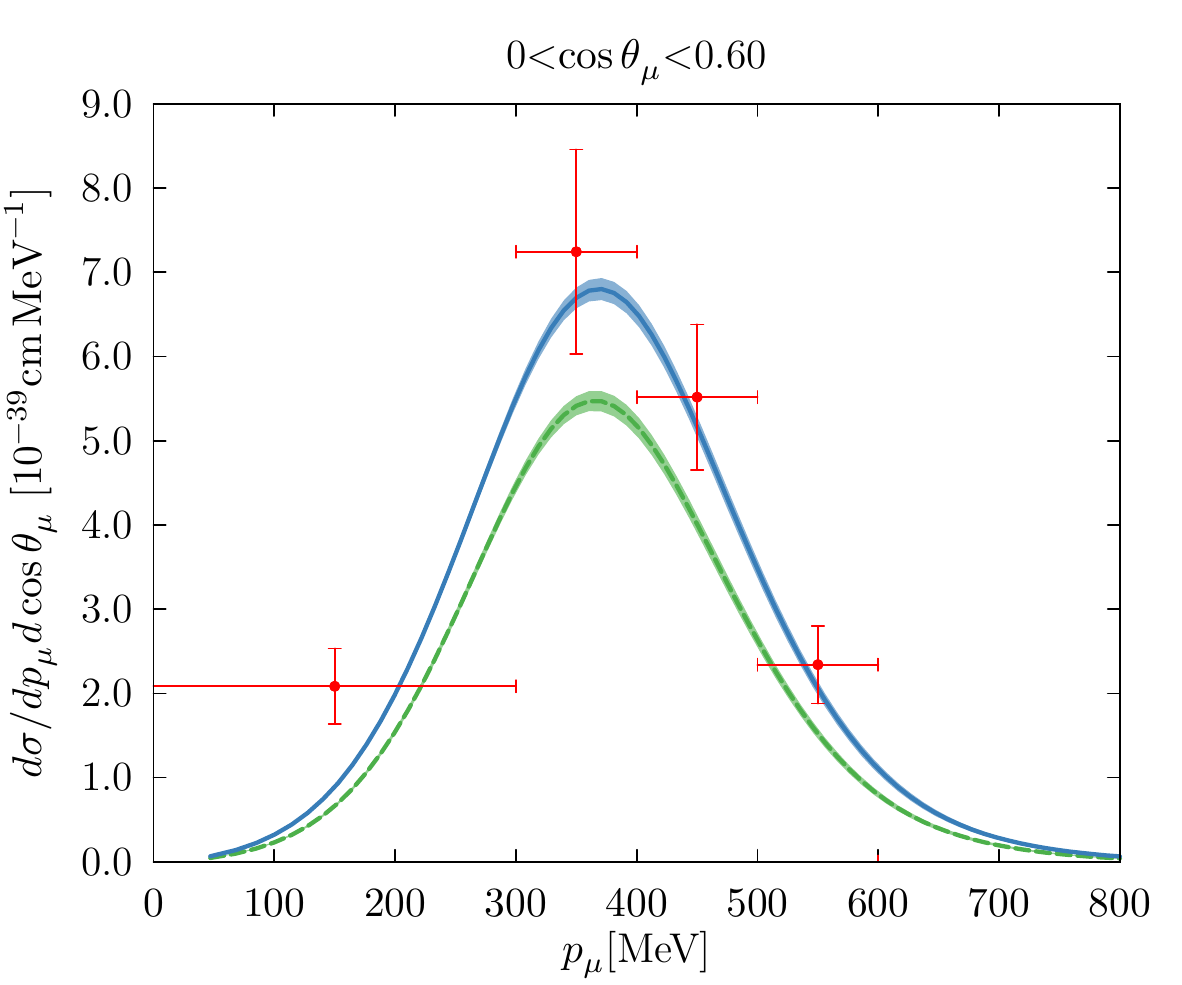}
\includegraphics[width=0.6\columnwidth]{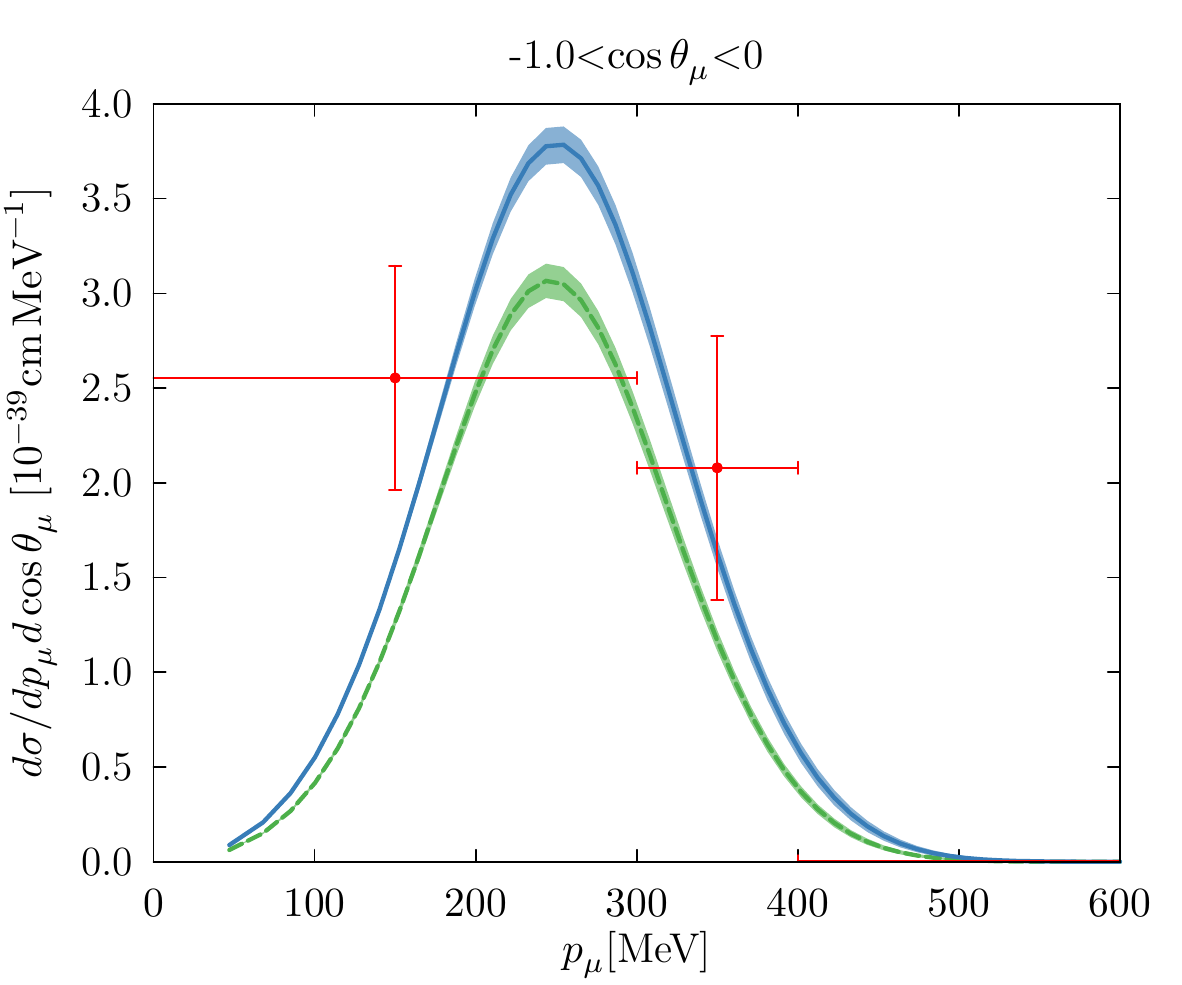}
\caption{T2K flux-folded double differential cross sections per target neutron for $\nu_\mu$-CCQE scattering on $^{12}$C, displayed as a function of the muon momentum $p_\mu$ for different ranges of $\cos\theta_\mu$. The experimental data and their shape uncertainties are from Ref.~\cite{Abe:2016tmq}.  Calculated cross sections are obtained with $\Lambda_A\,$=$\, 1.0$ GeV.}
\label{fig:T2K_nu}
\end{figure*}

Predictions for the flux-averaged cross sections on $^{12}$C corresponding to the two experiments
and obtained by including one-body only, and one- and two-body, currents are shown by, respectively,
dashed (green) and solid (blue) lines in Figs.~\ref{fig:MiniBooNE_nu}--\ref{fig:T2K_nu}.  The shaded areas result
from combining statistical errors associated with the GFMC evaluation of the Euclidean response functions,
uncertainties in the maximum-entropy inversion of them, and uncertainties due to extrapolation of the
response functions outside the calculated $(q,\omega)$ range, which is $100\,\, {\rm MeV} \leq q \leq 700$
MeV and $\omega$ from threshold to $\omega \lesssim q$.  This extrapolation is carried out by exploiting
the scaling property of the various response functions, as outlined at the end of the previous section.
The large cancellation between the dominant terms proportional to $v_{xx}\, R_{xx}$ and  $v_{xy}\, R_{xy}$
in antineutrino cross sections leads to somewhat broader error bands than for the neutrino
cross sections, for which those terms add up.  Furthermore, we note that the cross-section
scales in Figs.~\ref{fig:MiniBooNE_nu} and~\ref{fig:MiniBooNE_nubar} are different, those
for the $\overline{\nu}_\mu$-CCQE data being a factor of about 2 to 10 smaller than for the
$\nu$-CCQE data as the muon scattering angle increases from 0$^\circ$  to 90$^\circ$.

Overall, the MiniBooNE $\nu_\mu$ and $\overline{\nu}_\mu$, and T2K $\nu_\mu$,
data are in good agreement with theory, when including the contributions of
two-body currents.  This is especially noticeable in the case of the MiniBooNE
$\nu_\mu$ data at forward scattering angles.  However, the calculated cross
sections underestimate somewhat the MiniBooNE $\nu_\mu$ data at progressively
larger muon kinetic energy $T_\mu$ and backward scattering angles $\theta_\mu$,
and the $\overline{\nu}_\mu$ data at forward $\theta_\mu$ over the whole $T_\mu$
range.  By contrast, the full theory (with one- and two-body currents) appears to provide
a good description of the T2K $\nu_\mu$ data over the whole measured region.

For a given initial neutrino energy $E_\nu$, the calculated cross section is largest at the
muon energy $T_\mu$ corresponding to that of the quasielastic peak,
\begin{equation}
\label{eq:eqe}
T_\mu^{\rm qe}+m_\mu\approx\frac{E_\nu}{1+2\left(E_\nu/m\right)\, \sin^2\theta_\mu/2} \ ,
\end{equation}
where $m$ is nucleon mass, and on the r.h.s.~of the equation above we have
neglected the muon mass. The position of the quasielastic peak then moves to
the left, towards lower and lower $T^{\rm qe}_\mu$, as $\theta_\mu$ changes from
the forward to the backward hemisphere.  The general trend expected on the
basis of this simple picture is reflected in the calculation and data, even though
the cross sections in Figs.~\ref{fig:MiniBooNE_nu}-\ref{fig:T2K_nu} result from a
folding with the neutrino flux, which is far from being monochromatic.  Nevertheless,
the correlation between peak location in the flux-averaged cross sections and
$\theta_\mu$ remains.  For example, the T2K flux is largest at $E_\nu\approx 560$ MeV
and fairly narrow; hence, one would expect the T2K flux-averaged
cross section be peaked at the muon momentum $p_\mu^{\,{\rm qe}}\approx 550$ MeV for $\cos \theta_\mu\,$=$\, 1$,
and $p_\mu^{\,{\rm qe}}\approx 450$ MeV for $\cos \theta_\mu\,$=$\, 0.65$,
in reasonable accord with the data of Fig.~\ref{fig:T2K_nu}.

In Figs.~\ref{fig:MiniBooNE_nu} and~\ref{fig:MiniBooNE_nubar} 
we also present the flux-folded $\nu_\mu$ and $\overline{\nu}_\mu$ cross sections obtained in plane-wave-impulse-approximation (PWIA) 
for three different bins in $\cos\theta_\mu$ (corresponding to the forward, intermediate, and backward region) of the MiniBooNE data.  
We have adopted here the most naive (non-relativistic) formulation of PWIA based on the single-nucleon
momentum distribution rather than the spectral function.\footnote{It should be noted here that
{\it ab initio} calculations of the $^{12}$C spectral functions are not currently available.}
Hence, the PWIA response functions follow from
\begin{eqnarray}
\hspace{-0.5cm}R^{\rm PWIA}_{\alpha\beta} (q, \omega)\! &=&\!\int d{\bf p}\, N({\bf p}) \, x_{\alpha\beta}({\bf p},{\bf q}, \omega)  \nonumber\\
\hspace{-0.5cm}&&\hspace{-0.5cm}\times\,
\delta\!\left( \omega-\overline{E}-\frac{\left|{\bf p}+{\bf q}\right|^2}{2\, m} -\frac{p^2}{2\, m_{A-1} }\right) \ ,
\end{eqnarray} 
where the factors $x_{\alpha\beta}({\bf p},{\bf q},\omega)$ denote appropriate
combinations of the CC components (the same single-nucleon CC utilized in
the GFMC calculations), and $N({\bf p})$ is the nucleon momentum distribution in $^{12}$C
(as calculated in Ref.~\cite{Wiringa:2013ala}).  The effects of nuclear interactions are subsumed
in the single parameter $\overline{E}$, which can be interpreted as an average separation energy
(we take the value $\overline{E}\approx 20$ MeV).  The remaining terms in the $\delta$-function
are the final energies of the struck nucleon and recoiling ($A$--1) system of mass $m_{A-1}$.
From these $R^{\rm PWIA}_{\alpha\beta}$ we obtain the corresponding flux-folded cross sections
shown in Figs.~\ref{fig:MiniBooNE_nu} and ~\ref{fig:MiniBooNE_nubar} by the short-dashed (black)
line labeled PWIA.  Also shown in this figure
by the dot-dashed (purple) line (labeled PWIA-R)
are PWIA cross sections obtained by first fixing the nucleon electroweak form factor
entering $x_{\alpha\beta}({\bf p},{\bf q},\omega)$ at $Q^2_{\rm qe}$,
and then rescaling the various response functions by ratios of these
form factors, as indicated in Sec.~\ref{sec:sec2.b}.

A couple of comments are in order.  First, the cross sections in PWIA are to be compared
to those obtained with the GFMC method by including only one-body currents (curves
labeled GFMC 1b): they are found to be systematically larger than the GFMC predictions,
particularly at forward angles.
Furthermore, it appears that the (spurious) excess strength in the PWIA cross sections
(in the same forward-angle kinematics) matches the increase produced by two-body
currents in the GFMC calculations (difference between the GFMC 1b and GFMC 12b curves).
This should be viewed as accidental.

Second, the PWIA and PWIA-R cross sections are very close to each other, except in
the $\overline{\nu}$ case at backward angles.  In this kinematical regime
there are large cancelations between the dominant terms proportional to the transverse
and interference response functions; indeed, as $\theta_\mu$ changes from $0^\circ$ to
about $90^\circ$, the $\overline{\nu}$ cross section drops by an order of magnitude.
As already noted, these cancellations are also observed in the complete (GFMC 12b) calculation,
and lead to the rather broad uncertainty bands in Fig.~\ref{fig:MiniBooNE_nubar}.
Aside from this qualification, however, the closeness between the PWIA and PWIA-R results
provides corroboration for the validity of the rescaling procedure of the electroweak form factors,
needed to carry out the GFMC computation of the Euclidean response functions.

\section{Conclusions}
\label{sec:sec4}
We have reported on an {\it ab initio} study, based on realistic nuclear interactions
and electroweak currents, of neutrino (and antineutrino) inclusive scattering on
$^{12}$C in the CCQE regime of the MiniBooNE and T2K data.  Nuclear response
functions have been calculated with QMC methods and, therefore, within the description
of nuclear dynamics that we have adopted here, fully include the effects of many-body
correlations induced by the interactions in the initial and final states, and correctly
account for the important (constructive) interference between one- and two-body current
contributions.  This interference leads to a significant increase in the cross-section
results obtained in impulse approximation, and is important for bringing
theory into much better agreement with experiment. 

The nucleon and nucleon-to-$\Delta$ electroweak form factors entering
the currents have been taken from modern parameterizations of 
elastic electron scattering data on the nucleon and deuteron, and
neutrino scattering data on the proton and deuteron.  In particular,
the $Q^2$-dependence of the nucleon axial form factor $G_A(Q^2)$ is of a dipole
form with a cutoff $\Lambda_A\,$$\approx$$\,1$ GeV.  The nucleon-to-$\Delta$
axial coupling constant $g_A^*$ has been fixed by reproducing the Gamow-Teller
matrix element measured in tritium $\beta$ decay, while the $Q^2$-dependence of its
(transition) form factor $G_A^*(Q^2)$ has simply been assumed to be the same as that
of $G_A(Q^2)$, since no experimental information is currently available on $G_A^*(Q^2)$.

\begin{figure*}[]
\centering
\includegraphics[width=0.6\columnwidth]{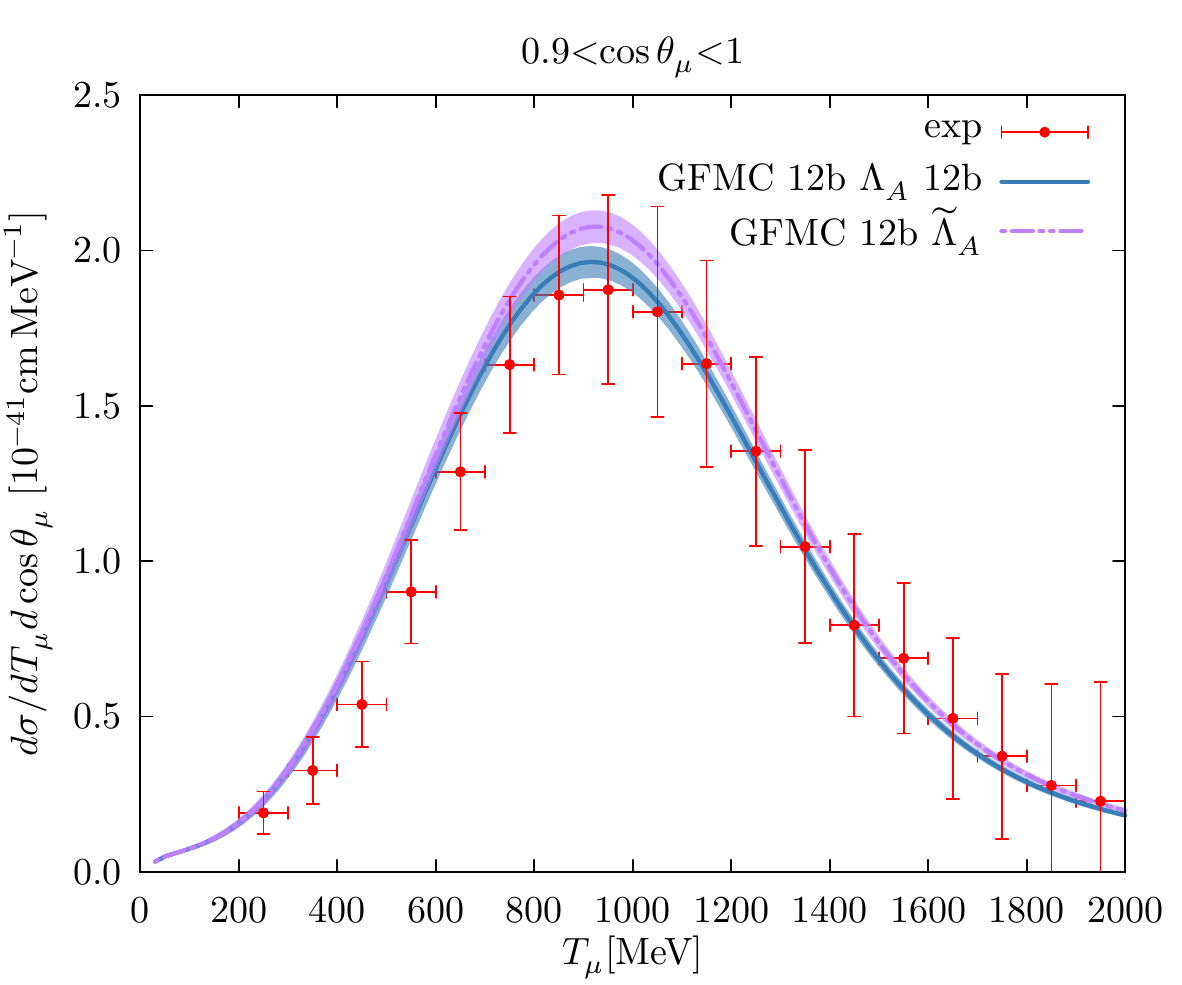}
\includegraphics[width=0.6\columnwidth]{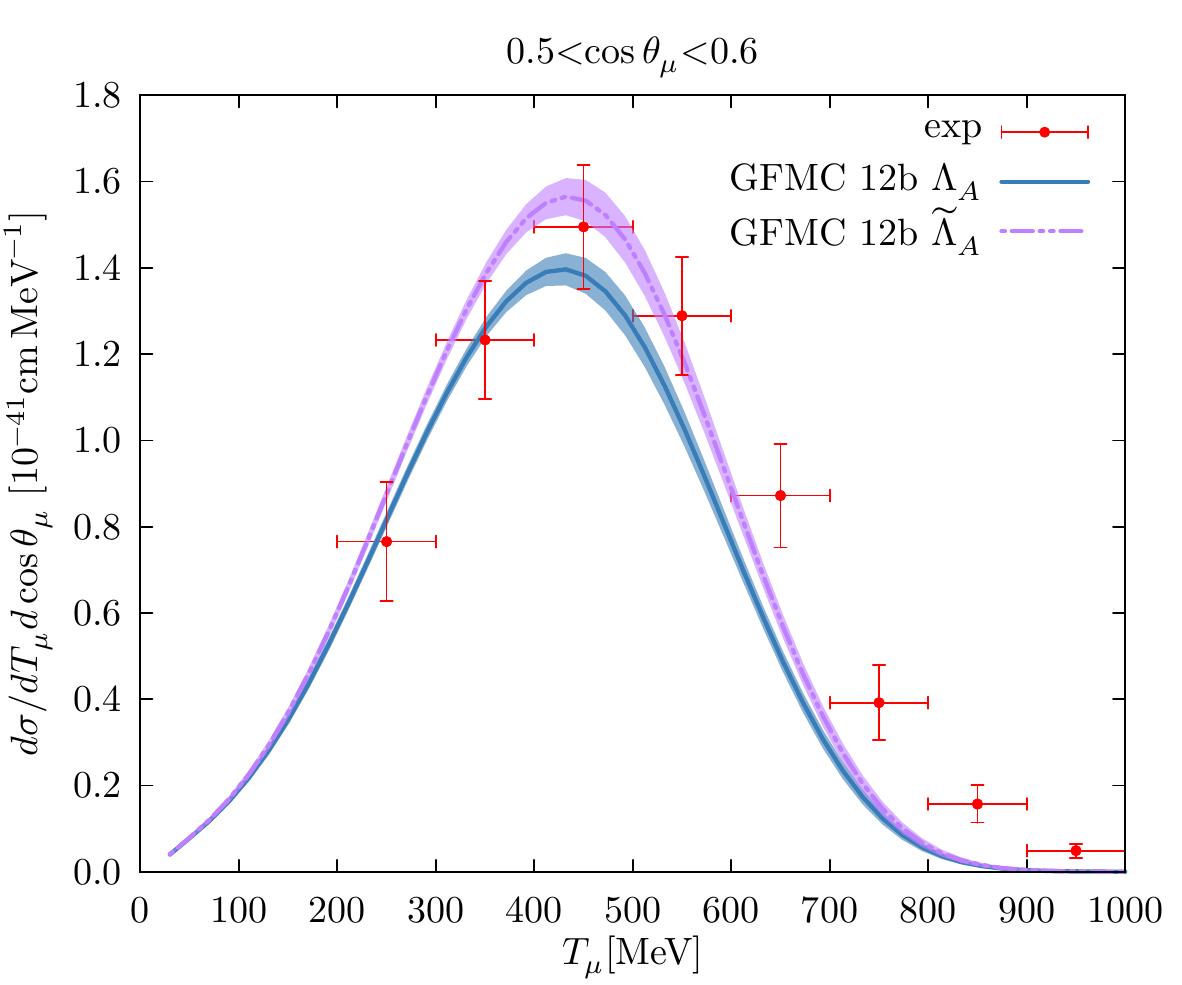}
\includegraphics[width=0.6\columnwidth]{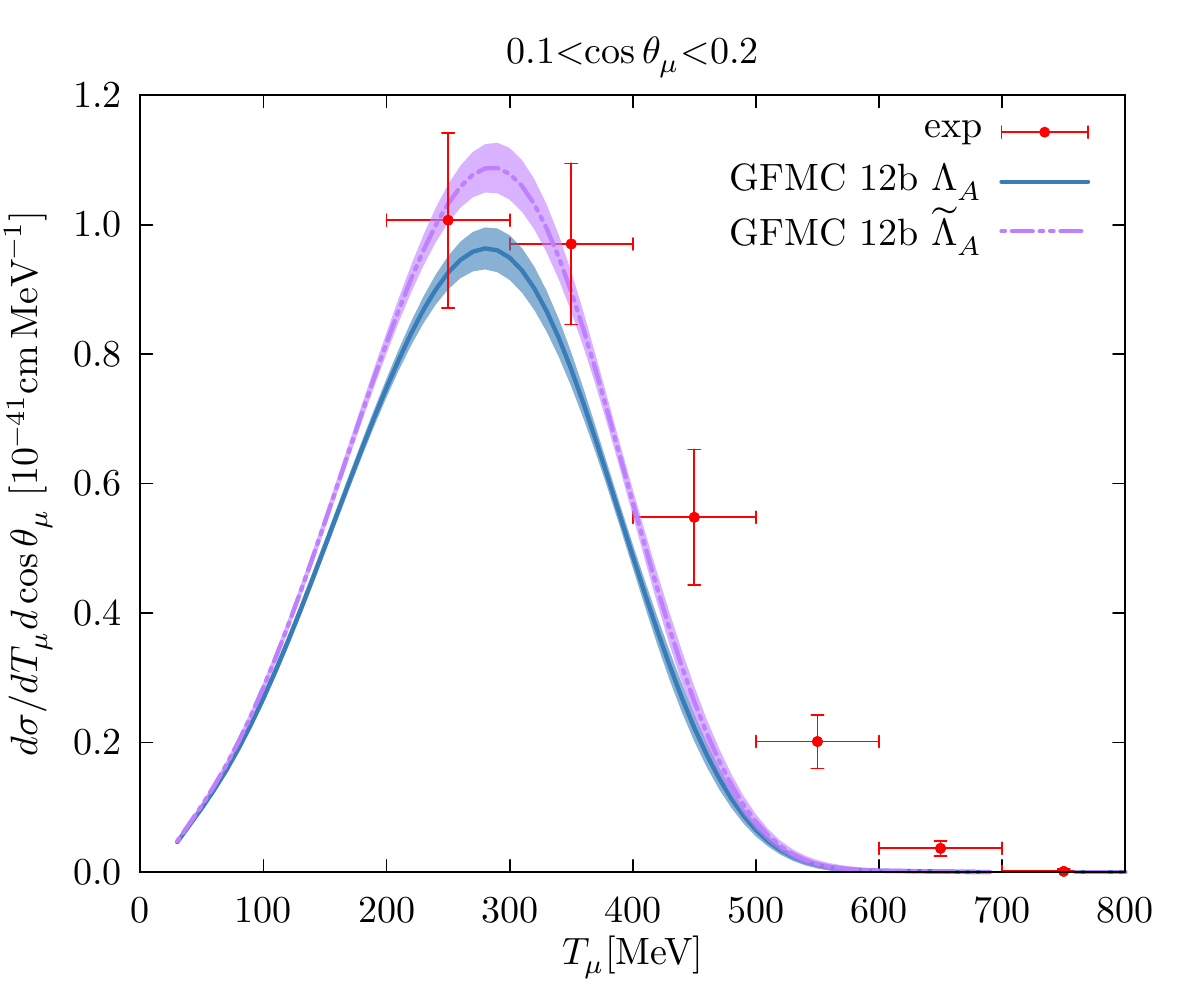}\\
\includegraphics[width=0.6\columnwidth]{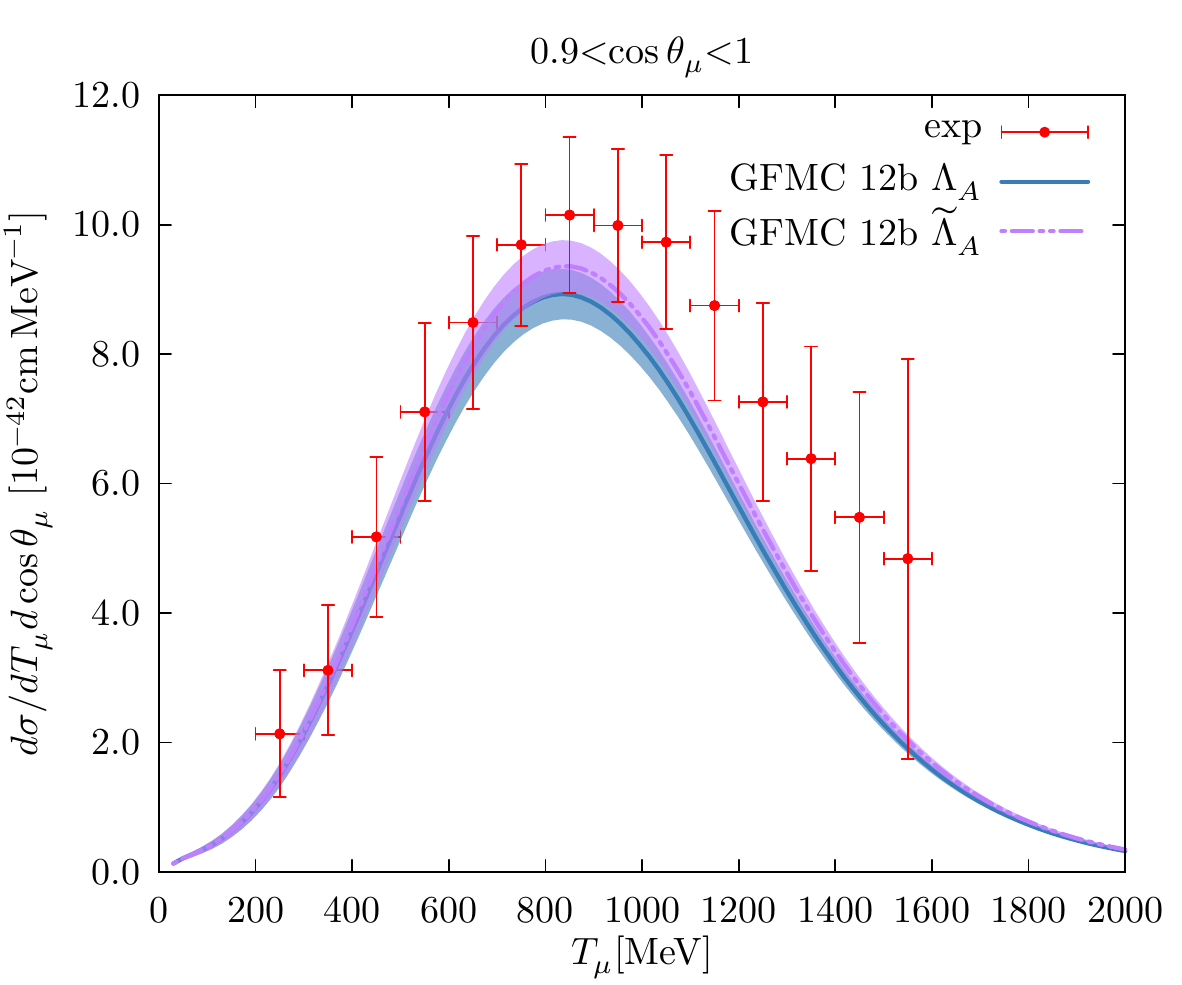}
\includegraphics[width=0.6\columnwidth]{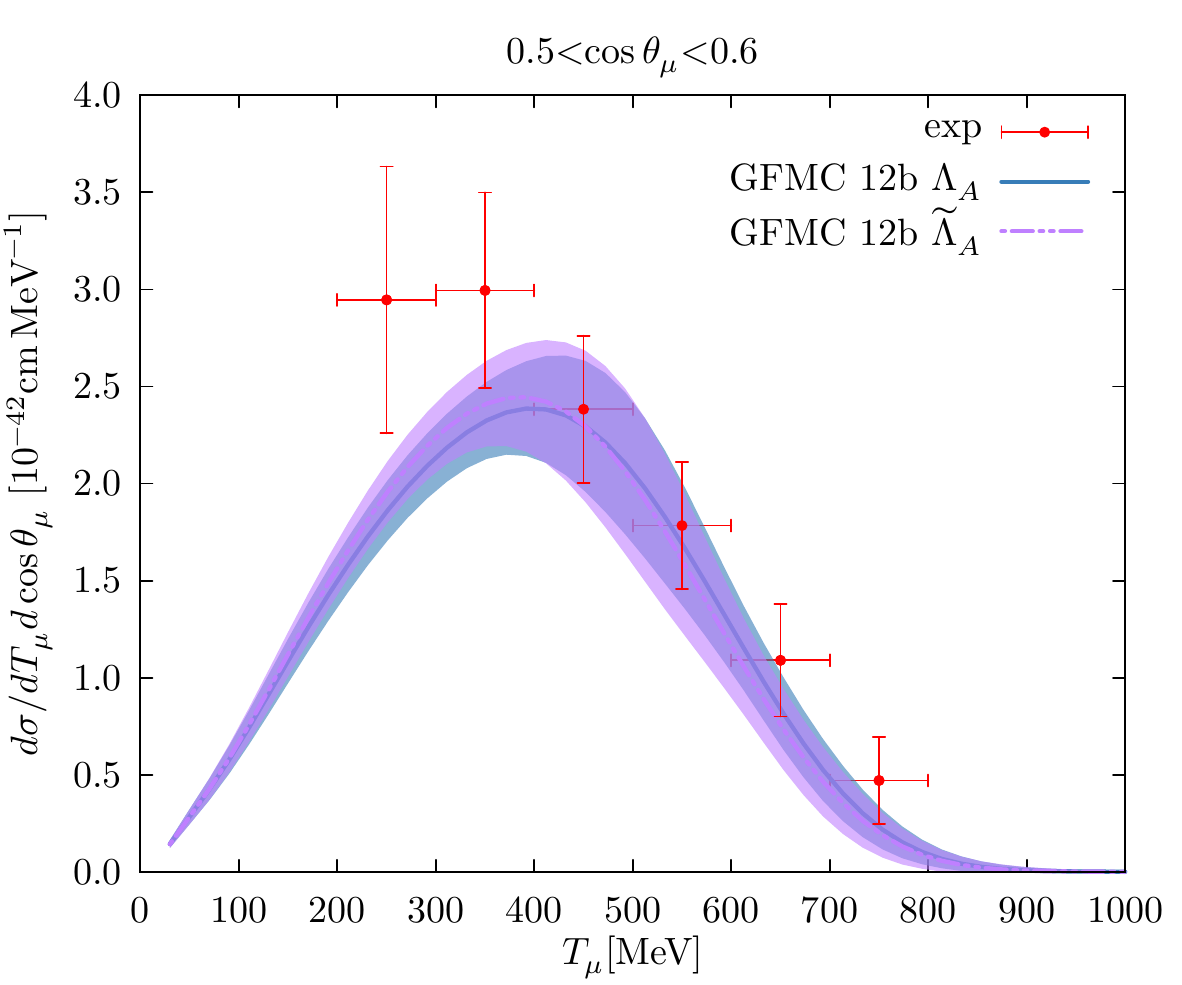}
\includegraphics[width=0.6\columnwidth]{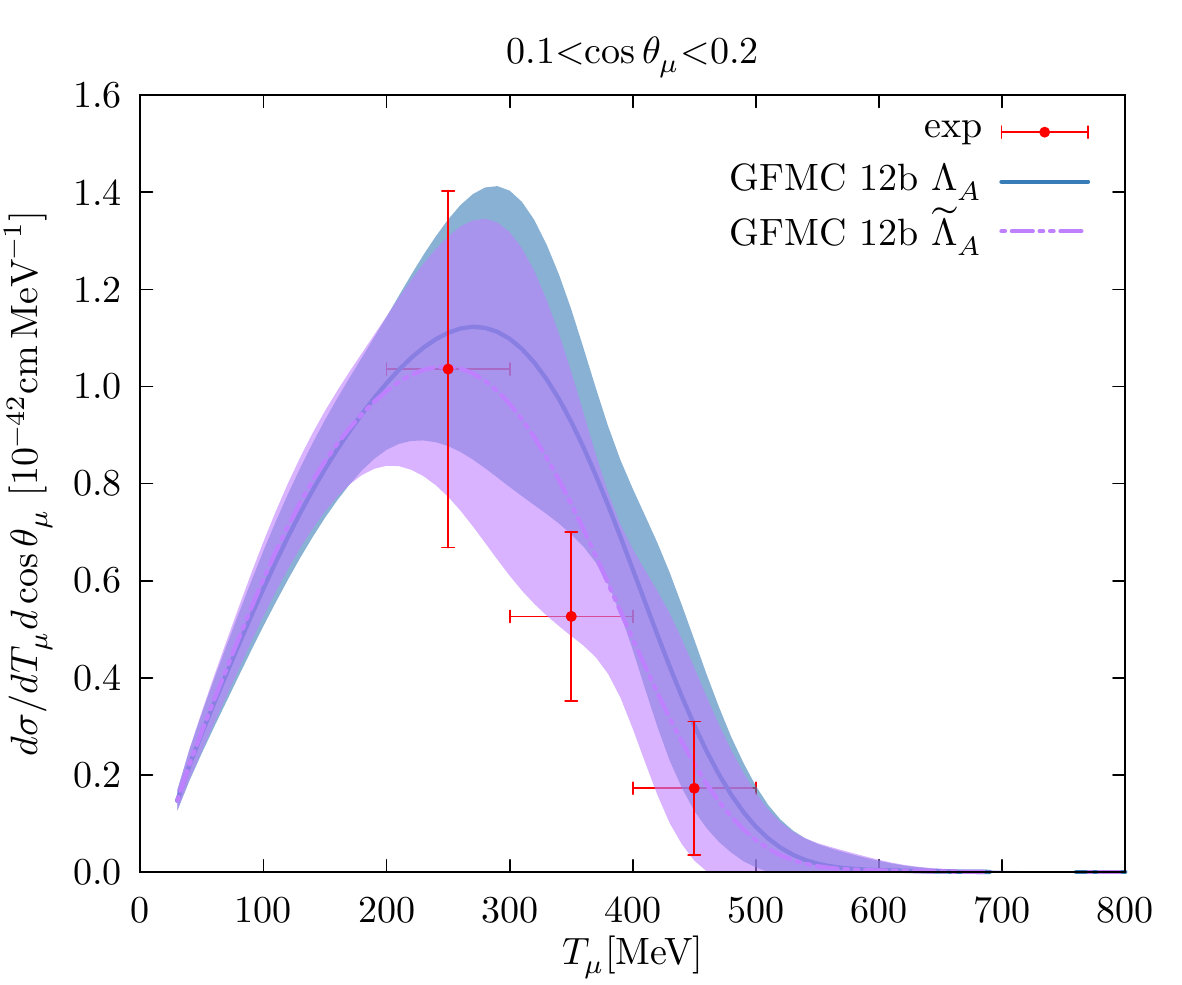}\\
\includegraphics[width=0.6\columnwidth]{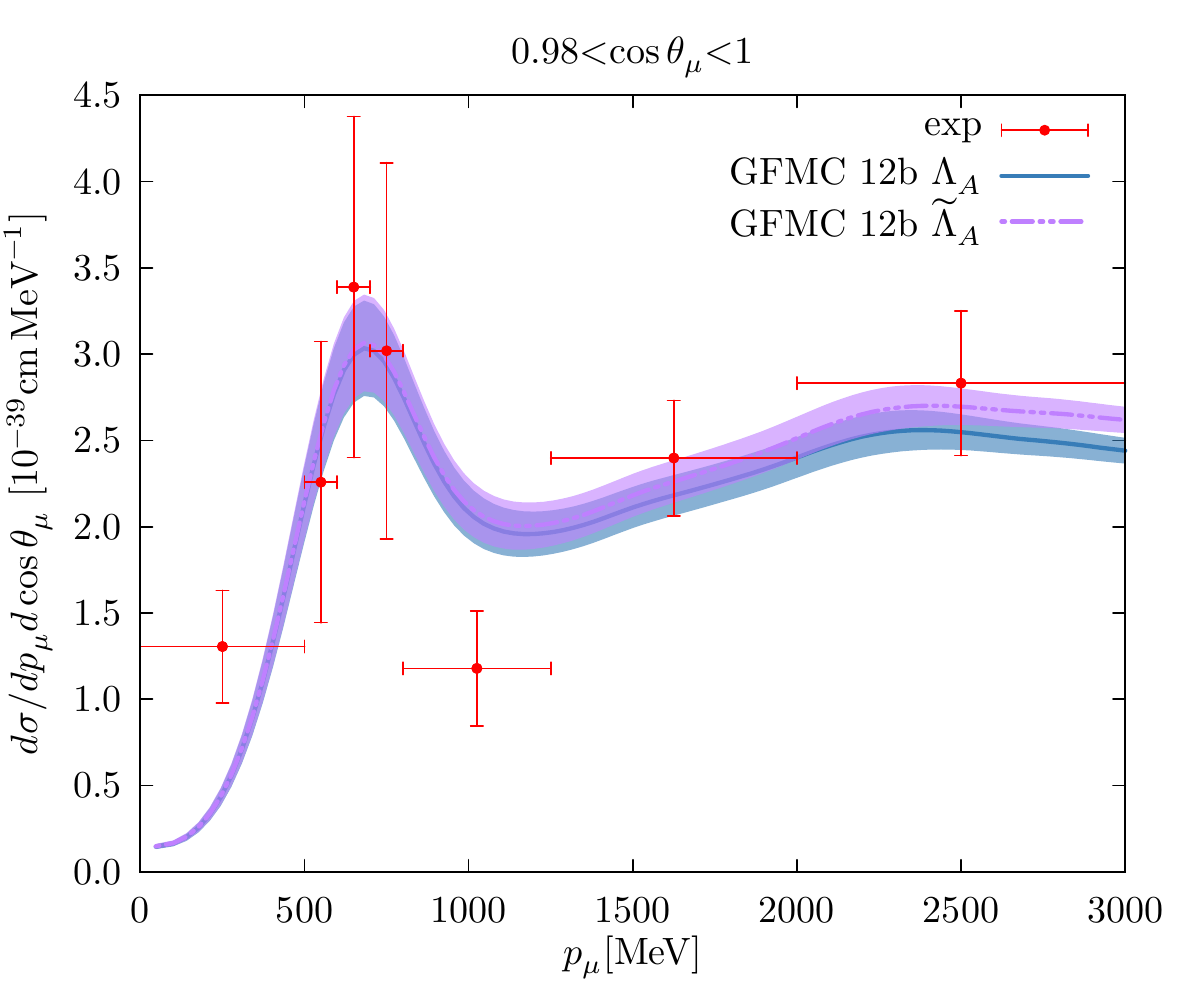}
\includegraphics[width=0.6\columnwidth]{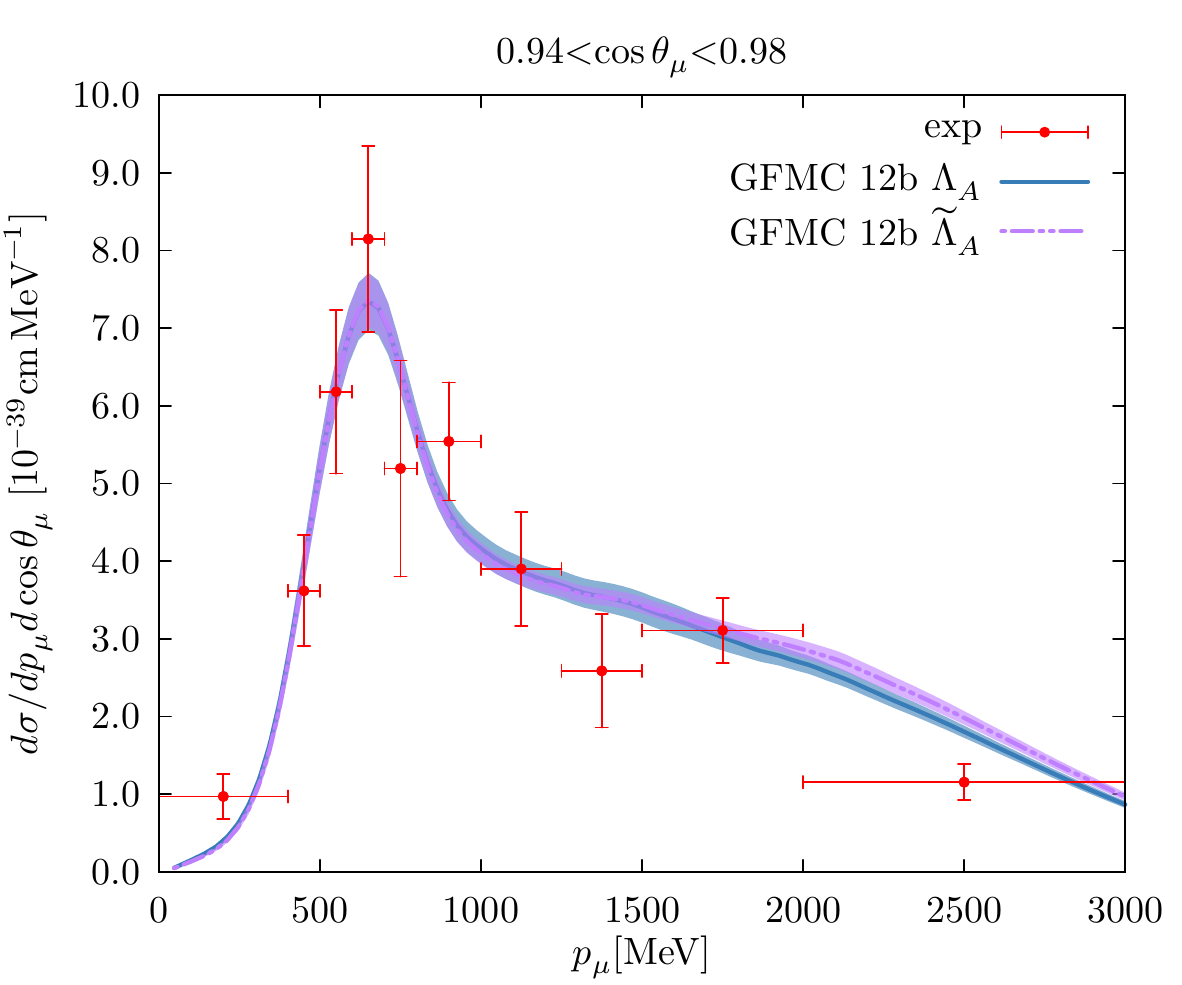}
\includegraphics[width=0.6\columnwidth]{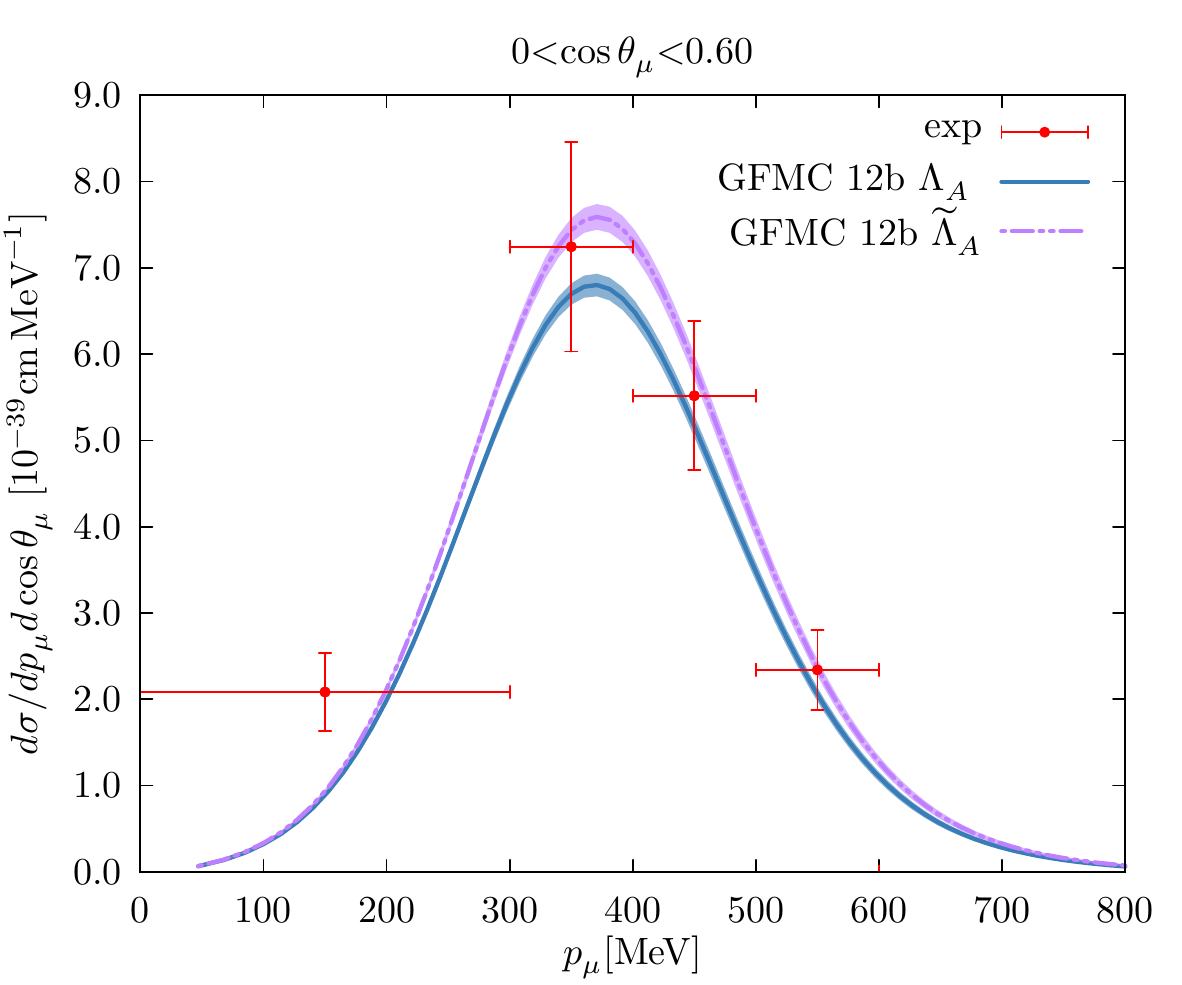}
\caption{The flux-folded GFMC cross sections for selected bins in $\cos\theta_\mu$,
obtained by replacing in the dipole parametrization the cutoff $\Lambda_A\,$$\approx$$\,1$ GeV
with the value $\widetilde{\Lambda}_A\,$$\approx$$\,1.15$ GeV,
more in line with a current LQCD determination~\cite{Bhattacharya:2019}.
The first two rows correspond to the MiniBooNE flux-folded $\nu_\mu$ and $\overline{\nu}_\mu$ CCQE cross sections, respectively;
the last row corresponds to the T2K $\nu_\mu$ CCQE data. 
In the theoretical curves the total one- plus two-body current contribution to the cross section is displayed.}
\label{fig:lqcd}
\end{figure*}

First-principles LQCD calculations of nucleon (and, possibly, nucleon-to-$\Delta$) electroweak
form factors could potentially have a significant impact on calculations of neutrino-nucleus cross
sections, since these form factors constitute essential inputs to the nuclear CC.  This is especially the case
for $G_A(Q^2)$ and the induced pseudoscalar form factor $G_P(Q^2)$, whose $Q^2$-dependence
is experimentally poorly known. In this context, it is interesting to note that recent
LQCD studies~\cite{Bhattacharya:2019,Jang:2019vkm,Park:2020axe} find the $Q^2$ fall-off of $G_A(Q^2)$ with
increasing $Q^2$ significantly less drastic than implied by the dipole behavior with
$\Lambda_A\,$$\approx$$\,1$ GeV.  They also find the nucleon isovector vector
form factors in agreement with experimental data
which are of course quite accurate.  These calculations suggest a larger value of $\Lambda_A$
may be appropriate. We investigate the implications of this finding by presenting
in Fig.~\ref{fig:lqcd} the flux-folded cross sections (for MiniBooNE and selected bins in $\cos\theta_\mu$),
obtained by replacing in the dipole parametrization the cutoff $\Lambda_A\,$$\approx$$\,1$ GeV
with the value $\widetilde{\Lambda}_A\,$$\approx$$\,1.15$ GeV.  As expected, this leads generally to an increase of the GFMC
predictions over the whole kinematical range.  Since the
dominant terms in the cross section proportional to the transverse and interference response
functions tend to cancel for $\overline{\nu}_\mu$, the magnitude of the increase turns
out to be more pronounced for $\nu_\mu$ than for $\overline{\nu}_\mu$---as a
matter of fact, the $\overline{\nu}_\mu$ cross sections are reduced
at backward angles ($0.1 \leq \cos\theta_\mu \leq 0.2$).  Overall, it appears that
the harder cutoff implied by the LQCD calculation of $G_A(Q^2)$ improves
the accord of theory with experiment, marginally for $\overline{\nu}_\mu$
and more substantially for $\nu_\mu$.  In view of the large errors and large normalization
uncertainties of the MiniBooNE and T2K data, however, we caution the reader from drawing
too definite conclusions from the present analysis. Indeed more
precise nucleon form factors can be obtained through further lattice QCD calculations or experiments on the nucleon and deuteron, respectively.

Of course, many challenges remain ahead, to mention just three: the inclusion
of relativity and pion-production mechanisms, and the treatment of heavier nuclei
(notably $^{40}$Ar). While some of these issues, for example the implementation
of relativistic dynamics via a relativistic Hamiltonian along the lines of Ref.~\cite{Carlson:1993zz}, could conceivably
be incorporated in the present GFMC approach, it is out of the question that
such an approach could be utilized to describe the $\Delta$-resonance
region of the cross section or, even more remotely, extended to nuclei with
mass number much larger than 12, at least for the foreseeable future.
In fact, it maybe unnecessary, as more approximate methods exist
to deal effectively with some of these challenges, including factorization
approaches based on one- and two-nucleon spectral functions~\cite{Rocco:2015cil,Rocco:2018mwt}
or on the short-time approximation of the nuclear many-body propagator~\cite{Pastore:2019urn}
for relativity and pion production, and auxiliary-field-diffusion Monte Carlo methods~\cite{Lonardoni:2018nob}
to describe the ground states of medium-weight nuclei.  
We are optimistic that the next few years will witness substantive progress in the further development and
implementation of these approximate methods to address the high-energy
region of the nuclear electroweak response.

Finally, factorization approaches can also be helpful in obtaining some information on exclusive final states.
For more complete treatment of these or, in fact, low energy peaks in the threshold region of the response
quantum computers could play a role, given sufficient size and sufficiently low error rates~\cite{Roggero:2019a,Roggero:2019b}.

\vspace{0.5cm}

\acknowledgments
The present research is supported by the U.S.~Department of Energy, Office
of Science, Office of Nuclear Physics, under contracts DE-AC02-06CH11357
(A.L.~and N.R.), DE-AC52-06NA25396 (S.G.~and J.C.), DE-AC05-06OR23177
(R.S.), and by the NUCLEI SciDAC and LANL LDRD programs.
Under an award of computer time provided by the INCITE program,
this work used resources of the Argonne Leadership Computing
Facility at Argonne National Laboratory, which is supported by the
Office of Science of the U.S.~Department of Energy under contract
DE-AC02-06CH11357.  It also used resources 
provided by Los Alamos Open Supercomputing, by the Argonne LCRC,
and by the National Energy Research Scientific Computing Center,
which is supported by the Office of Science of the U.S.~Department
of Energy under contract DE-AC02-05CH11231.

\bibliography{biblio}

\begin{thebibliography}{75}%
\makeatletter
\providecommand \@ifxundefined [1]{%
 \@ifx{#1\undefined}
}%
\providecommand \@ifnum [1]{%
 \ifnum #1\expandafter \@firstoftwo
 \else \expandafter \@secondoftwo
 \fi
}%
\providecommand \@ifx [1]{%
 \ifx #1\expandafter \@firstoftwo
 \else \expandafter \@secondoftwo
 \fi
}%
\providecommand \natexlab [1]{#1}%
\providecommand \enquote  [1]{``#1''}%
\providecommand \bibnamefont  [1]{#1}%
\providecommand \bibfnamefont [1]{#1}%
\providecommand \citenamefont [1]{#1}%
\providecommand \href@noop [0]{\@secondoftwo}%
\providecommand \href [0]{\begingroup \@sanitize@url \@href}%
\providecommand \@href[1]{\@@startlink{#1}\@@href}%
\providecommand \@@href[1]{\endgroup#1\@@endlink}%
\providecommand \@sanitize@url [0]{\catcode `\\12\catcode `\$12\catcode
  `\&12\catcode `\#12\catcode `\^12\catcode `\_12\catcode `\%12\relax}%
\providecommand \@@startlink[1]{}%
\providecommand \@@endlink[0]{}%
\providecommand \url  [0]{\begingroup\@sanitize@url \@url }%
\providecommand \@url [1]{\endgroup\@href {#1}{\urlprefix }}%
\providecommand \urlprefix  [0]{URL }%
\providecommand \Eprint [0]{\href }%
\providecommand \doibase [0]{http://dx.doi.org/}%
\providecommand \selectlanguage [0]{\@gobble}%
\providecommand \bibinfo  [0]{\@secondoftwo}%
\providecommand \bibfield  [0]{\@secondoftwo}%
\providecommand \translation [1]{[#1]}%
\providecommand \BibitemOpen [0]{}%
\providecommand \bibitemStop [0]{}%
\providecommand \bibitemNoStop [0]{.\EOS\space}%
\providecommand \EOS [0]{\spacefactor3000\relax}%
\providecommand \BibitemShut  [1]{\csname bibitem#1\endcsname}%
\let\auto@bib@innerbib\@empty
\bibitem [{\citenamefont {Alvarez-Ruso}\ \emph {et~al.}(2018)\citenamefont
  {Alvarez-Ruso}, \citenamefont {Sajjad~Athar}, \citenamefont {Barbaro},
  \citenamefont {Cherdack}, \citenamefont {Christy}, \citenamefont {Coloma},
  \citenamefont {Donnelly}, \citenamefont {Dytman}, \citenamefont {de~Gouvea},
  \citenamefont {Hill} \emph {et~al.}}]{Alvarez-Ruso:2017oui}%
  \BibitemOpen
  \bibfield  {author} {\bibinfo {author} {\bibfnamefont {L.}~\bibnamefont
  {Alvarez-Ruso}}, \bibinfo {author} {\bibfnamefont {M.}~\bibnamefont
  {Sajjad~Athar}}, \bibinfo {author} {\bibfnamefont {M.~B.}\ \bibnamefont
  {Barbaro}}, \bibinfo {author} {\bibfnamefont {D.}~\bibnamefont {Cherdack}},
  \bibinfo {author} {\bibfnamefont {M.~E.}\ \bibnamefont {Christy}}, \bibinfo
  {author} {\bibfnamefont {P.}~\bibnamefont {Coloma}}, \bibinfo {author}
  {\bibfnamefont {T.~W.}\ \bibnamefont {Donnelly}}, \bibinfo {author}
  {\bibfnamefont {S.}~\bibnamefont {Dytman}}, \bibinfo {author} {\bibfnamefont
  {A.}~\bibnamefont {de~Gouvea}}, \bibinfo {author} {\bibfnamefont {R.~J.}\
  \bibnamefont {Hill}},  \emph {et~al.},\ }\href {\doibase
  10.1016/j.ppnp.2018.01.006} {\bibfield  {journal} {\bibinfo  {journal} {Prog.
  Part. Nucl. Phys.}\ }\textbf {\bibinfo {volume} {100}},\ \bibinfo {pages} {1}
  (\bibinfo {year} {2018})},\ \Eprint {http://arxiv.org/abs/1706.03621}
  {arXiv:1706.03621 [hep-ph]} \BibitemShut {NoStop}%
\bibitem [{\citenamefont {Van~Orden}(1978)}]{VanOrden:1978}%
  \BibitemOpen
  \bibfield  {author} {\bibinfo {author} {\bibfnamefont {J.~W.}\ \bibnamefont
  {Van~Orden}},\ }\emph {\bibinfo {title} {{Deep Inelastic Electron Scattering
  from Nuclei}}},\ \href@noop {} {Ph.D. thesis},\ \bibinfo  {school} {Standford
  University} (\bibinfo {year} {1978})\BibitemShut {NoStop}%
\bibitem [{\citenamefont {Alberico}\ \emph {et~al.}(1988)\citenamefont
  {Alberico}, \citenamefont {Molinari}, \citenamefont {Donnelly}, \citenamefont
  {Kronenberg},\ and\ \citenamefont {Van~Orden}}]{Alberico:1988bv}%
  \BibitemOpen
  \bibfield  {author} {\bibinfo {author} {\bibfnamefont {W.~M.}\ \bibnamefont
  {Alberico}}, \bibinfo {author} {\bibfnamefont {A.}~\bibnamefont {Molinari}},
  \bibinfo {author} {\bibfnamefont {T.~W.}\ \bibnamefont {Donnelly}}, \bibinfo
  {author} {\bibfnamefont {E.~L.}\ \bibnamefont {Kronenberg}}, \ and\ \bibinfo
  {author} {\bibfnamefont {J.~W.}\ \bibnamefont {Van~Orden}},\ }\href {\doibase
  10.1103/PhysRevC.38.1801} {\bibfield  {journal} {\bibinfo  {journal} {Phys.
  Rev. C}\ }\textbf {\bibinfo {volume} {38}},\ \bibinfo {pages} {1801}
  (\bibinfo {year} {1988})}\BibitemShut {NoStop}%
\bibitem [{\citenamefont {Van~Orden}\ and\ \citenamefont
  {Donnelly}(1981)}]{VanOrden:1980tg}%
  \BibitemOpen
  \bibfield  {author} {\bibinfo {author} {\bibfnamefont {J.~W.}\ \bibnamefont
  {Van~Orden}}\ and\ \bibinfo {author} {\bibfnamefont {T.~W.}\ \bibnamefont
  {Donnelly}},\ }\href {\doibase 10.1016/0003-4916(81)90038-5} {\bibfield
  {journal} {\bibinfo  {journal} {Annals Phys.}\ }\textbf {\bibinfo {volume}
  {131}},\ \bibinfo {pages} {451} (\bibinfo {year} {1981})}\BibitemShut
  {NoStop}%
\bibitem [{\citenamefont {Walecka}(1974)}]{Walecka:1974qa}%
  \BibitemOpen
  \bibfield  {author} {\bibinfo {author} {\bibfnamefont {J.~D.}\ \bibnamefont
  {Walecka}},\ }\href {\doibase 10.1016/0003-4916(74)90208-5} {\bibfield
  {journal} {\bibinfo  {journal} {Annals Phys.}\ }\textbf {\bibinfo {volume}
  {83}},\ \bibinfo {pages} {491} (\bibinfo {year} {1974})}\BibitemShut
  {NoStop}%
\bibitem [{\citenamefont {Amaro}\ \emph {et~al.}(2007)\citenamefont {Amaro},
  \citenamefont {Barbaro}, \citenamefont {Caballero}, \citenamefont
  {Donnelly},\ and\ \citenamefont {Udias}}]{Amaro:2006if}%
  \BibitemOpen
  \bibfield  {author} {\bibinfo {author} {\bibfnamefont {J.~E.}\ \bibnamefont
  {Amaro}}, \bibinfo {author} {\bibfnamefont {M.~B.}\ \bibnamefont {Barbaro}},
  \bibinfo {author} {\bibfnamefont {J.~A.}\ \bibnamefont {Caballero}}, \bibinfo
  {author} {\bibfnamefont {T.~W.}\ \bibnamefont {Donnelly}}, \ and\ \bibinfo
  {author} {\bibfnamefont {J.~M.}\ \bibnamefont {Udias}},\ }\href {\doibase
  10.1103/PhysRevC.75.034613} {\bibfield  {journal} {\bibinfo  {journal} {Phys.
  Rev. C}\ }\textbf {\bibinfo {volume} {75}},\ \bibinfo {pages} {034613}
  (\bibinfo {year} {2007})},\ \Eprint {http://arxiv.org/abs/nucl-th/0612056}
  {arXiv:nucl-th/0612056 [nucl-th]} \BibitemShut {NoStop}%
\bibitem [{\citenamefont {Amaro}\ \emph {et~al.}(2011)\citenamefont {Amaro},
  \citenamefont {Barbaro}, \citenamefont {Caballero}, \citenamefont
  {Donnelly},\ and\ \citenamefont {Udias}}]{Amaro:2011qb}%
  \BibitemOpen
  \bibfield  {author} {\bibinfo {author} {\bibfnamefont {J.~E.}\ \bibnamefont
  {Amaro}}, \bibinfo {author} {\bibfnamefont {M.~B.}\ \bibnamefont {Barbaro}},
  \bibinfo {author} {\bibfnamefont {J.~A.}\ \bibnamefont {Caballero}}, \bibinfo
  {author} {\bibfnamefont {T.~W.}\ \bibnamefont {Donnelly}}, \ and\ \bibinfo
  {author} {\bibfnamefont {J.~M.}\ \bibnamefont {Udias}},\ }\href {\doibase
  10.1103/PhysRevD.84.033004} {\bibfield  {journal} {\bibinfo  {journal} {Phys.
  Rev. D}\ }\textbf {\bibinfo {volume} {84}},\ \bibinfo {pages} {033004}
  (\bibinfo {year} {2011})},\ \Eprint {http://arxiv.org/abs/1104.5446}
  {arXiv:1104.5446 [nucl-th]} \BibitemShut {NoStop}%
\bibitem [{\citenamefont {González-Jiménez}\ \emph {et~al.}(2019)\citenamefont
  {González-Jiménez}, \citenamefont {Nikolakopoulos}, \citenamefont
  {Jachowicz},\ and\ \citenamefont {Udías}}]{Gonzalez-Jimenez:2019qhq}%
  \BibitemOpen
  \bibfield  {author} {\bibinfo {author} {\bibfnamefont {R.}~\bibnamefont
  {González-Jiménez}}, \bibinfo {author} {\bibfnamefont {A.}~\bibnamefont
  {Nikolakopoulos}}, \bibinfo {author} {\bibfnamefont {N.}~\bibnamefont
  {Jachowicz}}, \ and\ \bibinfo {author} {\bibfnamefont {J.~M.}\ \bibnamefont
  {Udías}},\ }\href {\doibase 10.1103/PhysRevC.100.045501} {\bibfield
  {journal} {\bibinfo  {journal} {Phys. Rev. C}\ }\textbf {\bibinfo {volume}
  {100}},\ \bibinfo {pages} {045501} (\bibinfo {year} {2019})},\ \Eprint
  {http://arxiv.org/abs/1904.10696} {arXiv:1904.10696 [nucl-th]} \BibitemShut
  {NoStop}%
\bibitem [{\citenamefont {Martini}\ \emph {et~al.}(2009)\citenamefont
  {Martini}, \citenamefont {Ericson}, \citenamefont {Chanfray},\ and\
  \citenamefont {Marteau}}]{Martini:2009uj}%
  \BibitemOpen
  \bibfield  {author} {\bibinfo {author} {\bibfnamefont {M.}~\bibnamefont
  {Martini}}, \bibinfo {author} {\bibfnamefont {M.}~\bibnamefont {Ericson}},
  \bibinfo {author} {\bibfnamefont {G.}~\bibnamefont {Chanfray}}, \ and\
  \bibinfo {author} {\bibfnamefont {J.}~\bibnamefont {Marteau}},\ }\href
  {\doibase 10.1103/PhysRevC.80.065501} {\bibfield  {journal} {\bibinfo
  {journal} {Phys. Rev. C}\ }\textbf {\bibinfo {volume} {80}},\ \bibinfo
  {pages} {065501} (\bibinfo {year} {2009})},\ \Eprint
  {http://arxiv.org/abs/0910.2622} {arXiv:0910.2622 [nucl-th]} \BibitemShut
  {NoStop}%
\bibitem [{\citenamefont {Martini}\ \emph {et~al.}(2010)\citenamefont
  {Martini}, \citenamefont {Ericson}, \citenamefont {Chanfray},\ and\
  \citenamefont {Marteau}}]{Martini:2010ex}%
  \BibitemOpen
  \bibfield  {author} {\bibinfo {author} {\bibfnamefont {M.}~\bibnamefont
  {Martini}}, \bibinfo {author} {\bibfnamefont {M.}~\bibnamefont {Ericson}},
  \bibinfo {author} {\bibfnamefont {G.}~\bibnamefont {Chanfray}}, \ and\
  \bibinfo {author} {\bibfnamefont {J.}~\bibnamefont {Marteau}},\ }\href
  {\doibase 10.1103/PhysRevC.81.045502} {\bibfield  {journal} {\bibinfo
  {journal} {Phys. Rev. C}\ }\textbf {\bibinfo {volume} {81}},\ \bibinfo
  {pages} {045502} (\bibinfo {year} {2010})},\ \Eprint
  {http://arxiv.org/abs/1002.4538} {arXiv:1002.4538 [hep-ph]} \BibitemShut
  {NoStop}%
\bibitem [{\citenamefont {Nieves}\ \emph {et~al.}(2004)\citenamefont {Nieves},
  \citenamefont {Amaro},\ and\ \citenamefont {Valverde}}]{Nieves:2004wx}%
  \BibitemOpen
  \bibfield  {author} {\bibinfo {author} {\bibfnamefont {J.}~\bibnamefont
  {Nieves}}, \bibinfo {author} {\bibfnamefont {J.~E.}\ \bibnamefont {Amaro}}, \
  and\ \bibinfo {author} {\bibfnamefont {M.}~\bibnamefont {Valverde}},\
  }\href@noop {} {\bibfield  {journal} {\bibinfo  {journal} {Phys. Rev. C}\
  }\textbf {\bibinfo {volume} {70}},\ \bibinfo {pages} {055503} (\bibinfo
  {year} {2004})}\BibitemShut {NoStop}%
\bibitem [{\citenamefont {Nieves}\ \emph {et~al.}(2006)\citenamefont {Nieves},
  \citenamefont {Valverde},\ and\ \citenamefont
  {Vicente~Vacas}}]{Nieves:2005rq}%
  \BibitemOpen
  \bibfield  {author} {\bibinfo {author} {\bibfnamefont {J.}~\bibnamefont
  {Nieves}}, \bibinfo {author} {\bibfnamefont {M.}~\bibnamefont {Valverde}}, \
  and\ \bibinfo {author} {\bibfnamefont {M.~J.}\ \bibnamefont
  {Vicente~Vacas}},\ }\href {\doibase 10.1103/PhysRevC.73.025504} {\bibfield
  {journal} {\bibinfo  {journal} {Phys. Rev. C}\ }\textbf {\bibinfo {volume}
  {73}},\ \bibinfo {pages} {025504} (\bibinfo {year} {2006})}\BibitemShut
  {NoStop}%
\bibitem [{\citenamefont {Nieves}\ \emph {et~al.}(2011)\citenamefont {Nieves},
  \citenamefont {Simo},\ and\ \citenamefont {Vacas}}]{Nieves:2011}%
  \BibitemOpen
  \bibfield  {author} {\bibinfo {author} {\bibfnamefont {J.}~\bibnamefont
  {Nieves}}, \bibinfo {author} {\bibfnamefont {I.}~\bibnamefont {Simo}}, \ and\
  \bibinfo {author} {\bibfnamefont {M.}~\bibnamefont {Vacas}},\ }\href
  {\doibase 10.1103/PhysRevC.83.045501} {\bibfield  {journal} {\bibinfo
  {journal} {Phys. Rev. C}\ }\textbf {\bibinfo {volume} {83}},\ \bibinfo
  {pages} {045501} (\bibinfo {year} {2011})}\BibitemShut {NoStop}%
\bibitem [{\citenamefont {{Barbaro}}\ \emph {et~al.}(2006)\citenamefont
  {{Barbaro}}, \citenamefont {{Amaro}}, \citenamefont {{Caballero}},
  \citenamefont {{Donnelly}}, \citenamefont {{Molinari}},\ and\ \citenamefont
  {{Sick}}}]{Barbaro:2006}%
  \BibitemOpen
  \bibfield  {author} {\bibinfo {author} {\bibfnamefont {M.~B.}\ \bibnamefont
  {{Barbaro}}}, \bibinfo {author} {\bibfnamefont {J.~E.}\ \bibnamefont
  {{Amaro}}}, \bibinfo {author} {\bibfnamefont {J.~A.}\ \bibnamefont
  {{Caballero}}}, \bibinfo {author} {\bibfnamefont {T.~W.}\ \bibnamefont
  {{Donnelly}}}, \bibinfo {author} {\bibfnamefont {A.}~\bibnamefont
  {{Molinari}}}, \ and\ \bibinfo {author} {\bibfnamefont {I.}~\bibnamefont
  {{Sick}}},\ }\href {\doibase 10.1016/j.nuclphysbps.2006.02.065} {\bibfield
  {journal} {\bibinfo  {journal} {Nuclear Physics B Proceedings Supplements}\
  }\textbf {\bibinfo {volume} {155}},\ \bibinfo {pages} {257} (\bibinfo {year}
  {2006})},\ \Eprint {http://arxiv.org/abs/nucl-th/0509022}
  {arXiv:nucl-th/0509022 [nucl-th]} \BibitemShut {NoStop}%
\bibitem [{\citenamefont {Megias}\ \emph {et~al.}(2013)\citenamefont {Megias},
  \citenamefont {Amaro}, \citenamefont {Barbaro}, \citenamefont {Caballero},\
  and\ \citenamefont {Donnelly}}]{Amaro:2013yna}%
  \BibitemOpen
  \bibfield  {author} {\bibinfo {author} {\bibfnamefont {G.~D.}\ \bibnamefont
  {Megias}}, \bibinfo {author} {\bibfnamefont {J.~E.}\ \bibnamefont {Amaro}},
  \bibinfo {author} {\bibfnamefont {M.~B.}\ \bibnamefont {Barbaro}}, \bibinfo
  {author} {\bibfnamefont {J.~A.}\ \bibnamefont {Caballero}}, \ and\ \bibinfo
  {author} {\bibfnamefont {T.~W.}\ \bibnamefont {Donnelly}},\ }\href {\doibase
  10.1016/j.physletb.2013.07.004} {\bibfield  {journal} {\bibinfo  {journal}
  {Phys. Lett. B}\ }\textbf {\bibinfo {volume} {725}},\ \bibinfo {pages} {170}
  (\bibinfo {year} {2013})},\ \Eprint {http://arxiv.org/abs/1305.6884}
  {arXiv:1305.6884 [nucl-th]} \BibitemShut {NoStop}%
\bibitem [{\citenamefont {Amaro}\ \emph {et~al.}(2012)\citenamefont {Amaro},
  \citenamefont {Barbaro}, \citenamefont {Caballero},\ and\ \citenamefont
  {Donnelly}}]{Amaro:2011aa}%
  \BibitemOpen
  \bibfield  {author} {\bibinfo {author} {\bibfnamefont {J.~E.}\ \bibnamefont
  {Amaro}}, \bibinfo {author} {\bibfnamefont {M.~B.}\ \bibnamefont {Barbaro}},
  \bibinfo {author} {\bibfnamefont {J.~A.}\ \bibnamefont {Caballero}}, \ and\
  \bibinfo {author} {\bibfnamefont {T.~W.}\ \bibnamefont {Donnelly}},\ }\href
  {\doibase 10.1103/PhysRevLett.108.152501} {\bibfield  {journal} {\bibinfo
  {journal} {Phys. Rev. Lett.}\ }\textbf {\bibinfo {volume} {108}},\ \bibinfo
  {pages} {152501} (\bibinfo {year} {2012})},\ \Eprint
  {http://arxiv.org/abs/1112.2123} {arXiv:1112.2123 [nucl-th]} \BibitemShut
  {NoStop}%
\bibitem [{\citenamefont {Gonzaléz-Jiménez}\ \emph {et~al.}(2014)\citenamefont
  {Gonzaléz-Jiménez}, \citenamefont {Megias}, \citenamefont {Barbaro},
  \citenamefont {Caballero},\ and\ \citenamefont
  {Donnelly}}]{Gonzalez-Jimenez:2014eqa}%
  \BibitemOpen
  \bibfield  {author} {\bibinfo {author} {\bibfnamefont {R.}~\bibnamefont
  {Gonzaléz-Jiménez}}, \bibinfo {author} {\bibfnamefont {G.~D.}\ \bibnamefont
  {Megias}}, \bibinfo {author} {\bibfnamefont {M.~B.}\ \bibnamefont {Barbaro}},
  \bibinfo {author} {\bibfnamefont {J.~A.}\ \bibnamefont {Caballero}}, \ and\
  \bibinfo {author} {\bibfnamefont {T.~W.}\ \bibnamefont {Donnelly}},\ }\href
  {\doibase 10.1103/PhysRevC.90.035501} {\bibfield  {journal} {\bibinfo
  {journal} {Phys. Rev. C}\ }\textbf {\bibinfo {volume} {90}},\ \bibinfo
  {pages} {035501} (\bibinfo {year} {2014})},\ \Eprint
  {http://arxiv.org/abs/1407.8346} {arXiv:1407.8346 [nucl-th]} \BibitemShut
  {NoStop}%
\bibitem [{\citenamefont {Megias}\ \emph {et~al.}(2016)\citenamefont {Megias},
  \citenamefont {Amaro}, \citenamefont {Barbaro}, \citenamefont {Caballero},\
  and\ \citenamefont {Donnelly}}]{Megias:2016lke}%
  \BibitemOpen
  \bibfield  {author} {\bibinfo {author} {\bibfnamefont {G.~D.}\ \bibnamefont
  {Megias}}, \bibinfo {author} {\bibfnamefont {J.~E.}\ \bibnamefont {Amaro}},
  \bibinfo {author} {\bibfnamefont {M.~B.}\ \bibnamefont {Barbaro}}, \bibinfo
  {author} {\bibfnamefont {J.~A.}\ \bibnamefont {Caballero}}, \ and\ \bibinfo
  {author} {\bibfnamefont {T.~W.}\ \bibnamefont {Donnelly}},\ }\href {\doibase
  10.1103/PhysRevD.94.013012} {\bibfield  {journal} {\bibinfo  {journal} {Phys.
  Rev. D}\ }\textbf {\bibinfo {volume} {94}},\ \bibinfo {pages} {013012}
  (\bibinfo {year} {2016})},\ \Eprint {http://arxiv.org/abs/1603.08396}
  {arXiv:1603.08396 [nucl-th]} \BibitemShut {NoStop}%
\bibitem [{\citenamefont {Megias}\ \emph {et~al.}(2019)\citenamefont {Megias},
  \citenamefont {Barbaro}, \citenamefont {Caballero}, \citenamefont {Amaro},
  \citenamefont {Donnelly}, \citenamefont {Ruiz~Simo},\ and\ \citenamefont
  {Van~Orden}}]{Megias:2017cuh}%
  \BibitemOpen
  \bibfield  {author} {\bibinfo {author} {\bibfnamefont {G.~D.}\ \bibnamefont
  {Megias}}, \bibinfo {author} {\bibfnamefont {M.~B.}\ \bibnamefont {Barbaro}},
  \bibinfo {author} {\bibfnamefont {J.~A.}\ \bibnamefont {Caballero}}, \bibinfo
  {author} {\bibfnamefont {J.~E.}\ \bibnamefont {Amaro}}, \bibinfo {author}
  {\bibfnamefont {T.~W.}\ \bibnamefont {Donnelly}}, \bibinfo {author}
  {\bibfnamefont {I.}~\bibnamefont {Ruiz~Simo}}, \ and\ \bibinfo {author}
  {\bibfnamefont {J.~W.}\ \bibnamefont {Van~Orden}},\ }\href {\doibase
  10.1088/1361-6471/aaf3ae} {\bibfield  {journal} {\bibinfo  {journal} {J.
  Phys. G}\ }\textbf {\bibinfo {volume} {46}},\ \bibinfo {pages} {015104}
  (\bibinfo {year} {2019})},\ \Eprint {http://arxiv.org/abs/1711.00771}
  {arXiv:1711.00771 [nucl-th]} \BibitemShut {NoStop}%
\bibitem [{\citenamefont {Dickhoff}\ and\ \citenamefont
  {Barbieri}(2004)}]{Dickhoff2004ppnp}%
  \BibitemOpen
  \bibfield  {author} {\bibinfo {author} {\bibfnamefont {W.~H.}\ \bibnamefont
  {Dickhoff}}\ and\ \bibinfo {author} {\bibfnamefont {C.}~\bibnamefont
  {Barbieri}},\ }\href {\doibase 10.1016/j.ppnp.2004.02.038} {\bibfield
  {journal} {\bibinfo  {journal} {Prog. Part. Nucl. Phys.}\ }\textbf {\bibinfo
  {volume} {52}},\ \bibinfo {pages} {377} (\bibinfo {year} {2004})},\ \Eprint
  {http://arxiv.org/abs/nucl-th/0402034} {arXiv:nucl-th/0402034 [nucl-th]}
  \BibitemShut {NoStop}%
\bibitem [{\citenamefont {Barbieri}(2014)}]{Barbieri2014QMBT}%
  \BibitemOpen
  \bibfield  {author} {\bibinfo {author} {\bibfnamefont {C.}~\bibnamefont
  {Barbieri}},\ }\href {http://stacks.iop.org/1742-6596/529/i=1/a=012005}
  {\bibfield  {journal} {\bibinfo  {journal} {Journal of Physics: Conference
  Series}\ }\textbf {\bibinfo {volume} {529}},\ \bibinfo {pages} {012005}
  (\bibinfo {year} {2014})}\BibitemShut {NoStop}%
\bibitem [{\citenamefont {Rocco}\ and\ \citenamefont
  {Barbieri}(2018)}]{Rocco:2018vbf}%
  \BibitemOpen
  \bibfield  {author} {\bibinfo {author} {\bibfnamefont {N.}~\bibnamefont
  {Rocco}}\ and\ \bibinfo {author} {\bibfnamefont {C.}~\bibnamefont
  {Barbieri}},\ }\href {\doibase 10.1103/PhysRevC.98.025501} {\bibfield
  {journal} {\bibinfo  {journal} {Phys. Rev. C}\ }\textbf {\bibinfo {volume}
  {98}},\ \bibinfo {pages} {025501} (\bibinfo {year} {2018})},\ \Eprint
  {http://arxiv.org/abs/1803.00825} {arXiv:1803.00825 [nucl-th]} \BibitemShut
  {NoStop}%
\bibitem [{\citenamefont {Barbieri}\ \emph {et~al.}(2019)\citenamefont
  {Barbieri}, \citenamefont {Rocco},\ and\ \citenamefont
  {Somà}}]{Barbieri:2019ual}%
  \BibitemOpen
  \bibfield  {author} {\bibinfo {author} {\bibfnamefont {C.}~\bibnamefont
  {Barbieri}}, \bibinfo {author} {\bibfnamefont {N.}~\bibnamefont {Rocco}}, \
  and\ \bibinfo {author} {\bibfnamefont {V.}~\bibnamefont {Somà}},\ }\href
  {\doibase 10.1103/PhysRevC.100.062501} {\bibfield  {journal} {\bibinfo
  {journal} {Phys. Rev. C}\ }\textbf {\bibinfo {volume} {100}},\ \bibinfo
  {pages} {062501} (\bibinfo {year} {2019})},\ \Eprint
  {http://arxiv.org/abs/1907.01122} {arXiv:1907.01122 [nucl-th]} \BibitemShut
  {NoStop}%
\bibitem [{\citenamefont {Benhar}\ \emph {et~al.}(1989)\citenamefont {Benhar},
  \citenamefont {Fabrocini},\ and\ \citenamefont {Fantoni}}]{Benhar:1989aw}%
  \BibitemOpen
  \bibfield  {author} {\bibinfo {author} {\bibfnamefont {O.}~\bibnamefont
  {Benhar}}, \bibinfo {author} {\bibfnamefont {A.}~\bibnamefont {Fabrocini}}, \
  and\ \bibinfo {author} {\bibfnamefont {S.}~\bibnamefont {Fantoni}},\ }\href
  {\doibase 10.1016/0375-9474(89)90374-6} {\bibfield  {journal} {\bibinfo
  {journal} {Nucl. Phys. A}\ }\textbf {\bibinfo {volume} {505}},\ \bibinfo
  {pages} {267} (\bibinfo {year} {1989})}\BibitemShut {NoStop}%
\bibitem [{\citenamefont {Benhar}\ \emph {et~al.}(1991)\citenamefont {Benhar},
  \citenamefont {Fabrocini}, \citenamefont {Fantoni}, \citenamefont {Miller},
  \citenamefont {Pandharipande},\ and\ \citenamefont {Sick}}]{Benhar:1991af}%
  \BibitemOpen
  \bibfield  {author} {\bibinfo {author} {\bibfnamefont {O.}~\bibnamefont
  {Benhar}}, \bibinfo {author} {\bibfnamefont {A.}~\bibnamefont {Fabrocini}},
  \bibinfo {author} {\bibfnamefont {S.}~\bibnamefont {Fantoni}}, \bibinfo
  {author} {\bibfnamefont {G.~A.}\ \bibnamefont {Miller}}, \bibinfo {author}
  {\bibfnamefont {V.~R.}\ \bibnamefont {Pandharipande}}, \ and\ \bibinfo
  {author} {\bibfnamefont {I.}~\bibnamefont {Sick}},\ }\href {\doibase
  10.1103/PhysRevC.44.2328} {\bibfield  {journal} {\bibinfo  {journal} {Phys.
  Rev. C}\ }\textbf {\bibinfo {volume} {44}},\ \bibinfo {pages} {2328}
  (\bibinfo {year} {1991})}\BibitemShut {NoStop}%
\bibitem [{\citenamefont {Benhar}\ \emph {et~al.}(1992)\citenamefont {Benhar},
  \citenamefont {Fabrocini},\ and\ \citenamefont {Fantoni}}]{Benhar:1992cnb}%
  \BibitemOpen
  \bibfield  {author} {\bibinfo {author} {\bibfnamefont {O.}~\bibnamefont
  {Benhar}}, \bibinfo {author} {\bibfnamefont {A.}~\bibnamefont {Fabrocini}}, \
  and\ \bibinfo {author} {\bibfnamefont {S.}~\bibnamefont {Fantoni}},\ }\href
  {\doibase 10.1016/0375-9474(92)90679-E} {\bibfield  {journal} {\bibinfo
  {journal} {Nucl. Phys. A}\ }\textbf {\bibinfo {volume} {550}},\ \bibinfo
  {pages} {201} (\bibinfo {year} {1992})}\BibitemShut {NoStop}%
\bibitem [{\citenamefont {Benhar}\ \emph {et~al.}(2008)\citenamefont {Benhar},
  \citenamefont {Day},\ and\ \citenamefont {Sick}}]{Benhar:2006wy}%
  \BibitemOpen
  \bibfield  {author} {\bibinfo {author} {\bibfnamefont {O.}~\bibnamefont
  {Benhar}}, \bibinfo {author} {\bibfnamefont {D.}~\bibnamefont {Day}}, \ and\
  \bibinfo {author} {\bibfnamefont {I.}~\bibnamefont {Sick}},\ }\href {\doibase
  10.1103/RevModPhys.80.189} {\bibfield  {journal} {\bibinfo  {journal} {Rev.
  Mod. Phys.}\ }\textbf {\bibinfo {volume} {80}},\ \bibinfo {pages} {189}
  (\bibinfo {year} {2008})},\ \Eprint {http://arxiv.org/abs/nucl-ex/0603029}
  {arXiv:nucl-ex/0603029 [nucl-ex]} \BibitemShut {NoStop}%
\bibitem [{\citenamefont {Rocco}\ \emph {et~al.}(2016)\citenamefont {Rocco},
  \citenamefont {Lovato},\ and\ \citenamefont {Benhar}}]{Rocco:2015cil}%
  \BibitemOpen
  \bibfield  {author} {\bibinfo {author} {\bibfnamefont {N.}~\bibnamefont
  {Rocco}}, \bibinfo {author} {\bibfnamefont {A.}~\bibnamefont {Lovato}}, \
  and\ \bibinfo {author} {\bibfnamefont {O.}~\bibnamefont {Benhar}},\ }\href
  {\doibase 10.1103/PhysRevLett.116.192501} {\bibfield  {journal} {\bibinfo
  {journal} {Phys. Rev. Lett.}\ }\textbf {\bibinfo {volume} {116}},\ \bibinfo
  {pages} {192501} (\bibinfo {year} {2016})},\ \Eprint
  {http://arxiv.org/abs/1512.07426} {arXiv:1512.07426 [nucl-th]} \BibitemShut
  {NoStop}%
\bibitem [{\citenamefont {Benhar}\ \emph {et~al.}(1994)\citenamefont {Benhar},
  \citenamefont {Fabrocini}, \citenamefont {Fantoni},\ and\ \citenamefont
  {Sick}}]{Benhar:1994hw}%
  \BibitemOpen
  \bibfield  {author} {\bibinfo {author} {\bibfnamefont {O.}~\bibnamefont
  {Benhar}}, \bibinfo {author} {\bibfnamefont {A.}~\bibnamefont {Fabrocini}},
  \bibinfo {author} {\bibfnamefont {S.}~\bibnamefont {Fantoni}}, \ and\
  \bibinfo {author} {\bibfnamefont {I.}~\bibnamefont {Sick}},\ }\href {\doibase
  10.1016/0375-9474(94)90920-2} {\bibfield  {journal} {\bibinfo  {journal}
  {Nucl.\ Phys.\ A}\ }\textbf {\bibinfo {volume} {579}},\ \bibinfo {pages}
  {493} (\bibinfo {year} {1994})}\BibitemShut {NoStop}%
\bibitem [{\citenamefont {Sick}\ \emph {et~al.}(1994)\citenamefont {Sick},
  \citenamefont {Fantoni}, \citenamefont {Fabrocini},\ and\ \citenamefont
  {Benhar}}]{Sick:1994vj}%
  \BibitemOpen
  \bibfield  {author} {\bibinfo {author} {\bibfnamefont {I.}~\bibnamefont
  {Sick}}, \bibinfo {author} {\bibfnamefont {S.}~\bibnamefont {Fantoni}},
  \bibinfo {author} {\bibfnamefont {A.}~\bibnamefont {Fabrocini}}, \ and\
  \bibinfo {author} {\bibfnamefont {O.}~\bibnamefont {Benhar}},\ }\href
  {\doibase 10.1016/0370-2693(94)91218-1} {\bibfield  {journal} {\bibinfo
  {journal} {Phys. Lett. B}\ }\textbf {\bibinfo {volume} {323}},\ \bibinfo
  {pages} {267} (\bibinfo {year} {1994})}\BibitemShut {NoStop}%
\bibitem [{\citenamefont {Carlson}\ and\ \citenamefont
  {Schiavilla}(1998)}]{Carlson:1998}%
  \BibitemOpen
  \bibfield  {author} {\bibinfo {author} {\bibfnamefont {J.}~\bibnamefont
  {Carlson}}\ and\ \bibinfo {author} {\bibfnamefont {R.}~\bibnamefont
  {Schiavilla}},\ }\href {\doibase 10.1103/RevModPhys.70.743} {\bibfield
  {journal} {\bibinfo  {journal} {Rev. Mod. Phys.}\ }\textbf {\bibinfo {volume}
  {70}},\ \bibinfo {pages} {743} (\bibinfo {year} {1998})}\BibitemShut
  {NoStop}%
\bibitem [{\citenamefont {Carlson}\ \emph {et~al.}(2015)\citenamefont
  {Carlson}, \citenamefont {Gandolfi}, \citenamefont {Pederiva}, \citenamefont
  {Pieper}, \citenamefont {Schiavilla}, \citenamefont {Schmidt},\ and\
  \citenamefont {Wiringa}}]{Carlson:2015}%
  \BibitemOpen
  \bibfield  {author} {\bibinfo {author} {\bibfnamefont {J.}~\bibnamefont
  {Carlson}}, \bibinfo {author} {\bibfnamefont {S.}~\bibnamefont {Gandolfi}},
  \bibinfo {author} {\bibfnamefont {F.}~\bibnamefont {Pederiva}}, \bibinfo
  {author} {\bibfnamefont {S.~C.}\ \bibnamefont {Pieper}}, \bibinfo {author}
  {\bibfnamefont {R.}~\bibnamefont {Schiavilla}}, \bibinfo {author}
  {\bibfnamefont {K.~E.}\ \bibnamefont {Schmidt}}, \ and\ \bibinfo {author}
  {\bibfnamefont {R.~B.}\ \bibnamefont {Wiringa}},\ }\href {\doibase
  10.1103/RevModPhys.87.1067} {\bibfield  {journal} {\bibinfo  {journal} {Rev.
  Mod. Phys.}\ }\textbf {\bibinfo {volume} {87}},\ \bibinfo {pages} {1067}
  (\bibinfo {year} {2015})}\BibitemShut {NoStop}%
\bibitem [{\citenamefont {Wiringa}\ \emph {et~al.}(1995)\citenamefont
  {Wiringa}, \citenamefont {Stoks},\ and\ \citenamefont
  {Schiavilla}}]{Wiringa:1995}%
  \BibitemOpen
  \bibfield  {author} {\bibinfo {author} {\bibfnamefont {R.~B.}\ \bibnamefont
  {Wiringa}}, \bibinfo {author} {\bibfnamefont {V.~G.~J.}\ \bibnamefont
  {Stoks}}, \ and\ \bibinfo {author} {\bibfnamefont {R.}~\bibnamefont
  {Schiavilla}},\ }\href {\doibase 10.1103/PhysRevC.51.38} {\bibfield
  {journal} {\bibinfo  {journal} {Phys. Rev. C}\ }\textbf {\bibinfo {volume}
  {51}},\ \bibinfo {pages} {38} (\bibinfo {year} {1995})}\BibitemShut {NoStop}%
\bibitem [{\citenamefont {Pieper}(2008)}]{Pieper:2008}%
  \BibitemOpen
  \bibfield  {author} {\bibinfo {author} {\bibfnamefont {S.~C.}\ \bibnamefont
  {Pieper}},\ }\href {\doibase 10.1063/1.2932280} {\bibfield  {journal}
  {\bibinfo  {journal} {AIP Conf. Proc.}\ }\textbf {\bibinfo {volume} {1011}},\
  \bibinfo {pages} {143} (\bibinfo {year} {2008})}\BibitemShut {NoStop}%
\bibitem [{\citenamefont {Baker}\ \emph {et~al.}(1981)\citenamefont {Baker},
  \citenamefont {Cnops}, \citenamefont {Connolly}, \citenamefont {Kahn},
  \citenamefont {Kirk}, \citenamefont {Murtagh}, \citenamefont {Palmer},
  \citenamefont {Samios},\ and\ \citenamefont {Tanaka}}]{Baker:1981su}%
  \BibitemOpen
  \bibfield  {author} {\bibinfo {author} {\bibfnamefont {N.~J.}\ \bibnamefont
  {Baker}}, \bibinfo {author} {\bibfnamefont {A.~M.}\ \bibnamefont {Cnops}},
  \bibinfo {author} {\bibfnamefont {P.~L.}\ \bibnamefont {Connolly}}, \bibinfo
  {author} {\bibfnamefont {S.~A.}\ \bibnamefont {Kahn}}, \bibinfo {author}
  {\bibfnamefont {H.~G.}\ \bibnamefont {Kirk}}, \bibinfo {author}
  {\bibfnamefont {M.~J.}\ \bibnamefont {Murtagh}}, \bibinfo {author}
  {\bibfnamefont {R.~B.}\ \bibnamefont {Palmer}}, \bibinfo {author}
  {\bibfnamefont {N.~P.}\ \bibnamefont {Samios}}, \ and\ \bibinfo {author}
  {\bibfnamefont {M.}~\bibnamefont {Tanaka}},\ }\href {\doibase
  10.1103/PhysRevD.23.2499} {\bibfield  {journal} {\bibinfo  {journal} {Phys.
  Rev.}\ }\textbf {\bibinfo {volume} {D23}},\ \bibinfo {pages} {2499} (\bibinfo
  {year} {1981})}\BibitemShut {NoStop}%
\bibitem [{\citenamefont {Miller}\ \emph {et~al.}(1982)\citenamefont {Miller}
  \emph {et~al.}}]{Miller:1982qi}%
  \BibitemOpen
  \bibfield  {author} {\bibinfo {author} {\bibfnamefont {K.~L.}\ \bibnamefont
  {Miller}} \emph {et~al.},\ }\href {\doibase 10.1103/PhysRevD.26.537}
  {\bibfield  {journal} {\bibinfo  {journal} {Phys. Rev.}\ }\textbf {\bibinfo
  {volume} {D26}},\ \bibinfo {pages} {537} (\bibinfo {year}
  {1982})}\BibitemShut {NoStop}%
\bibitem [{\citenamefont {Kitagaki}\ \emph {et~al.}(1983)\citenamefont
  {Kitagaki} \emph {et~al.}}]{Kitagaki:1983px}%
  \BibitemOpen
  \bibfield  {author} {\bibinfo {author} {\bibfnamefont {T.}~\bibnamefont
  {Kitagaki}} \emph {et~al.},\ }\href {\doibase 10.1103/PhysRevD.28.436}
  {\bibfield  {journal} {\bibinfo  {journal} {Phys. Rev.}\ }\textbf {\bibinfo
  {volume} {D28}},\ \bibinfo {pages} {436} (\bibinfo {year}
  {1983})}\BibitemShut {NoStop}%
\bibitem [{\citenamefont {Ahrens}\ \emph {et~al.}(1987)\citenamefont {Ahrens}
  \emph {et~al.}}]{Ahrens:1986xe}%
  \BibitemOpen
  \bibfield  {author} {\bibinfo {author} {\bibfnamefont {L.~A.}\ \bibnamefont
  {Ahrens}} \emph {et~al.},\ }\href {\doibase 10.1103/PhysRevD.35.785}
  {\bibfield  {journal} {\bibinfo  {journal} {Phys. Rev.}\ }\textbf {\bibinfo
  {volume} {D35}},\ \bibinfo {pages} {785} (\bibinfo {year}
  {1987})}\BibitemShut {NoStop}%
\bibitem [{\citenamefont {Bhattacharya}\ \emph {et~al.}(2019)\citenamefont
  {Bhattacharya}, \citenamefont {Gupta},\ and\ \citenamefont
  {Yoon}}]{Bhattacharya:2019}%
  \BibitemOpen
  \bibfield  {author} {\bibinfo {author} {\bibfnamefont {T.}~\bibnamefont
  {Bhattacharya}}, \bibinfo {author} {\bibfnamefont {R.}~\bibnamefont {Gupta}},
  \ and\ \bibinfo {author} {\bibfnamefont {B.}~\bibnamefont {Yoon}},\ }in\
  \href@noop {} {\emph {\bibinfo {booktitle} {Proceedings of Science: 37th
  International Symposium on Lattice Field Theory}}}\ (\bibinfo {year} {2019})\
  p.\ \bibinfo {pages} {247}\BibitemShut {NoStop}%
\bibitem [{\citenamefont {Park}\ \emph {et~al.}(2020)\citenamefont {Park},
  \citenamefont {Bhattacharya}, \citenamefont {Gupta}, \citenamefont {Jang},
  \citenamefont {Joo}, \citenamefont {Lin},\ and\ \citenamefont
  {Yoon}}]{Park:2020axe}%
  \BibitemOpen
  \bibfield  {author} {\bibinfo {author} {\bibfnamefont {S.}~\bibnamefont
  {Park}}, \bibinfo {author} {\bibfnamefont {T.}~\bibnamefont {Bhattacharya}},
  \bibinfo {author} {\bibfnamefont {R.}~\bibnamefont {Gupta}}, \bibinfo
  {author} {\bibfnamefont {Y.-C.}\ \bibnamefont {Jang}}, \bibinfo {author}
  {\bibfnamefont {B.}~\bibnamefont {Joo}}, \bibinfo {author} {\bibfnamefont
  {H.-W.}\ \bibnamefont {Lin}}, \ and\ \bibinfo {author} {\bibfnamefont
  {B.}~\bibnamefont {Yoon}},\ }in\ \href@noop {} {\emph {\bibinfo {booktitle}
  {{37th International Symposium on Lattice Field Theory (Lattice 2019) Wuhan,
  Hubei, China, June 16-22, 2019}}}}\ (\bibinfo {year} {2020})\ \Eprint
  {http://arxiv.org/abs/2002.02147} {arXiv:2002.02147 [hep-lat]} \BibitemShut
  {NoStop}%
\bibitem [{\citenamefont {Shen}\ \emph {et~al.}(2012)\citenamefont {Shen},
  \citenamefont {Marcucci}, \citenamefont {Carlson}, \citenamefont {Gandolfi},\
  and\ \citenamefont {Schiavilla}}]{Shen:2012}%
  \BibitemOpen
  \bibfield  {author} {\bibinfo {author} {\bibfnamefont {G.}~\bibnamefont
  {Shen}}, \bibinfo {author} {\bibfnamefont {L.}~\bibnamefont {Marcucci}},
  \bibinfo {author} {\bibfnamefont {J.}~\bibnamefont {Carlson}}, \bibinfo
  {author} {\bibfnamefont {S.}~\bibnamefont {Gandolfi}}, \ and\ \bibinfo
  {author} {\bibfnamefont {R.}~\bibnamefont {Schiavilla}},\ }\href {\doibase
  10.1103/PhysRevC.86.035503} {\bibfield  {journal} {\bibinfo  {journal} {Phys.
  Rev. C}\ }\textbf {\bibinfo {volume} {86}},\ \bibinfo {pages} {035503}
  (\bibinfo {year} {2012})}\BibitemShut {NoStop}%
\bibitem [{\citenamefont {Marcucci}\ \emph {et~al.}(2001)\citenamefont
  {Marcucci}, \citenamefont {Schiavilla}, \citenamefont {Viviani},
  \citenamefont {Kievsky}, \citenamefont {Rosati},\ and\ \citenamefont
  {Beacom}}]{Marcucci:2000xy}%
  \BibitemOpen
  \bibfield  {author} {\bibinfo {author} {\bibfnamefont {L.~E.}\ \bibnamefont
  {Marcucci}}, \bibinfo {author} {\bibfnamefont {R.}~\bibnamefont
  {Schiavilla}}, \bibinfo {author} {\bibfnamefont {M.}~\bibnamefont {Viviani}},
  \bibinfo {author} {\bibfnamefont {A.}~\bibnamefont {Kievsky}}, \bibinfo
  {author} {\bibfnamefont {S.}~\bibnamefont {Rosati}}, \ and\ \bibinfo {author}
  {\bibfnamefont {J.~F.}\ \bibnamefont {Beacom}},\ }\href {\doibase
  10.1103/PhysRevC.63.015801} {\bibfield  {journal} {\bibinfo  {journal} {Phys.
  Rev. C}\ }\textbf {\bibinfo {volume} {63}},\ \bibinfo {pages} {015801}
  (\bibinfo {year} {2001})},\ \Eprint {http://arxiv.org/abs/nucl-th/0006005}
  {arXiv:nucl-th/0006005 [nucl-th]} \BibitemShut {NoStop}%
\bibitem [{\citenamefont {Marcucci}\ \emph {et~al.}(2005)\citenamefont
  {Marcucci}, \citenamefont {Viviani}, \citenamefont {Schiavilla},
  \citenamefont {Kievsky},\ and\ \citenamefont {Rosati}}]{Marcucci:2005zc}%
  \BibitemOpen
  \bibfield  {author} {\bibinfo {author} {\bibfnamefont {L.~E.}\ \bibnamefont
  {Marcucci}}, \bibinfo {author} {\bibfnamefont {M.}~\bibnamefont {Viviani}},
  \bibinfo {author} {\bibfnamefont {R.}~\bibnamefont {Schiavilla}}, \bibinfo
  {author} {\bibfnamefont {A.}~\bibnamefont {Kievsky}}, \ and\ \bibinfo
  {author} {\bibfnamefont {S.}~\bibnamefont {Rosati}},\ }\href {\doibase
  10.1103/PhysRevC.72.014001} {\bibfield  {journal} {\bibinfo  {journal} {Phys.
  Rev. C}\ }\textbf {\bibinfo {volume} {72}},\ \bibinfo {pages} {014001}
  (\bibinfo {year} {2005})},\ \Eprint {http://arxiv.org/abs/nucl-th/0502048}
  {arXiv:nucl-th/0502048 [nucl-th]} \BibitemShut {NoStop}%
\bibitem [{\citenamefont {Lovato}\ \emph {et~al.}(2013)\citenamefont {Lovato},
  \citenamefont {Gandolfi}, \citenamefont {Butler}, \citenamefont {Carlson},
  \citenamefont {Lusk}, \citenamefont {Pieper},\ and\ \citenamefont
  {Schiavilla}}]{Lovato:2013}%
  \BibitemOpen
  \bibfield  {author} {\bibinfo {author} {\bibfnamefont {A.}~\bibnamefont
  {Lovato}}, \bibinfo {author} {\bibfnamefont {S.}~\bibnamefont {Gandolfi}},
  \bibinfo {author} {\bibfnamefont {R.}~\bibnamefont {Butler}}, \bibinfo
  {author} {\bibfnamefont {J.}~\bibnamefont {Carlson}}, \bibinfo {author}
  {\bibfnamefont {E.}~\bibnamefont {Lusk}}, \bibinfo {author} {\bibfnamefont
  {S.~C.}\ \bibnamefont {Pieper}}, \ and\ \bibinfo {author} {\bibfnamefont
  {R.}~\bibnamefont {Schiavilla}},\ }\href {\doibase
  10.1103/PhysRevLett.111.092501} {\bibfield  {journal} {\bibinfo  {journal}
  {Phys. Rev. Lett.}\ }\textbf {\bibinfo {volume} {111}},\ \bibinfo {pages}
  {092501} (\bibinfo {year} {2013})}\BibitemShut {NoStop}%
\bibitem [{\citenamefont {Lovato}\ \emph {et~al.}(2016)\citenamefont {Lovato},
  \citenamefont {Gandolfi}, \citenamefont {Carlson}, \citenamefont {Pieper},\
  and\ \citenamefont {Schiavilla}}]{Lovato:2016gkq}%
  \BibitemOpen
  \bibfield  {author} {\bibinfo {author} {\bibfnamefont {A.}~\bibnamefont
  {Lovato}}, \bibinfo {author} {\bibfnamefont {S.}~\bibnamefont {Gandolfi}},
  \bibinfo {author} {\bibfnamefont {J.}~\bibnamefont {Carlson}}, \bibinfo
  {author} {\bibfnamefont {S.~C.}\ \bibnamefont {Pieper}}, \ and\ \bibinfo
  {author} {\bibfnamefont {R.}~\bibnamefont {Schiavilla}},\ }\href {\doibase
  10.1103/PhysRevLett.117.082501} {\bibfield  {journal} {\bibinfo  {journal}
  {Phys. Rev. Lett.}\ }\textbf {\bibinfo {volume} {117}},\ \bibinfo {pages}
  {082501} (\bibinfo {year} {2016})},\ \Eprint
  {http://arxiv.org/abs/1605.00248} {arXiv:1605.00248 [nucl-th]} \BibitemShut
  {NoStop}%
\bibitem [{\citenamefont {Aguilar-Arevalo}\ \emph {et~al.}(2010)\citenamefont
  {Aguilar-Arevalo} \emph {et~al.}}]{AguilarArevalo:2010zc}%
  \BibitemOpen
  \bibfield  {author} {\bibinfo {author} {\bibfnamefont {A.~A.}\ \bibnamefont
  {Aguilar-Arevalo}} \emph {et~al.} (\bibinfo {collaboration} {MiniBooNE}),\
  }\href {\doibase 10.1103/PhysRevD.81.092005} {\bibfield  {journal} {\bibinfo
  {journal} {Phys. Rev. D}\ }\textbf {\bibinfo {volume} {81}},\ \bibinfo
  {pages} {092005} (\bibinfo {year} {2010})},\ \Eprint
  {http://arxiv.org/abs/1002.2680} {arXiv:1002.2680 [hep-ex]} \BibitemShut
  {NoStop}%
\bibitem [{\citenamefont {Aguilar-Arevalo}\ \emph {et~al.}(2013)\citenamefont
  {Aguilar-Arevalo} \emph {et~al.}}]{Aguilar-Arevalo:2013dva}%
  \BibitemOpen
  \bibfield  {author} {\bibinfo {author} {\bibfnamefont {A.~A.}\ \bibnamefont
  {Aguilar-Arevalo}} \emph {et~al.} (\bibinfo {collaboration} {MiniBooNE}),\
  }\href {\doibase 10.1103/PhysRevD.88.032001} {\bibfield  {journal} {\bibinfo
  {journal} {Phys. Rev. D}\ }\textbf {\bibinfo {volume} {88}},\ \bibinfo
  {pages} {032001} (\bibinfo {year} {2013})},\ \Eprint
  {http://arxiv.org/abs/1301.7067} {arXiv:1301.7067 [hep-ex]} \BibitemShut
  {NoStop}%
\bibitem [{\citenamefont {Abe}\ \emph {et~al.}(2016)\citenamefont {Abe} \emph
  {et~al.}}]{Abe:2016tmq}%
  \BibitemOpen
  \bibfield  {author} {\bibinfo {author} {\bibfnamefont {K.}~\bibnamefont
  {Abe}} \emph {et~al.} (\bibinfo {collaboration} {T2K}),\ }\href {\doibase
  10.1103/PhysRevD.93.112012} {\bibfield  {journal} {\bibinfo  {journal} {Phys.
  Rev. D}\ }\textbf {\bibinfo {volume} {93}},\ \bibinfo {pages} {112012}
  (\bibinfo {year} {2016})},\ \Eprint {http://arxiv.org/abs/1602.03652}
  {arXiv:1602.03652 [hep-ex]} \BibitemShut {NoStop}%
\bibitem [{\citenamefont {Leitner}\ and\ \citenamefont
  {Mosel}(2010)}]{Leitner:2010kp}%
  \BibitemOpen
  \bibfield  {author} {\bibinfo {author} {\bibfnamefont {T.}~\bibnamefont
  {Leitner}}\ and\ \bibinfo {author} {\bibfnamefont {U.}~\bibnamefont
  {Mosel}},\ }\href {\doibase 10.1103/PhysRevC.81.064614} {\bibfield  {journal}
  {\bibinfo  {journal} {Phys. Rev. C}\ }\textbf {\bibinfo {volume} {81}},\
  \bibinfo {pages} {064614} (\bibinfo {year} {2010})},\ \Eprint
  {http://arxiv.org/abs/1004.4433} {arXiv:1004.4433 [nucl-th]} \BibitemShut
  {NoStop}%
\bibitem [{\citenamefont {Nakamura}\ \emph {et~al.}(2002)\citenamefont
  {Nakamura}, \citenamefont {Sato}, \citenamefont {Ando}, \citenamefont {Park},
  \citenamefont {Myhrer}, \citenamefont {Gudkov},\ and\ \citenamefont
  {Kubodera}}]{Nakamura:2002jg}%
  \BibitemOpen
  \bibfield  {author} {\bibinfo {author} {\bibfnamefont {S.}~\bibnamefont
  {Nakamura}}, \bibinfo {author} {\bibfnamefont {T.}~\bibnamefont {Sato}},
  \bibinfo {author} {\bibfnamefont {S.}~\bibnamefont {Ando}}, \bibinfo {author}
  {\bibfnamefont {T.~S.}\ \bibnamefont {Park}}, \bibinfo {author}
  {\bibfnamefont {F.}~\bibnamefont {Myhrer}}, \bibinfo {author} {\bibfnamefont
  {V.~P.}\ \bibnamefont {Gudkov}}, \ and\ \bibinfo {author} {\bibfnamefont
  {K.}~\bibnamefont {Kubodera}},\ }\href {\doibase
  10.1016/S0375-9474(02)00993-4} {\bibfield  {journal} {\bibinfo  {journal}
  {Nucl. Phys. A}\ }\textbf {\bibinfo {volume} {707}},\ \bibinfo {pages} {561}
  (\bibinfo {year} {2002})},\ \Eprint {http://arxiv.org/abs/nucl-th/0201062}
  {arXiv:nucl-th/0201062 [nucl-th]} \BibitemShut {NoStop}%
\bibitem [{\citenamefont {Nakamura}\ and\ \citenamefont {Group}(2010)}]{PDG}%
  \BibitemOpen
  \bibfield  {author} {\bibinfo {author} {\bibfnamefont {K.}~\bibnamefont
  {Nakamura}}\ and\ \bibinfo {author} {\bibfnamefont {P.~D.}\ \bibnamefont
  {Group}},\ }\href@noop {} {\bibfield  {journal} {\bibinfo  {journal} {Journal
  of Physics G: Nuclear and Particle Physics}\ }\textbf {\bibinfo {volume}
  {37}},\ \bibinfo {pages} {075021} (\bibinfo {year} {2010})}\BibitemShut
  {NoStop}%
\bibitem [{\citenamefont {Piarulli}\ \emph {et~al.}(2015)\citenamefont
  {Piarulli}, \citenamefont {Girlanda}, \citenamefont {Schiavilla},
  \citenamefont {P\'erez}, \citenamefont {Amaro},\ and\ \citenamefont
  {Arriola}}]{Piarulli:2015}%
  \BibitemOpen
  \bibfield  {author} {\bibinfo {author} {\bibfnamefont {M.}~\bibnamefont
  {Piarulli}}, \bibinfo {author} {\bibfnamefont {L.}~\bibnamefont {Girlanda}},
  \bibinfo {author} {\bibfnamefont {R.}~\bibnamefont {Schiavilla}}, \bibinfo
  {author} {\bibfnamefont {R.~N.}\ \bibnamefont {P\'erez}}, \bibinfo {author}
  {\bibfnamefont {J.~E.}\ \bibnamefont {Amaro}}, \ and\ \bibinfo {author}
  {\bibfnamefont {E.~R.}\ \bibnamefont {Arriola}},\ }\href {\doibase
  10.1103/PhysRevC.91.024003} {\bibfield  {journal} {\bibinfo  {journal} {Phys.
  Rev. C}\ }\textbf {\bibinfo {volume} {91}},\ \bibinfo {pages} {024003}
  (\bibinfo {year} {2015})}\BibitemShut {NoStop}%
\bibitem [{\citenamefont {Pieper}\ \emph {et~al.}(2001)\citenamefont {Pieper},
  \citenamefont {Pandharipande}, \citenamefont {Wiringa},\ and\ \citenamefont
  {Carlson}}]{Pieper:2001ap}%
  \BibitemOpen
  \bibfield  {author} {\bibinfo {author} {\bibfnamefont {S.~C.}\ \bibnamefont
  {Pieper}}, \bibinfo {author} {\bibfnamefont {V.~R.}\ \bibnamefont
  {Pandharipande}}, \bibinfo {author} {\bibfnamefont {R.~B.}\ \bibnamefont
  {Wiringa}}, \ and\ \bibinfo {author} {\bibfnamefont {J.}~\bibnamefont
  {Carlson}},\ }\href {\doibase 10.1103/PhysRevC.64.014001} {\bibfield
  {journal} {\bibinfo  {journal} {Phys. Rev. C}\ }\textbf {\bibinfo {volume}
  {64}},\ \bibinfo {pages} {014001} (\bibinfo {year} {2001})},\ \Eprint
  {http://arxiv.org/abs/nucl-th/0102004} {arXiv:nucl-th/0102004 [nucl-th]}
  \BibitemShut {NoStop}%
\bibitem [{\citenamefont {Lovato}\ \emph {et~al.}(2019)\citenamefont {Lovato},
  \citenamefont {Rocco},\ and\ \citenamefont {Schiavilla}}]{Lovato:2019fiw}%
  \BibitemOpen
  \bibfield  {author} {\bibinfo {author} {\bibfnamefont {A.}~\bibnamefont
  {Lovato}}, \bibinfo {author} {\bibfnamefont {N.}~\bibnamefont {Rocco}}, \
  and\ \bibinfo {author} {\bibfnamefont {R.}~\bibnamefont {Schiavilla}},\
  }\href {\doibase 10.1103/PhysRevC.100.035502} {\bibfield  {journal} {\bibinfo
   {journal} {Phys. Rev. C}\ }\textbf {\bibinfo {volume} {100}},\ \bibinfo
  {pages} {035502} (\bibinfo {year} {2019})},\ \Eprint
  {http://arxiv.org/abs/1903.08078} {arXiv:1903.08078 [nucl-th]} \BibitemShut
  {NoStop}%
\bibitem [{\citenamefont {Kelly}(2004)}]{Kelly:2004hm}%
  \BibitemOpen
  \bibfield  {author} {\bibinfo {author} {\bibfnamefont {J.~J.}\ \bibnamefont
  {Kelly}},\ }\href {\doibase 10.1103/PhysRevC.70.068202} {\bibfield  {journal}
  {\bibinfo  {journal} {Phys. Rev. C}\ }\textbf {\bibinfo {volume} {70}},\
  \bibinfo {pages} {068202} (\bibinfo {year} {2004})}\BibitemShut {NoStop}%
\bibitem [{\citenamefont {Andreev}\ \emph {et~al.}(2007)\citenamefont
  {Andreev}, \citenamefont {Banks}, \citenamefont {Case}, \citenamefont
  {Chitwood}, \citenamefont {Clayton}, \citenamefont {Crowe}, \citenamefont
  {Deutsch}, \citenamefont {Egger}, \citenamefont {Freedman}, \citenamefont
  {Ganzha}, \citenamefont {Gorringe}, \citenamefont {Gray}, \citenamefont
  {Hertzog}, \citenamefont {Hildebrandt}, \citenamefont {Kammel}, \citenamefont
  {Kiburg}, \citenamefont {Knaack}, \citenamefont {Kravtsov}, \citenamefont
  {Krivshich}, \citenamefont {Lauss}, \citenamefont {Lynch}, \citenamefont
  {Maev}, \citenamefont {Maev}, \citenamefont {Mulhauser}, \citenamefont
  {\"Ozben}, \citenamefont {Petitjean}, \citenamefont {Petrov}, \citenamefont
  {Prieels}, \citenamefont {Schapkin}, \citenamefont {Semenchuk}, \citenamefont
  {Soroka}, \citenamefont {Tishchenko}, \citenamefont {Vasilyev}, \citenamefont
  {Vorobyov}, \citenamefont {Vznuzdaev},\ and\ \citenamefont
  {Winter}}]{And07-all}%
  \BibitemOpen
  \bibfield  {author} {\bibinfo {author} {\bibfnamefont {V.~A.}\ \bibnamefont
  {Andreev}}, \bibinfo {author} {\bibfnamefont {T.~I.}\ \bibnamefont {Banks}},
  \bibinfo {author} {\bibfnamefont {T.~A.}\ \bibnamefont {Case}}, \bibinfo
  {author} {\bibfnamefont {D.~B.}\ \bibnamefont {Chitwood}}, \bibinfo {author}
  {\bibfnamefont {S.~M.}\ \bibnamefont {Clayton}}, \bibinfo {author}
  {\bibfnamefont {K.~M.}\ \bibnamefont {Crowe}}, \bibinfo {author}
  {\bibfnamefont {J.}~\bibnamefont {Deutsch}}, \bibinfo {author} {\bibfnamefont
  {J.}~\bibnamefont {Egger}}, \bibinfo {author} {\bibfnamefont {S.~J.}\
  \bibnamefont {Freedman}}, \bibinfo {author} {\bibfnamefont {V.~A.}\
  \bibnamefont {Ganzha}}, \bibinfo {author} {\bibfnamefont {T.}~\bibnamefont
  {Gorringe}}, \bibinfo {author} {\bibfnamefont {F.~E.}\ \bibnamefont {Gray}},
  \bibinfo {author} {\bibfnamefont {D.~W.}\ \bibnamefont {Hertzog}}, \bibinfo
  {author} {\bibfnamefont {M.}~\bibnamefont {Hildebrandt}}, \bibinfo {author}
  {\bibfnamefont {P.}~\bibnamefont {Kammel}}, \bibinfo {author} {\bibfnamefont
  {B.}~\bibnamefont {Kiburg}}, \bibinfo {author} {\bibfnamefont
  {S.}~\bibnamefont {Knaack}}, \bibinfo {author} {\bibfnamefont {P.~A.}\
  \bibnamefont {Kravtsov}}, \bibinfo {author} {\bibfnamefont {A.~G.}\
  \bibnamefont {Krivshich}}, \bibinfo {author} {\bibfnamefont {B.}~\bibnamefont
  {Lauss}}, \bibinfo {author} {\bibfnamefont {K.~L.}\ \bibnamefont {Lynch}},
  \bibinfo {author} {\bibfnamefont {E.~M.}\ \bibnamefont {Maev}}, \bibinfo
  {author} {\bibfnamefont {O.~E.}\ \bibnamefont {Maev}}, \bibinfo {author}
  {\bibfnamefont {F.}~\bibnamefont {Mulhauser}}, \bibinfo {author}
  {\bibfnamefont {C.~S.}\ \bibnamefont {\"Ozben}}, \bibinfo {author}
  {\bibfnamefont {C.}~\bibnamefont {Petitjean}}, \bibinfo {author}
  {\bibfnamefont {G.~E.}\ \bibnamefont {Petrov}}, \bibinfo {author}
  {\bibfnamefont {R.}~\bibnamefont {Prieels}}, \bibinfo {author} {\bibfnamefont
  {G.~N.}\ \bibnamefont {Schapkin}}, \bibinfo {author} {\bibfnamefont {G.~G.}\
  \bibnamefont {Semenchuk}}, \bibinfo {author} {\bibfnamefont {M.~A.}\
  \bibnamefont {Soroka}}, \bibinfo {author} {\bibfnamefont {V.}~\bibnamefont
  {Tishchenko}}, \bibinfo {author} {\bibfnamefont {A.~A.}\ \bibnamefont
  {Vasilyev}}, \bibinfo {author} {\bibfnamefont {A.~A.}\ \bibnamefont
  {Vorobyov}}, \bibinfo {author} {\bibfnamefont {M.~E.}\ \bibnamefont
  {Vznuzdaev}}, \ and\ \bibinfo {author} {\bibfnamefont {P.}~\bibnamefont
  {Winter}} (\bibinfo {collaboration} {MuCap Collaboration}),\ }\href {\doibase
  10.1103/PhysRevLett.99.032002} {\bibfield  {journal} {\bibinfo  {journal}
  {Phys. Rev. Lett.}\ }\textbf {\bibinfo {volume} {99}},\ \bibinfo {pages}
  {032002} (\bibinfo {year} {2007})}\BibitemShut {NoStop}%
\bibitem [{\citenamefont {Bernard}\ \emph {et~al.}(1994)\citenamefont
  {Bernard}, \citenamefont {Kaiser},\ and\ \citenamefont {Meissner}}]{Ber94}%
  \BibitemOpen
  \bibfield  {author} {\bibinfo {author} {\bibfnamefont {V.}~\bibnamefont
  {Bernard}}, \bibinfo {author} {\bibfnamefont {N.}~\bibnamefont {Kaiser}}, \
  and\ \bibinfo {author} {\bibfnamefont {U.-G.}\ \bibnamefont {Meissner}},\
  }\href {\doibase 10.1103/PhysRevD.50.6899} {\bibfield  {journal} {\bibinfo
  {journal} {Phys. Rev. D}\ }\textbf {\bibinfo {volume} {50}},\ \bibinfo
  {pages} {6899} (\bibinfo {year} {1994})}\BibitemShut {NoStop}%
\bibitem [{\citenamefont {Bernard}\ \emph {et~al.}(2002)\citenamefont
  {Bernard}, \citenamefont {Elouadrhiri},\ and\ \citenamefont
  {Meissner}}]{Bernard:2001rs}%
  \BibitemOpen
  \bibfield  {author} {\bibinfo {author} {\bibfnamefont {V.}~\bibnamefont
  {Bernard}}, \bibinfo {author} {\bibfnamefont {L.}~\bibnamefont
  {Elouadrhiri}}, \ and\ \bibinfo {author} {\bibfnamefont {U.-G.}\ \bibnamefont
  {Meissner}},\ }\href {\doibase 10.1088/0954-3899/28/1/201} {\bibfield
  {journal} {\bibinfo  {journal} {J. Phys. G}\ }\textbf {\bibinfo {volume}
  {28}},\ \bibinfo {pages} {R1} (\bibinfo {year} {2002})},\ \Eprint
  {http://arxiv.org/abs/hep-ph/0107088} {arXiv:hep-ph/0107088 [hep-ph]}
  \BibitemShut {NoStop}%
\bibitem [{\citenamefont {Carlson}(1986)}]{Carlson:1985mm}%
  \BibitemOpen
  \bibfield  {author} {\bibinfo {author} {\bibfnamefont {C.~E.}\ \bibnamefont
  {Carlson}},\ }\href {\doibase 10.1103/PhysRevD.34.2704} {\bibfield  {journal}
  {\bibinfo  {journal} {Phys. Rev. D}\ }\textbf {\bibinfo {volume} {34}},\
  \bibinfo {pages} {2704} (\bibinfo {year} {1986})}\BibitemShut {NoStop}%
\bibitem [{\citenamefont {Carlson}\ and\ \citenamefont
  {Schiavilla}(1992)}]{Carlson:1992}%
  \BibitemOpen
  \bibfield  {author} {\bibinfo {author} {\bibfnamefont {J.}~\bibnamefont
  {Carlson}}\ and\ \bibinfo {author} {\bibfnamefont {R.}~\bibnamefont
  {Schiavilla}},\ }\href {\doibase 10.1103/PhysRevLett.68.3682} {\bibfield
  {journal} {\bibinfo  {journal} {Phys. Rev. Lett.}\ }\textbf {\bibinfo
  {volume} {68}},\ \bibinfo {pages} {3682} (\bibinfo {year}
  {1992})}\BibitemShut {NoStop}%
\bibitem [{\citenamefont {Carlson}\ and\ \citenamefont
  {Schiavilla}(1994)}]{Carlson:1994zz}%
  \BibitemOpen
  \bibfield  {author} {\bibinfo {author} {\bibfnamefont {J.}~\bibnamefont
  {Carlson}}\ and\ \bibinfo {author} {\bibfnamefont {R.}~\bibnamefont
  {Schiavilla}},\ }\href {\doibase 10.1103/PhysRevC.49.R2880} {\bibfield
  {journal} {\bibinfo  {journal} {Phys. Rev. C}\ }\textbf {\bibinfo {volume}
  {49}},\ \bibinfo {pages} {R2880} (\bibinfo {year} {1994})}\BibitemShut
  {NoStop}%
\bibitem [{\citenamefont {Lovato}\ \emph {et~al.}(2015)\citenamefont {Lovato},
  \citenamefont {Gandolfi}, \citenamefont {Carlson}, \citenamefont {Pieper},\
  and\ \citenamefont {Schiavilla}}]{Lovato:2015}%
  \BibitemOpen
  \bibfield  {author} {\bibinfo {author} {\bibfnamefont {A.}~\bibnamefont
  {Lovato}}, \bibinfo {author} {\bibfnamefont {S.}~\bibnamefont {Gandolfi}},
  \bibinfo {author} {\bibfnamefont {J.}~\bibnamefont {Carlson}}, \bibinfo
  {author} {\bibfnamefont {S.~C.}\ \bibnamefont {Pieper}}, \ and\ \bibinfo
  {author} {\bibfnamefont {R.}~\bibnamefont {Schiavilla}},\ }\href {\doibase
  10.1103/PhysRevC.91.062501} {\bibfield  {journal} {\bibinfo  {journal} {Phys.
  Rev. C}\ }\textbf {\bibinfo {volume} {91}},\ \bibinfo {pages} {062501}
  (\bibinfo {year} {2015})}\BibitemShut {NoStop}%
\bibitem [{\citenamefont {Lovato}\ \emph {et~al.}(2018)\citenamefont {Lovato},
  \citenamefont {Gandolfi}, \citenamefont {Carlson}, \citenamefont {Lusk},
  \citenamefont {Pieper},\ and\ \citenamefont {Schiavilla}}]{Lovato:2017cux}%
  \BibitemOpen
  \bibfield  {author} {\bibinfo {author} {\bibfnamefont {A.}~\bibnamefont
  {Lovato}}, \bibinfo {author} {\bibfnamefont {S.}~\bibnamefont {Gandolfi}},
  \bibinfo {author} {\bibfnamefont {J.}~\bibnamefont {Carlson}}, \bibinfo
  {author} {\bibfnamefont {E.}~\bibnamefont {Lusk}}, \bibinfo {author}
  {\bibfnamefont {S.~C.}\ \bibnamefont {Pieper}}, \ and\ \bibinfo {author}
  {\bibfnamefont {R.}~\bibnamefont {Schiavilla}},\ }\href {\doibase
  10.1103/PhysRevC.97.022502} {\bibfield  {journal} {\bibinfo  {journal} {Phys.
  Rev. C}\ }\textbf {\bibinfo {volume} {97}},\ \bibinfo {pages} {022502}
  (\bibinfo {year} {2018})},\ \Eprint {http://arxiv.org/abs/1711.02047}
  {arXiv:1711.02047 [nucl-th]} \BibitemShut {NoStop}%
\bibitem [{\citenamefont {Lovato}\ \emph {et~al.}(2014)\citenamefont {Lovato},
  \citenamefont {Gandolfi}, \citenamefont {Carlson}, \citenamefont {Pieper},\
  and\ \citenamefont {Schiavilla}}]{Lovato:2014}%
  \BibitemOpen
  \bibfield  {author} {\bibinfo {author} {\bibfnamefont {A.}~\bibnamefont
  {Lovato}}, \bibinfo {author} {\bibfnamefont {S.}~\bibnamefont {Gandolfi}},
  \bibinfo {author} {\bibfnamefont {J.}~\bibnamefont {Carlson}}, \bibinfo
  {author} {\bibfnamefont {S.~C.}\ \bibnamefont {Pieper}}, \ and\ \bibinfo
  {author} {\bibfnamefont {R.}~\bibnamefont {Schiavilla}},\ }\href {\doibase
  10.1103/PhysRevLett.112.182502} {\bibfield  {journal} {\bibinfo  {journal}
  {Phys. Rev. Lett.}\ }\textbf {\bibinfo {volume} {112}},\ \bibinfo {pages}
  {182502} (\bibinfo {year} {2014})}\BibitemShut {NoStop}%
\bibitem [{\citenamefont {Rocco}\ \emph {et~al.}(2017)\citenamefont {Rocco},
  \citenamefont {Alvarez-Ruso}, \citenamefont {Lovato},\ and\ \citenamefont
  {Nieves}}]{Rocco:2017hmh}%
  \BibitemOpen
  \bibfield  {author} {\bibinfo {author} {\bibfnamefont {N.}~\bibnamefont
  {Rocco}}, \bibinfo {author} {\bibfnamefont {L.}~\bibnamefont {Alvarez-Ruso}},
  \bibinfo {author} {\bibfnamefont {A.}~\bibnamefont {Lovato}}, \ and\ \bibinfo
  {author} {\bibfnamefont {J.}~\bibnamefont {Nieves}},\ }\href {\doibase
  10.1103/PhysRevC.96.015504} {\bibfield  {journal} {\bibinfo  {journal} {Phys.
  Rev. C}\ }\textbf {\bibinfo {volume} {96}},\ \bibinfo {pages} {015504}
  (\bibinfo {year} {2017})},\ \Eprint {http://arxiv.org/abs/1701.05151}
  {arXiv:1701.05151 [nucl-th]} \BibitemShut {NoStop}%
\bibitem [{\citenamefont {Amaro}\ \emph {et~al.}(2005)\citenamefont {Amaro},
  \citenamefont {Barbaro}, \citenamefont {Caballero}, \citenamefont {Donnelly},
  \citenamefont {Molinari},\ and\ \citenamefont {Sick}}]{Amaro:2004bs}%
  \BibitemOpen
  \bibfield  {author} {\bibinfo {author} {\bibfnamefont {J.~E.}\ \bibnamefont
  {Amaro}}, \bibinfo {author} {\bibfnamefont {M.~B.}\ \bibnamefont {Barbaro}},
  \bibinfo {author} {\bibfnamefont {J.~A.}\ \bibnamefont {Caballero}}, \bibinfo
  {author} {\bibfnamefont {T.~W.}\ \bibnamefont {Donnelly}}, \bibinfo {author}
  {\bibfnamefont {A.}~\bibnamefont {Molinari}}, \ and\ \bibinfo {author}
  {\bibfnamefont {I.}~\bibnamefont {Sick}},\ }\href {\doibase
  10.1103/PhysRevC.71.015501} {\bibfield  {journal} {\bibinfo  {journal} {Phys.
  Rev. C}\ }\textbf {\bibinfo {volume} {71}},\ \bibinfo {pages} {015501}
  (\bibinfo {year} {2005})},\ \Eprint {http://arxiv.org/abs/nucl-th/0409078}
  {arXiv:nucl-th/0409078 [nucl-th]} \BibitemShut {NoStop}%
\bibitem [{\citenamefont {{Caballero}}\ \emph {et~al.}(2007)\citenamefont
  {{Caballero}}, \citenamefont {{Amaro}}, \citenamefont {{Barbaro}},
  \citenamefont {{Donnelly}},\ and\ \citenamefont
  {{Ud{\'{\i}}as}}}]{Caballero:2007}%
  \BibitemOpen
  \bibfield  {author} {\bibinfo {author} {\bibfnamefont {J.~A.}\ \bibnamefont
  {{Caballero}}}, \bibinfo {author} {\bibfnamefont {J.~E.}\ \bibnamefont
  {{Amaro}}}, \bibinfo {author} {\bibfnamefont {M.~B.}\ \bibnamefont
  {{Barbaro}}}, \bibinfo {author} {\bibfnamefont {T.~W.}\ \bibnamefont
  {{Donnelly}}}, \ and\ \bibinfo {author} {\bibfnamefont {J.~M.}\ \bibnamefont
  {{Ud{\'{\i}}as}}},\ }\href {\doibase 10.1016/j.physletb.2007.08.018}
  {\bibfield  {journal} {\bibinfo  {journal} {Physics Letters B}\ }\textbf
  {\bibinfo {volume} {653}},\ \bibinfo {pages} {366} (\bibinfo {year}
  {2007})},\ \Eprint {http://arxiv.org/abs/0705.1429} {arXiv:0705.1429
  [nucl-th]} \BibitemShut {NoStop}%
\bibitem [{\citenamefont {Pastore}\ \emph {et~al.}(2019)\citenamefont
  {Pastore}, \citenamefont {Carlson}, \citenamefont {Gandolfi}, \citenamefont
  {Schiavilla},\ and\ \citenamefont {Wiringa}}]{Pastore:2019urn}%
  \BibitemOpen
  \bibfield  {author} {\bibinfo {author} {\bibfnamefont {S.}~\bibnamefont
  {Pastore}}, \bibinfo {author} {\bibfnamefont {J.}~\bibnamefont {Carlson}},
  \bibinfo {author} {\bibfnamefont {S.}~\bibnamefont {Gandolfi}}, \bibinfo
  {author} {\bibfnamefont {R.}~\bibnamefont {Schiavilla}}, \ and\ \bibinfo
  {author} {\bibfnamefont {R.~B.}\ \bibnamefont {Wiringa}},\ }\href@noop {} {\
  (\bibinfo {year} {2019})},\ \Eprint {http://arxiv.org/abs/1909.06400}
  {arXiv:1909.06400 [nucl-th]} \BibitemShut {NoStop}%
\bibitem [{\citenamefont {Wiringa}\ \emph {et~al.}(2014)\citenamefont
  {Wiringa}, \citenamefont {Schiavilla}, \citenamefont {Pieper},\ and\
  \citenamefont {Carlson}}]{Wiringa:2013ala}%
  \BibitemOpen
  \bibfield  {author} {\bibinfo {author} {\bibfnamefont {R.~B.}\ \bibnamefont
  {Wiringa}}, \bibinfo {author} {\bibfnamefont {R.}~\bibnamefont {Schiavilla}},
  \bibinfo {author} {\bibfnamefont {S.~C.}\ \bibnamefont {Pieper}}, \ and\
  \bibinfo {author} {\bibfnamefont {J.}~\bibnamefont {Carlson}},\ }\href
  {\doibase 10.1103/PhysRevC.89.024305} {\bibfield  {journal} {\bibinfo
  {journal} {Phys. Rev. C}\ }\textbf {\bibinfo {volume} {89}},\ \bibinfo
  {pages} {024305} (\bibinfo {year} {2014})},\ \Eprint
  {http://arxiv.org/abs/1309.3794} {arXiv:1309.3794 [nucl-th]} \BibitemShut
  {NoStop}%
\bibitem [{\citenamefont {Jang}\ \emph {et~al.}(2020)\citenamefont {Jang},
  \citenamefont {Gupta}, \citenamefont {Yoon},\ and\ \citenamefont
  {Bhattacharya}}]{Jang:2019vkm}%
  \BibitemOpen
  \bibfield  {author} {\bibinfo {author} {\bibfnamefont {Y.-C.}\ \bibnamefont
  {Jang}}, \bibinfo {author} {\bibfnamefont {R.}~\bibnamefont {Gupta}},
  \bibinfo {author} {\bibfnamefont {B.}~\bibnamefont {Yoon}}, \ and\ \bibinfo
  {author} {\bibfnamefont {T.}~\bibnamefont {Bhattacharya}},\ }\href {\doibase
  10.1103/PhysRevLett.124.072002} {\bibfield  {journal} {\bibinfo  {journal}
  {Phys. Rev. Lett.}\ }\textbf {\bibinfo {volume} {124}},\ \bibinfo {pages}
  {072002} (\bibinfo {year} {2020})},\ \Eprint
  {http://arxiv.org/abs/1905.06470} {arXiv:1905.06470 [hep-lat]} \BibitemShut
  {NoStop}%
\bibitem [{\citenamefont {Carlson}\ \emph {et~al.}(1993)\citenamefont
  {Carlson}, \citenamefont {Pandharipande},\ and\ \citenamefont
  {Schiavilla}}]{Carlson:1993zz}%
  \BibitemOpen
  \bibfield  {author} {\bibinfo {author} {\bibfnamefont {J.}~\bibnamefont
  {Carlson}}, \bibinfo {author} {\bibfnamefont {V.~R.}\ \bibnamefont
  {Pandharipande}}, \ and\ \bibinfo {author} {\bibfnamefont {R.}~\bibnamefont
  {Schiavilla}},\ }\href {\doibase 10.1103/PhysRevC.47.484} {\bibfield
  {journal} {\bibinfo  {journal} {Phys. Rev. C}\ }\textbf {\bibinfo {volume}
  {47}},\ \bibinfo {pages} {484} (\bibinfo {year} {1993})}\BibitemShut
  {NoStop}%
\bibitem [{\citenamefont {Rocco}\ \emph {et~al.}(2019)\citenamefont {Rocco},
  \citenamefont {Barbieri}, \citenamefont {Benhar}, \citenamefont {De~Pace},\
  and\ \citenamefont {Lovato}}]{Rocco:2018mwt}%
  \BibitemOpen
  \bibfield  {author} {\bibinfo {author} {\bibfnamefont {N.}~\bibnamefont
  {Rocco}}, \bibinfo {author} {\bibfnamefont {C.}~\bibnamefont {Barbieri}},
  \bibinfo {author} {\bibfnamefont {O.}~\bibnamefont {Benhar}}, \bibinfo
  {author} {\bibfnamefont {A.}~\bibnamefont {De~Pace}}, \ and\ \bibinfo
  {author} {\bibfnamefont {A.}~\bibnamefont {Lovato}},\ }\href {\doibase
  10.1103/PhysRevC.99.025502} {\bibfield  {journal} {\bibinfo  {journal} {Phys.
  Rev. C}\ }\textbf {\bibinfo {volume} {99}},\ \bibinfo {pages} {025502}
  (\bibinfo {year} {2019})},\ \Eprint {http://arxiv.org/abs/1810.07647}
  {arXiv:1810.07647 [nucl-th]} \BibitemShut {NoStop}%
\bibitem [{\citenamefont {Lonardoni}\ \emph {et~al.}(2018)\citenamefont
  {Lonardoni}, \citenamefont {Gandolfi}, \citenamefont {Lynn}, \citenamefont
  {Petrie}, \citenamefont {Carlson}, \citenamefont {Schmidt},\ and\
  \citenamefont {Schwenk}}]{Lonardoni:2018nob}%
  \BibitemOpen
  \bibfield  {author} {\bibinfo {author} {\bibfnamefont {D.}~\bibnamefont
  {Lonardoni}}, \bibinfo {author} {\bibfnamefont {S.}~\bibnamefont {Gandolfi}},
  \bibinfo {author} {\bibfnamefont {J.~E.}\ \bibnamefont {Lynn}}, \bibinfo
  {author} {\bibfnamefont {C.}~\bibnamefont {Petrie}}, \bibinfo {author}
  {\bibfnamefont {J.}~\bibnamefont {Carlson}}, \bibinfo {author} {\bibfnamefont
  {K.~E.}\ \bibnamefont {Schmidt}}, \ and\ \bibinfo {author} {\bibfnamefont
  {A.}~\bibnamefont {Schwenk}},\ }\href {\doibase 10.1103/PhysRevC.97.044318}
  {\bibfield  {journal} {\bibinfo  {journal} {Phys. Rev. C}\ }\textbf {\bibinfo
  {volume} {97}},\ \bibinfo {pages} {044318} (\bibinfo {year} {2018})},\
  \Eprint {http://arxiv.org/abs/1802.08932} {arXiv:1802.08932 [nucl-th]}
  \BibitemShut {NoStop}%
\bibitem [{\citenamefont {Roggero}\ and\ \citenamefont
  {Carlson}(2019)}]{Roggero:2019a}%
  \BibitemOpen
  \bibfield  {author} {\bibinfo {author} {\bibfnamefont {A.}~\bibnamefont
  {Roggero}}\ and\ \bibinfo {author} {\bibfnamefont {J.}~\bibnamefont
  {Carlson}},\ }\href {\doibase 10.1103/PhysRevC.100.034610} {\bibfield
  {journal} {\bibinfo  {journal} {Phys. Rev. C}\ }\textbf {\bibinfo {volume}
  {100}},\ \bibinfo {pages} {034610} (\bibinfo {year} {2019})}\BibitemShut
  {NoStop}%
\bibitem [{\citenamefont {Roggero}\ \emph {et~al.}(2019)\citenamefont
  {Roggero}, \citenamefont {Li}, \citenamefont {Carlson}, \citenamefont
  {Gupta},\ and\ \citenamefont {Perdue}}]{Roggero:2019b}%
  \BibitemOpen
  \bibfield  {author} {\bibinfo {author} {\bibfnamefont {A.}~\bibnamefont
  {Roggero}}, \bibinfo {author} {\bibfnamefont {A.~C.~Y.}\ \bibnamefont {Li}},
  \bibinfo {author} {\bibfnamefont {J.}~\bibnamefont {Carlson}}, \bibinfo
  {author} {\bibfnamefont {R.}~\bibnamefont {Gupta}}, \ and\ \bibinfo {author}
  {\bibfnamefont {G.~N.}\ \bibnamefont {Perdue}},\ }\href@noop {} {\  (\bibinfo
  {year} {2019})},\ \Eprint {http://arxiv.org/abs/1911.06368} {arXiv:1911.06368
  [quant-ph]} \BibitemShut {NoStop}%
\end{thebibliography}%

\end{document}